\documentclass[12pt,a4paper]{report}
\usepackage{mathpazo}
\usepackage{amsfonts}
\usepackage[intlimits]{amsmath}
\usepackage{mathrsfs}
\usepackage{fancyhdr}
\usepackage{epsfig}

\newenvironment{narrowmargin}{
\begin{list}{}
    {
    \setlength{\topsep}{0pt}
    \setlength{\leftmargin}{1cm}
    \setlength{\rightmargin}{1cm}
    \setlength{\listparindent}{\parindent}
    \setlength{\itemindent}{\parindent}
    \setlength{\parsep}{\parskip}
    }\item[]}
{\end{list}}

\newcommand{\plaqa}{\setlength{\unitlength}{.5cm}
  \raisebox{-.2cm}{
  \begin{picture}(1.2,1.2)(-.6,-.6)
  \put(-.5,-.5){\line(1,0){1}}
  \put(.5,-.5){\line(0,1){1}}
  \put(.5,.5){\line(-1,0){1}}
  \put(-.5,.5){\line(0,-1){1}}
  \put(-.5,-.5){\circle*{.2}}
  \put(-.5,.5){\circle*{.2}}
  \put(.5,-.5){\circle*{.2}}
  \put(.5,.5){\circle*{.2}}
  \end{picture}}}

\newcommand{\nagyplaqa}{\setlength{\unitlength}{.5cm}
  \raisebox{-.35cm}{
  \begin{picture}(2.2,2.2)(-1.1,-1.1)
  \put(-1,-1){\line(1,0){2}}
  \put(-1,1){\line(1,0){2}}
  \put(-1,-1){\line(0,1){2}}
  \put(1,-1){\line(0,1){2}}
  \multiput(-1,-1)(1,0){3}{\circle*{.2}}
  \multiput(-1,1)(1,0){3}{\circle*{.2}}
  \put(-1,0){\circle*{0.2}}
  \put(1,0){\circle*{0.2}}
  \end{picture}}}

\def\tr{{\rm Tr\,}}
\def\det{{\rm det\,}}
\def\re{{\rm Re\,}}
\def\im{{\rm Im\,}}
\def\adj{{\rm adj\,}}
\def\diag{{\rm diag\,}}
\def\sign{{\rm sign\,}}
\def\dim{{\rm dim\,}}
\def\cn{{\rm cn}}
\def\sn{{\rm sn}}
\def\dn{{\rm dn}}

\def\R{{\mathbb R}}
\def\Q{{\mathbb H}}
\def\C{{\mathbb C}}
\def\Z{{\mathbb Z}}
\def\GL{GL(k,\mathbb C)}

\def\gl{gl(k,\mathbb C)}
\def\Gc{G_{\mathbb C}}
\def\gc{{\mathfrak g}_{\mathbb C}}
\def\cpone{{\mathbb C \rm P^1}}
\def\cp{\mathbb C \rm P^2}
\def\cpthree{\mathbb C \rm P^3}
\def\gotg{{\mathfrak g}}
\def\nyil{\, \longrightarrow \,}

\def\bea{\begin{eqnarray}}
\def\eea{\end{eqnarray}}
\def\nn{\nonumber}
\def\half{\frac{1}{2}}
\def\vesszo{{\, \prime}}
\def\kahler{k\"ahler }
\def\Kahler{K\"ahler }
\def\d{\partial}

\def\otwo{{\mathscr O}(2)}
\def\ok{{\mathscr O}(k)}
\def\ahat{{\hat{A}}}
\def\atilde{{\tilde{A}}}
\def\dash{\textendash\, }
\def\lmatrix{\left(\begin{array}}
\def\rmatrix{\end{array}\right)}
\def\dirac{{\,\slash\!\!\!\!D}}

\def\curly{\mathscr}
\def\x{{\mbox {\boldmath$x$}}}
\def\u{{\mbox {\boldmath$u$}}}
\def\y{{\mbox {\boldmath$y$}}}
\def\e{{\mbox {\boldmath$e$}}}
\def\n{{\mbox {\boldmath$n$}}}
\def\bomega{{\mbox {\boldmath$\omega$}}}
\def\bahat{{\mbox {\boldmath$\hat{A}$}}}
\def\batilde{{\mbox {\boldmath$\tilde{A}$}}}
\def\k{{\rm k}}
\def\cy{{\rm y}}
\def\bcy{{\rm\bf y}}

\begin{document}
\pagenumbering{gobble}

\begin{center}{\Large\bf Multi-calorons and their moduli}\\
\vspace{0.5cm}
D\'aniel N\'ogr\'adi\\
\vspace{0.5cm}
{\em Institute Lorentz for Theoretical Physics, University
of Leiden,\\ P.O. Box 9506, 2300 RA Leiden, The Netherlands}\\
\vspace{1cm}
PhD thesis\\
\vspace{0.5cm}
\end{center}

\begin{narrowmargin}
Pure Yang-Mills instantons are considered on $S^1\times \R^3$ \dash so-called calorons. The holonomy \dash or Polyakov loop
around the thermal $S^1$
at spatial infinity \dash is assumed to be a non-centre element of the gauge group $SU(n)$
as most appropriate for QCD applications in the confined phase. It is shown that
a charge $k$ caloron can be seen as a collection of $nk$ massive magnetic monopoles \dash coming in $n$ types \dash each carrying fractional
topological charge. This interpretation offers
a physically appealing way of introducing monopole degrees of freedom into pure gluodynamics: as constituents of
finite temperature instantons.

Using the Nahm transform
an elaborate treatment is given for arbitrary topological charge and new exact and explicit solutions are found for $SU(2)$ and charge 2.
The $k$ zero-modes of the Dirac operator in the background of a charge $k$ caloron are computed, and are shown to `hop' between the $n$
types of monopoles as a function of the temporal boundary condition. The abelian limit \dash where it is assumed
that the massive field components can be dropped \dash is analysed in great detail and is shown that the abelian charge
distribution of each monopole type coincides with the corresponding fermion zero-mode density.

The $4nk$ dimensional hyper\kahler
moduli space is identified as an algebraic variety, defined by four matrices obeying a constraint, modulo a natural
adjoint action. This moduli space is proved to be the same as the moduli space of stable holomorphic bundles
over the complex projective plane which are trivial on two complex lines.
A description is given for its twistor space which allows for the computation of
the exact hyper\kahler metric \dash at least in principle. 

Finally, lattice gauge theoretic applications are mentioned and is explicitly demonstrated
how to obtain calorons on the lattice using the method of cooling.
\end{narrowmargin}

\newpage

\tableofcontents
\newpage
\pagenumbering{arabic}

\chapter{Introduction}
\label{introduction}

More than 30 years after its formulation quantum chromodynamics is still not
solved, yet there is overwhelming evidence for its correctness. One of the most important phenomena of
QCD is quark confinement. It is poorly understood in terms of first principles and
yet this phenomenon is vital for our understanding of basic properties of hadronic matter and its
interactions. The primary reason
for the lack of understanding for this non-perturbative phenomenon is the fact that we do not know
how to describe the true vacuum of QCD, i.e.\ we do not know which are the ``good'' degrees of freedom to study
the dynamics in the strongly coupled infrared regime.

On the other hand perturbative calculations work reliably for short distances due to asymptotic freedom. In fact
non-abelian gauge theories are the only known examples of asymptotically free quantum field theories in four dimensions. In this
regime the free fields serve as ``good'' degrees of freedom and quantum fluctuations around the unique perturbative vacuum
are under precise calculational control.

Quark confinement is a phenomenon that is present in QCD for small enough temperatures.
As the temperature is increased a phase transition occurs at a critical value which separates the
confined and deconfined phases. Above the critical temperature the quarks are liberated \dash deconfined \dash and
together with the gluons form a quark-gluon plasma. The
actual mechanism that causes the quarks to confine below the phase transition is believed to be the same as
the mechanism at zero temperature. Even though it is not directly obvious how the limit of zero temperature
affects the microscopic description, robust phenomena such as confinement ought to survive the limit. For this very reason,
if the presence of finite temperature in QCD made it easier to address some non-perturbative
effects present at zero temperature as well, there is no reason not to formulate the theory at
non-vanishing temperature.
\newpage
\section{Gauge theories}
\label{gaugetheories}

Not only QCD but essentially all the fundamental interactions, as we know them at present, are successfully described
by gauge theories. Despite their simple formulation there are several inherent puzzling features. The basic field
of a gauge theory is not observable and does not have any physical meaning. If one wants to correct
for this and use variables which have clear physical meaning and in particular are gauge invariant then the theory quickly
becomes utterly complicated. Being forced to use gauge dependent variables leads to a whole zoo of possible computational
schemes each corresponding to different gauges. However, as gauge invariance {\em{is}} important and any physical answer should refer to only 
gauge invariant quantities, additional computations are neccessary to check the gauge independence of the results.

The seemingly innocent looking Lagrangian \dash presented below \dash hides the complexities underlying non-abelian
gauge theories. The innocent look is partly due to the fact that if some reasonable assumptions, such as
gauge invariance, locality and Lorentz invariance are imposed on a theory describing spin 1 fields then 
the Lagrangian is essentially unique and natural. 

For gauge group $SU(n)$ the dynamical variables are the 4 components of an $n\times n$ anti-hermitian matrix valued field $A_\mu(x)$,
the gauge potential. 
In terms of the field strength $F_{\mu\nu} = \d_\mu A_\nu - \d_\nu A_\mu + [ A_\mu,A_\nu]$ the Lagrangian of pure Yang-Mills theory is
\bea
\label{lagrangian}
{\curly L} = -\frac{1}{2\,g_{\rm{YM}}^2} \tr F_{\mu\nu}^2\, ,
\eea
where $g_{\rm{YM}}$ is the dimensionless coupling constant and for simplicity we assume a metric of Euclidean signature. Note
that the minus sign is included in order to make the action non-negative. The equations of motion that follow from the Lagrangian are
\bea
\label{eom}
D_\mu F_{\mu\nu} = 0\, ,
\eea
where $D_\mu = \d_\mu + A_\mu$ acts in the adjoint representation. Here we are considering a theory without matter,
which would otherwise give a source term to the right hand side.

The action, or equivalently the equations of motion, are invariant under gauge transformations. A gauge group valued field
$g(x)$ acts on the gauge potential and correspondingly on the field strength as
\bea
\label{gaugetransformation}
A_\mu &\nyil& g A_\mu g^{-1} - \d_\mu g g^{-1}\nn\\
F_{\mu\nu} &\nyil& g F_{\mu\nu} g^{-1}\, .
\eea

Another invariance of classical Yang-Mills theory in 4 dimensions is conformal symmetry. If the metric is rescaled
by an arbitrary spacetime dependent factor, the action does not change. This is because the Lagrangian
in the presence of an arbitrary \dash but still of Euclidean signature \dash metric $g_{\mu\nu}$ is
\bea
\label{metricaction}
{\curly L} = -\frac{1}{2\,g_{\rm{YM}}^2} \sqrt{\det g}\, g^{\mu\rho} g^{\nu\sigma}\tr F_{\mu\nu} F_{\rho\sigma}\,,
\eea
and if the metric is rescaled by any local factor $\lambda(x)$ as $g_{\mu\nu}\to \lambda g_{\mu\nu}$ then the
inverses change according to $g^{\mu\nu}\to \lambda^{-1} g^{\mu\nu}$, which is exactly cancelled by the change
in the volume factor $\sqrt{\det g}\to\lambda^2 \sqrt{\det g}$.

The above analysis was classical and conformal symmetry is destroyed by quantum fluctuations, it is even
broken on the perturbative level. Physics is not the same on all scales. The coupling constant changes
with the energy scale with which the system is probed, more specifically with the momentum transfer involved
\bea
\label{betafunction}
\mu \frac{d g_{\rm{YM}}}{d\mu} = \beta(g_{\rm{YM}})\,,
\eea
where $\mu$ is the renormalization scale. For non-abelian gauge theories the $\beta$-function
is negative at small coupling, resulting in a decrease of the coupling constant at large energies. This phenomenon is called asymptotic freedom and is
a direct consequence of the self-interaction of the gluons, that is of the non-abelian nature of the theory
\cite{Gross:1973id, Politzer:1973fx}.

The Lagrangian of non-abelian gauge theories we have presented is not the full Lagrangian of QCD, only of its bosonic sector.
The Dirac fermion fields $\psi$ are in the fundamental representation of the gauge group and come in $n_f$ flavours. Each flavour $f$ transforms
as $\psi_f \to g \psi_f$ under gauge transformations and it is easy to see that the full Lagrangian
\bea
\label{qcdlagrangian}
{\curly L} = -\half \tr F_{\mu\nu}^2 + \sum_f \bar{\psi}_f (\dirac-m_f) \psi_f
\eea
is also gauge invariant. The parameters $m_f$ are giving bare masses to each flavour
and $\dirac$ is the hermitian covariant Dirac operator. As long as the number of flavours is small enough the anti-screening
of charge due to the self-interaction of the gluons is over compensating the usual screening present also in abelian
theories and the $\beta$-function remains negative for small coupling.

In the massless limit a new symmetry emerges \dash at least classically. The infinitesimal transformation by
an $n_f\times n_f$ anti-hermitian matrix $\omega$,
\bea
\label{chiral}
\delta\psi_f &=& \omega_{ff^\prime} \,\gamma_5\, \psi_{f^\prime}
\eea
leaves the action (\ref{qcdlagrangian}) invariant if $m_f=0$. This chiral $U(n_f)$ symmetry
is, however, broken as a result of the quantum dynamics. More precisely, the axial $U(1)$ of $U(n_f)$ is
broken by instantons through an anomaly as we will see in the next section and the remaining $SU(n_f)$ group is broken spontaneously.
The order parameter
of the phase transition associated to the spontaneous breaking is the chiral condensate
$\langle\bar{\psi} \psi\rangle$, where an averaging over flavour is implicit. Even though classically and to every
finite order of perturbation theory it remains zero, in the full quantum theory $\langle\bar{\psi}\psi\rangle\neq 0$. A formula
due to Banks and Casher relates the chiral condensate to the spectral density $\rho(\lambda)$ of the Dirac operator around zero eigenvalue,
\bea
\label{bankscasher}
\langle\bar{\psi} \psi\rangle=\pi \lim_{m\to0} \lim_{V\to\infty} \frac{\rho(0)}{V}\,,
\eea
if the theory is formulated in finite volume $V$ and with non-vanishing masses $m$ for each flavour \cite{Banks:1979yr}. The
order of the two limits is important, first the thermodynamic limit should be taken, followed by the chiral limit. The
spectral density $\rho(\lambda) d\lambda$ counts the average number of eigenvalues of $\dirac$ between $\lambda$ and $\lambda+d\lambda$
and thus the Banks-Casher formula relates the chiral condensate to the low-lying spectrum of the Dirac operator.
We will see in the next section that instantons dramatically affect the spectrum of the Dirac operator, in particular
they give rise to zero-modes and hence are of significant importance for the phenomenology of chiral symmetry breaking.

Due to the running of the coupling constant a dimensionful parameter, $\Lambda_{QCD}$, has to emerge in the theory.
This new parameter is essentially the constant of integration that naturally appears when solving (\ref{betafunction})
and fixing $\Lambda_{QCD}$ fully specifies the theory with no adjustable parameters. In particular the coupling constant will also be fixed by
the $\beta$-function equation (\ref{betafunction}). This fashion of
trading a dimensionless coupling for a dimensionful one is called dimensional transmutation.

As we will be concerned with certain classical solutions of Yang-Mills theory, dimensional transmutation does not play a role
and we will put $g_{YM} = 1$.

\section{Instantons}
\label{instantons}

Topological excitations are special gauge configurations in Yang-Mills theory \cite{Belavin:1975fg, 'tHooft:1976fv}.
They are required to have finite action and
be stable minima of the action functional. As a result they are solutions of the equations of motion.
However, the requirement of stability puts further constraints on them besides eq.\ (\ref{eom}) and a quick
inspection of the following trick provides us with such a constraint
\bea
\label{trick}
S &=& -\frac{1}{2} \int d^4x \tr F_{\mu\nu}^2 = -\frac{1}{4}\int d^4x \tr \left( F_{\mu\nu} \pm \tilde{F}_{\mu\nu} \right)^2 \pm
\frac{1}{2}\int d^4x \tr F_{\mu\nu} \tilde{F}_{\mu\nu} =\nn\\
 &=& -\frac{1}{4}\int d^4x \tr \left( F_{\mu\nu} \pm \tilde{F}_{\mu\nu} \right)^2 \pm 8\pi^2 k\, ,
\eea
where $\tilde{F}_{\mu\nu} = \frac{1}{2} \varepsilon_{\mu\nu\rho\sigma} F_{\rho\sigma}$ stands for the dual field strength and we have
introduced the topological charge
\bea
\label{k}
k=\frac{1}{16\pi^2}\int d^4x \tr F_{\mu\nu} \tilde{F}_{\mu\nu}\, .
\eea
For the prototypical example of spacetime being $\R^4$ it is an integer once
the action is required to be finite. In this case
the field strength must go to zero at infinity and hence the gauge field must be a pure gauge $A_\mu = U^{-1} \d_\mu U$, where
$U$ is only defined on the boundary $S^3$. Such $S^3\to SU(n)$ mappings are classified up to homotopy by
an integer which is exactly given by (\ref{k}).

It follows from the above trick that $8\pi^2 |k| \leq S$, and equality is achieved if and only if
\bea
\label{selfduality}
F_{\mu\nu}=\pm \tilde{F}_{\mu\nu}\, ,
\eea
which are the celebrated (anti)self-duality equations depending on the $\pm$ sign. They are really three equations,
\bea
\label{comp}
F_{01} = \pm F_{23}\, , \qquad F_{02} = \pm F_{31}\,,\qquad F_{03} = \pm F_{12}\, .
\eea
If a configuration is (anti)self-dual then it automatically satisfies the equations of motion, however the converse
is in general not true. Self-dual configurations are called instantons if $k>0$ and anti-instantons if $k<0$.
For reviews, see \cite{Coleman:1978ae, Vainshtein:1981wh, Dorey:2002ik, Diakonov:2002fq}.

There is an alternative definition in terms of chiral fermions that is useful. Using the representation
\bea
\label{gammamatrices}
\gamma_\mu = \lmatrix{cc} 0 & -i\sigma_\mu \\ i\bar{\sigma}_\mu & 0 \rmatrix\,,
\eea
for the Dirac $\gamma$-matrices the covariant Dirac operator becomes
\bea
\label{diracoperatordecomp}
\dirac = i \gamma_\mu D_\mu = \lmatrix{cc} 0 & D \\ D^\dagger & 0 \rmatrix\,,
\eea
where we have introduced the chiral and anti-chiral Dirac operators (also called Weyl operators)
$D=\sigma_\mu D_\mu$ and $D^\dagger=-\bar{\sigma}_\mu D_\mu$. Here the $\sigma_\mu$ are the basic
quaternions; our notation is summarized at the beginning of chapter \ref{selfdual}. In this representation
the chirality operator is
\bea
\label{gamma5}
\gamma_5 = \lmatrix{rr} -1 & 0 \\ 0 & 1 \rmatrix\,,
\eea
where the blocks are $2\times 2$. It is easy to check that
\bea
\label{selfdualityfermions}
D^\dagger D = - D_\mu D_\mu -\half \bar{\eta}_{\mu\nu} F_{\mu\nu}\,,
\eea
with the familiar anti-self-dual 't Hooft tensor $\bar{\eta}_{\mu\nu}=\bar{\eta}_{\mu\nu}^i\sigma_i$. Since the contraction of a self-dual
and an anti-self-dual tensor vanishes, we have the following alternative definition: a gauge field is self-dual if and only
if the corresponding $D^\dagger D$ operator is a real quaternion and in particular commutes with the quaternions.
In this case it equals the negative of the covariant Laplacian.
For anti-self-dual fields the definition is similar with the role of $D$ and $D^\dagger$ interchanged.

Whether or not a gauge field is (anti)self-dual, its topological charge is always given by the formula (\ref{k}) and
is always an integer as long as the field strength falls off faster than $1/x^2$ for large $x$.
A non-vanishing topological charge has dramatic effect on the spectrum of the Dirac operator and will be
described below.

Since the Dirac operator anti-commutes with the chirality operator, $\{\dirac,\gamma_5\}=0$, it follows
that its real non-zero eigenvalues come in pairs. If
$\lambda$ is an eigenvalue with eigenmode $\psi$ then $\gamma_5 \psi$ is an eigenmode with eigenvalue $-\lambda$.
Also we see that if $\psi$ is a zero-mode then so is $\gamma_5\psi$ and then the combinations
$\half(1\pm\gamma_5)\psi$ are also zero-modes and are eigenmodes of $\gamma_5$ as well with eigenvalue $\pm1$. Thus in the space of normalizable zero-modes
the basis vectors can be chosen with definite chirality. Denote by $n_\pm$ the number of normalizable
zero-modes with chirality $\pm$. It follows from the explicit forms (\ref{diracoperatordecomp}-\ref{gamma5})
that in terms of the chiral and anti-chiral Dirac operators, 
$n_+ (n_-)$ is the number of normalizable zero-modes of $D\,(D^\dagger)$.

We now wish to demonstrate the sum rule $k=n_- - n_+$. To this end we
consider the quantum field theory of a massive fermion coupled to a classical gauge field \cite{Coleman:1978ae}. The action is,
\bea
\label{classicalgauge}
S= \int d^4 x\, \bar{\psi} (\dirac -m )\psi,
\eea
where the gauge field in $\dirac$ is treated classically, thus only $\psi(x)$ and $\bar{\psi}(x)$ are integrated over
to compute expectation values. The simplest chiral Ward identity in this theory states that
\bea
\label{ward}
\d_\mu \langle j_\mu^{\,5}(x)\rangle =- m \langle \bar{\psi} \gamma_5 \psi(x)\rangle  -\frac{1}{16\pi^2} \tr \tilde{F} F (x)\,,
\eea
where $j_\mu^{\,5} = \bar{\psi} \gamma_\mu \gamma_5 \psi$ is the current associated to the axial $U(1)$ transformation,
see (\ref{chiral}). The first term on the right hand side is due to
explicit chiral symmetry breaking by non-zero mass and 
the second term is the famous Adler-Bell-Jackiw anomaly \cite{Adler:1968tw, Bell:1969ts} present even in the massless limit. Integrating the
Ward identity over all of spacetime gives
\bea
\label{intward}
m \langle\int d^4 x\, \bar{\psi} \gamma_5 \psi\rangle = -k\,,
\eea
because the left hand side in (\ref{ward}) is a total derivative and since we are dealing with a 
massive theory no contribution can come from the boundary. The vacuum expectation value \dash or more precisely
the expectation value in the presence of a classical gauge field \dash on the left hand side can be computed
using the fact that
\bea
\label{propagator}
\langle \psi(x) \bar{\psi}(y) \rangle = - \frac{1}{\dirac - m}(x-y)
\eea
is the known propagator, which gives immediately
\bea
\label{vev}
\langle\int d^4 x\, \bar{\psi} \gamma_5 \psi\rangle = - \tr \left( \gamma_5 \frac{1}{\dirac - m}\right)\,.
\eea
Now we have seen that for non-zero eigenvalues the eigenmodes $\psi$ and $\gamma_5 \psi$ belong to different
eigenvalues, hence are orthogonal. As a result the evaluation of the trace is conveniently done in the
basis of eigenmodes and only the zero-modes contribute. Due to $\gamma_5$ in the trace those with chirality (+)
contribute 1, those with chirality $(-)$ contribute $-1$ leading to
\bea
\label{npluszminusz}
\langle\int d^4 x\, \bar{\psi} \gamma_5 \psi\rangle = \frac{n_+-n_-}{m},
\eea
which together with (\ref{intward}) is the desired result $k=n_--n_+$, also called the Atiyah-Singer index theorem \cite{Atiyah:1968mp,Atiyah:1971rm}.
It holds for any gauge field, whether or not it is a solution. We now specialize to instantons.

We have seen that for instantons the covariant Laplacian factorizes,
\bea
\label{factorization}
D^\dagger D = - D_\mu D_\mu
\eea
and that the number of normalizable zero-modes of $D\,(D^\dagger)$ is $n_+\,(n_-)$. Suppose that $n_+>0$. In this
case there is a normalizable zero-mode $\psi$ for $D$, $D\psi = 0$. Applying $D^\dagger$ to both sides, multiplying
by $\psi^\dagger$ and then integrating over spacetime gives
\bea
\label{laplaceintegrate}
0 = \int d^4x\,\psi^\dagger D^\dagger D \psi = \int d^4x\, \psi^\dagger (-D_\mu D_\mu) \psi = \int d^4x\, D_\mu \psi^\dagger D_\mu \psi\,,
\eea
which is only possible if $\psi$ is covariantly constant, contradicting its normalizability. Hence $n_+ =0$
and the index theorem for instantons states that $D$ has no normalizable zero-modes whereas $D^\dagger$ has
as many as the topological charge of the underlying gauge field. This result will be heavily used.

The Banks-Casher formula (\ref{bankscasher}) relates the chiral condensate to the low-lying spectrum of
the Dirac operator hence it is not surprising that instantons play a crucial role in the dynamics of
chiral symmetry breaking.

The nature of the fermionic zero-modes can be illustrated by the plots of the exact solutions.
The field strength square of a generic charge $k$ instanton looks like $k$ lumps. The
$k$ zero-mode densities $\psi^\dagger \psi$ are such that \dash after choosing an appropriate basis \dash
they peak roughly at the location of the lumps.
This harmony between the fermionic and bosonic degrees of freedom is expressed by
saying that each basic charge 1 instanton carries its own zero-mode.

\section{Finite temperature}
\label{finitetemperature}

In the previous section spacetime was $\R^4$ and had Euclidean signature. This is appropriate for a Wick rotated
Minkowski spacetime or for a finite temperature system in the limit of zero temperature. For truly finite temperature
one has to consider the imaginary time being periodic with period $1/k_BT$ where $k_B$
is Boltzmann's constant and $T$ is the temperature. Hence the manifold will be $S^1\times \R^3$
over which the (anti)self-duality equations will be studied and (anti)self-dual configurations
will be called calorons. For a comprehensive discussion of instantons in finite temperature
QCD, see \cite{Gross:1980br}, for a review of caloron solutions, see \cite{Bruckmann:2003yq}.

The motivation is the desire to understand or at least be able to say something about the phase transition of QCD.
The order parameter is the vacuum expectation value $\langle p(\x) \rangle$ of the trace of the Polyakov loop
\bea
\label{polyakovloop}
p(\x) = \frac{1}{n} \tr {\rm P} \exp \int_0^{1/k_BT} A_0(t,\x) dt\,.
\eea

For high enough temperatures \dash in the deconfined phase \dash $\langle p(\x) \rangle$ is close to one of the $n^{th}$ roots of unity, each
representing a vacuum. Clearly, the choice of any specific one out of the $n$ possibilities breaks
the $\Z_n$ symmetry associated to cyclically permuting the $n$ vacua. If $p(\x)$ is close to a root of unity then
its length is fluctuating around its maximal value $|p(\x)|\sim1$, which is only possible if the eigenvalues of $p(\x)$ are all close
to being the same.

On the other hand for low temperatures $\langle p(\x) \rangle = 0$ and the $\Z_n$-symmetry is restored. The fact
that $p(\x)$ is fluctuating around zero means that the length of the Polyakov loop is minimal. A simple exercise reveals
that this is only possible if the eigenvalues are close to being as different as possible. Let us denote
the eigenvalues by $\exp{(2\pi i \mu_A)}$, ordered as $\mu_1\leq\mu_2\leq\cdot\cdot\cdot\leq\mu_n$, then
\bea
\label{simpleexcercise}
|p(\x)|^2 = \frac{2}{n^2}\sum_{A<B} \cos 2\pi(\mu_A-\mu_B) + \frac{1}{n}\,.
\eea
Clearly, two coinciding eigenvalues give a large contribution to $|p(\x)|^2$ as $\cos 2\pi(\mu_A-\mu_B)$ attains
its maximum for $\mu_A=\mu_B$. The fluctuations in the $\Z_n$-symmetric \dash or confined \dash phase are such that the eigenvalues
repel each other.

This is our motivation for studying topological excitations \dash calorons \dash in a Polyakov loop background with
no coincident eigenvalues. Since the path ordered exponential around a closed loop is also called a holonomy, such a
Polyakov loop is also referred to as having non-trivial holonomy.

The mechanism of confinement at zero temperature should be the same as at finite temperature $T$ as
long as $T<T_c$. In addition in the real world definitely $T > 0$ and in most circumstances $T<T_c$, except
maybe at RHIC or in the early universe. Hence understanding confinement for finite temperature is perhaps not
enough to collect $\$1.000.000$, but is sufficient to understand confinement in
the real world \cite{milka}. 
Undoubtedly, studying topologically non-trivial solutions of classical Yang-Mills theory at finite temperature
will not solve or in any way explain this non-perturbative phenomenon as for large coupling or low temperature
semi-classical arguments are insufficient.
Our motivation is solely to reveal the true degrees of freedom in the
topologically non-trivial sector of QCD at non-zero temperature. We find that in the confined phase instantons dissociate into magnetic monopoles
changing the character of the basic topological object present in the QCD vacuum.

Just as instantons play a crucial non-perturbative role at zero temperature, we believe that
a similar role is played by the constituent monopoles that take the place of instantons at finite temperature and especially
in the confined phase. The true quantum dynamics of these monopoles or the
quantum dynamics of any degree of freedom for that matter is beyond
our considerations as we are working in the semi-classical regime, but we would like to stress that isolating
the ``good'' variables is the first step in formulating a dynamical model. 

Having non-trivial holonomy is essential for arriving at
massive constituent monopoles and the arguments presented above are in favour of such a scenario. One
note, however, is in order when discussing the dynamical importance of configurations with
non-trivial holonomy. It was observed that the one-loop correction to the action of configurations 
with a non-trivial asymptotic value of the Polyakov loop gives rise to an 
infinite action barrier and hence these configurations were considered irrelevant \cite{Gross:1980br}.  
However, the infinity simply arises due to the integration over the finite 
energy {\em density} induced by the perturbative fluctuations in the background
of a non-trivial Polyakov loop \cite{Weiss:1980rj}. The proper setting would therefore
rather be to calculate the non-perturbative contribution of calorons \dash with a 
given asymptotic value of the Polyakov loop \dash to this energy density, as was 
first successfully implemented in supersymmetric theories \cite{Davies:1999uw}, where 
the perturbative contribution vanishes. The resulting effective potential has a minimum where the trace of 
the Polyakov loop vanishes, i.e.\ at maximal non-trivial holonomy.

In a recent study at high temperatures, where one presumably can trust the semi-classical approximation, the non-perturbative contribution of the
monopole constituents was computed \cite{Diakonov:2004jn}. More precisely, the effective potential
due to the one-loop determinant in a caloron background was computed. When 
added to the perturbative contribution with its minima at center 
elements, a local minimum develops where the trace of the Polyakov loop 
vanishes, deepening further for decreasing temperature. This gives support 
for a phase in which the center symmetry, broken in the high temperature 
phase, is restored and provides an indication that the monopole constituents 
might be the relevant degrees of freedom in the confined phase.

Monopole based models in the spirit of a dual superconductor that are conjectured to
lead to confinement were introduced long ago \cite{Mandelstam:1974pi,'tHooft:1977hy}. Traditionally, monopoles
are static objects in non-abelian Higgs models, but such
a Higgs field is absent in QCD. Another possibility is abelian projection \cite{'tHooft:1981ht, 'tHooft:1982ns}
but in this approach magnetic monopoles
enter essentially as gauge singularities and their physical interpretation \dash i.e.\ gauge independence \dash is not so clear.
The alternative we offer to introduce monopoles into QCD, through the constituents
of finite temperature instantons, is gauge invariant and physically appealing. 

\section{Collective coordinates}
\label{collective}

The set of collective coordinates that may enter the most general instanton or caloron solution
carries important information about their physical interpretation.
Some of the parameters are interpreted as gauge orientations or phases, some others
as scales and locations \cite{Bernard:1977nr}. Exploring the whole moduli space is necessary to identify the role of every parameter.

What we find is that the $4nk$ dimensional moduli space of instantons which consists of $(4n-5)k$ gauge
orientations, $k$ 4-dimensional locations and $k$ scales is traded at finite temperature for $nk$ 3-dimensional locations and $nk$
phases. The interpreation of these parameters is clear, they describe $nk$ magnetic monopoles. A more
detailed analysis of the moduli space confirms
that this is not just an arbitrary juggling with the possible ways of factoring the number $4nk$, but
really a physically sensible reorganization of the collective coordinates takes place. In particular
a charge $k$ object is not an approximate superposition of $k$ charge 1 basic objects, but rather
an approximate superposition of $nk$ objects each with fractional topological charge.

There is an apparent puzzle that seems unavoidable following our discussion of the chiral fermion zero-modes
in the instanton background. We have seen that a generic charge $k$ instanton can be thought of as an approximate
superposition of $k$ charge 1 instantons each carrying a fermion zero-mode. If at finite temperature we have
$nk$ basic objects in a charge $k$ configuration
then how can chiral zero-modes be supported on all of them if the index theorem still dictates
that only $k$ zero-modes exist? The answer is that the $nk$ monopoles come in $n$ distinct types,
each corresponding to a $U(1)$ subgroup. In addition, the presence of finite
temperature necessitates a choice of boundary condition for the zero-modes in the compact time direction,
they can be chosen to be periodic up to an arbitrary phase, $\exp(-2\pi i z)$. According to the
Callias index theorem \cite{Callias:1977kg}, for a given choice of this phase,
say $\mu_A < z < \mu_{A+1}$, the $k$ zero-modes localize to the $k$ monopoles of type $A$ only. Whenever $\exp(2\pi iz)$
passes an eigenvalue of the Polyakov loop, the zero-modes hop from one type of monopole to the next, eventually
visiting all of them. For $z=\mu_A$ the zero-modes delocalize or spread over both types $A-1$ and $A$.

It should also be noted that for finite temperature field theory there is a canonical choice, namely that the fermions
are anti-periodic. Thus from a physical point of view $n-1$ out of the $n$ possible boundary conditions are non-physical
and perhaps this was part of the reason why constituent monopoles were not seen in lattice gauge theoretical studies in
the past. If one, however, performs simulations with the non-physical boundary conditions as well, the behaviour predicted
by the exact solutions is revealed. In this sense the zero-modes are used as probes of the underlying gauge configuration
rather than as dynamical, physical fermions.

The same comment applies to supersymmetric gauge theory compactified on $S^1\times\R^3$.\\
There is also a canonical
choice of boundary condition in this case, the (adjoint or fundamental) fermions should be periodic in order to preserve supersymmetry.
This is the right choice for computing the caloron contribution to the gluino condensate for instance in ${\curly N} = 1$
super Yang-Mills theory \cite{Davies:1999uw}. The non-physical fermions are nevertheless still there and
can be used for diagnostic purposes but do not play a role dynamically.

\section{Outline}
\label{outline}

In the following chapter we will present the well-known results on self-dual Yang-Mills fields over
flat spaces $T^p\times \R^q$ for $p+q\leq4$. The organizing principle is Nahm's duality
on the 4-torus that maps $U(n)$ instantons of topological charge $k$ on $T^4$ to $U(k)$ instantons of charge $n$
on the dual torus $\hat{T}^4$ with periods inverted. By sending some of the periods to infinity
or shrinking them to zero this approach puts the ADHM construction, BPS monopoles, 
vortices, calorons, etc. into a common framework. This will be discussed in section \ref{compactification}.

Chapter \ref{multicaloronsolutions} deals with the application of Nahm's transform for calorons with
arbitrary topological charge. Our rather general results are specialized to various limiting cases
in section \ref{asymptoticregions} that include BPS monopoles and the abelian limit when the non-abelian
cores of the massive monopoles are shrunk to zero size. Both the general formulae and the limiting
behaviour are exlicitly spelled out for $SU(2)$ and charge 2 in section \ref{charge2}.

The fermionic sector, concretely the zero-modes of the Dirac operator in the caloron
background, is investigated in chapter \ref{diracoperator}. It is shown how the zero-modes can be used
to probe the monopole content of the caloron field by varying their boundary condition in the compact
temperature direction. The abelian limit of the previous chapter is employed to achieve maximal
localization for the zero-modes. Again, our general results are made explicit for charge 2.

We also show
that upon large separation between the constituents the zero-modes ``see'' point-like monopoles.
The exact results from the previous chapter are used to resolve the singularity structure of the
abelian limit and we obtain the exact zero-modes as well.

Chapter \ref{twistors} explores the moduli space of multi-calorons. The essential tool is Euclidean twistor theory
as applied to hyper\kahler geometry. We derive an explicit parametrization in terms of
finite dimensional matrices. Using this a correspondence is established between the caloron moduli space
and the moduli space of stable holomorphic bundles over the projective plane which are trivial on two
projective lines. We also construct the corresponding twistor space which encodes the hyper\kahler
metric of the moduli. We show that upon large separation the moduli space becomes $nk$ copies of $S^1\times \R^3$ each
describing a charge 1 BPS monopole. This observation lends
support for our constituent monopole picture for arbitrary rank and charge from a geometrical point of view.

Lattice gauge theory is the natural framework to study non-perturbative phenomena in QCD and
should be decisive on dynamical questions such as the dynamical importance of our caloron solutions. Lattice
aspects of our analytical work is presented in chapter \ref{lattice}. Monte-Carlo
simulations are performed to demonstrate the confinement \dash deconfinement phase transition for $SU(2)$ and
exploratory investigations are done in the confined phase in search of calorons.

Finally, in chapter \ref{concluding} we end with a number of concluding remarks.

\chapter{Self-dual Yang-Mills fields}
\label{selfdual}

This chapter will outline the general structure of the self-duality equation $F_{\mu\nu} = \tilde{F}_{\mu\nu}$
and its various limits. Starting
with Nahm's tranformation on the 4-torus we show how to obtain a whole web of interrelated systems by shrinking some of
the periods to zero or by sending them to infinity.
The most general form of Nahm's duality considered here will relate self-duality on $\R^p\times T^q$
and $\R^{4-p-q}\times \hat{T}^q$ for $p+q\leq4$ 
where $\hat{T}^q$ is the dual torus with periods inverted. In particular this includes the algebraic ADHM
construction of instantons on $\R^4$, BPS monopoles on $\R^3$ and their relation to Nahm's equation,
calorons on $S^1\times \R^3$ and their relation to Nahm's equation with periodic boundary conditions,
vortices on $\R^2$, etc. For more detailed geometrical aspects see chapter \ref{twistors}.

The gauge group is limited to be $SU(n)$ (or $U(n)$), generalization to other classical groups are possible. In
fact some of the computations will be done in the $Sp(n)$ series for $SU(2)=Sp(1)$.

Our conventions are that for the quaternions $\sigma_\mu$ we use $(\sigma_0,\sigma_j) =(1, -i\tau_j)$
as well as $(\bar{\sigma}_0,\bar{\sigma}_j)=(1,i\tau_j)$ where 
$\tau_j$ are the usual Pauli matrices, for 't Hooft's self-dual and anti-self-dual tensors we define
\bea
\label{thoofttensors}
\eta_{\mu\nu} &=& \eta^j_{\mu\nu}\sigma_j = \half\left(\sigma_\mu \bar{\sigma}_\nu - \sigma_\nu \bar{\sigma}_\mu\right)\nn\\
\bar{\eta}_{\mu\nu} &=& \bar{\eta}^j_{\mu\nu}\sigma_j = \half \left(\bar{\sigma}_\mu \sigma_\nu - \bar{\sigma}_\nu \sigma_\mu\right)\,.
\eea
We have the identities
\bea
\label{sigmaidentities}
\sigma_\mu \bar{\sigma}_\nu + \sigma_\nu \bar{\sigma}_\mu &=& 2\delta_{\mu\nu} \nn\\
\bar{\sigma}_\mu \sigma_\nu + \bar{\sigma}_\nu \sigma_\mu &=& 2\delta_{\mu\nu} \nn\\
\bar{\sigma}_\mu \sigma_j \sigma_\nu - \bar{\sigma}_\nu \sigma_j \sigma_\mu &=& 2 \bar{\eta}^j_{\mu\nu}\,.
\eea
For a quaternion
${\rm Re}/\im$ will always mean the quaternionic real/imaginary part, that is for $q=q_\mu\sigma_\mu$ we have
$\re q = q_0$ and $\im q = q_i \sigma_i$. If a quantity, say $A_\mu$, has 4 components then without the $\mu$ index
we will always mean the corresponding quaternion $A = A_\mu \sigma_\mu$. Spatial vectors $x_i, e_i,\ldots$ with 3 indices will be bold,
$\x, \e,\ldots$ and their dot product will simply be written $\x\u, \y^2$, etc.

The various indices will be such that $\alpha, \beta, \ldots$ have values 1, 2 and are used for chiral spinors,
$a,b,c,\ldots$ are the dual gauge indices and are running from 1 to $k$, and $A,B,C,\ldots$ are the indices for
$SU(n)$ and are running from 1 to $n$.

\section{Nahm duality on the 4-torus}
\label{nahmduality}

When formulated on $T^4$ the Nahm transform \cite{Nahm:1979yw, Nahm} assigns to every generic $U(n)$ instanton of topological charge $k$, another
instanton with topological charge $n$ and gauge group $U(k)$ but living on the dual 4-torus $\hat{T}^4$ whose
periods are inverted \cite{Braam:1988qk}. A nice feature of this duality
is the constructive nature of it. Once the $U(n)$ instanton is given with charge $k$, there is a recipe
to construct the corresponding $U(k)$ instanton of charge $n$ although the actual computations
can be cumbersome. Another property is that applying it twice
gives back the original instanton, in other words the Nahm transform squares to one. In addition, since
the moduli space of instantons on $T^4$ carries a natural hyper\kahler structure, one can show that
the transformation is a hyper\kahler isometry \cite{Braam:1988qk}.

Considering various limits of the periods of $T^4$ one ends up with correspondences between objects
in a variety of dimensions and it is hoped that the magical properties of the Nahm transform survive these limits.
Maybe it is worth a note that some of the analytical properties have not been rigorously proved for all cases, but
from a physical point of view the principle is clear. The particular case of the caloron has actually
been dealt with in a mathematically sound way in \cite{Nye:2000eg, Nye:2001hf} and henceforth we will not bother with rigorous proofs.

Let us start with a $U(n)$ instanton gauge field $A_\mu(x)$ of charge $k$ on the 4-torus $T^4$ with
4 periods $2\pi L_\mu$. One can modify the gauge field in such a way that self-duality is not violated
by adding a flat factor, $A_\mu(x) \to A_\mu(x) - 2\pi i z_\mu$ where the $z_\mu$ are numbers. Indeed, such
a shift does not affect the curvature. It is possible to change $z_\mu$ to $z_\mu + n_\mu/L_\mu$ for any integers $n_\mu$
by applying a periodic $U(1)$ gauge transformation, so it is best to
think of the $z_\mu$ variables as parametrizing the dual torus $\hat{T}^4$ with periods $\hat{L}_\mu = 1/L_\mu$.

Now consider the chiral and anti-chiral Dirac operators in the fundamental representation
$D_z = \sigma_\mu \left(D_\mu-2\pi i z_\mu\right)$ and
$D_z^\dagger = -\bar{\sigma}_\mu \left(D_\mu - 2\pi i z_\mu\right)$ with $D_\mu=\d_\mu+A_\mu$. Generically
$D_z$ will have no normalizable zero-modes, whereas $D_z^\dagger$ will have $k$ of them according to
the index theorem. \footnote{Naturally, there is an obvious symmetry between $D_z$ and $D_z^\dagger$ once
the gauge field is changed from self-dual to anti-self-dual.} Denote by $\psi_z(x)\,$ the $2n\times k$
matrix of linearly independent orthonormal zero-modes
of $D_z^\dagger$,
\bea
\label{zeromodes}
D_z^\dagger \psi_z = 0\, , \qquad \int_{T^4} d^4 x\,\psi_z^\dagger \psi_z = 1\,,
\eea
with a $k\times k$ identity matrix on the right hand side.
The $\psi_z(x)$ parametrically depend on the $z_\mu$ variables and it is possible to define
\bea
\label{dualA}
\hat{A}_\mu(z) = \int_{T^4} d^4x \,\psi_z^\dagger \frac{\d}{\d z_\mu}\, \psi_z
\eea
as a $k\times k$ gauge field on the dual torus which will be referred to as the dual gauge field. Gauge transformations
arise because there is a $z$-dependent $U(k)$ choice in the $\psi_z$ matrix of zero-modes. The transformation
$\psi_z(x)\to\psi_z(x) g^{-1}(z)$ for unitary $g(z)$ induces
\bea
\label{dualgaugeinduce}
\hat{A}_\mu \nyil g \hat{A}_\mu\, g^{-1} - \frac{\d g}{\d z_\mu} \, g^{-1}\,.
\eea

It is easy to see through integration by parts that $\hat{A}_\mu$
is anti-hermitian and it is in fact also self-dual. In order to see this notice that
\bea
\label{sameargument}
D_z^\dagger D_z = - (D_\mu-2\pi i z_\mu)(D_\mu-2\pi i z_\mu) - \half \bar{\eta}_{\mu\nu} F_{\mu\nu}=-(D_\mu-2\pi i z_\mu)(D_\mu-2\pi i z_\mu)\,,
\eea
because the original field
strength is self-dual and drops out when contracted with the anti-self-dual 't Hooft tensor.
This argument is the same as our fermionic characterisation of instantons in (\ref{selfdualityfermions}).
Thus ${D_z}^\dagger D_z$ is a real quaternionic
operator and one can introduce its Green function or inverse as an $n\times n$
matrix $f_z(x,y)$ by
\bea
\label{inversegreendef}
D_z^\dagger D_z\,f_z(x,y) = \delta(x-y)\,,
\eea
where $\delta(x-y)$ is the periodic Dirac delta on $T^4$. In terms of the Green function the field strength of $\hat{A}_\mu$ is \cite{Braam:1988qk}
\bea
\label{dualF}
\hat{F}_{\mu\nu}(z) = 8\pi^2 \int_{T^4\times T^4} d^4x\, d^4y\, \psi_z^\dagger(x) f_z(x,y) \eta_{\mu\nu} \psi_z(y)\, ,
\eea
which is clearly self-dual.

It can be shown that the topological charge of $\hat{A}_\mu$ is $n$ and also that if $A_\mu(x)$ is
gauge transformed then $\hat{A}_\mu(z)$ does not change. This means that the
map $A_\mu \to \hat{A}_\mu$ is a map between gauge equivalence classes of $U(n)$ instantons of charge $k$ on $T^4$
and $U(k)$ instantons of charge $n$ on $\hat{T}^4$, which happens to be a hyper\kahler isometry. Also it holds that applying it twice
gives back the original instanton, hence the Nahm transformation is an involution.

\section{Decompactification and dimensional reduction}
\label{compactification}

The Nahm transformation on the 4-torus serves as a rich source for a whole web of interrelated models.
One way of obtaining them is decompactifying some of the periods or shrinking them to zero. In the current
section we will review self-duality over flat spaces that can be obtained as such limits of $T^4$.

If a period tends to infinity then the dual period shrinks to zero and the dual torus
is dimensionally reduced. In addition the dual field strength will not be exactly self-dual,
its anti-self-dual part will not equal zero but to some source term coming from the boundary
which will be non-trivial in the presence of non-compact directions. More specifically, the
formula (\ref{dualF}) will have an additional contribution coming from integration over the
boundary which is absent for the compact case $T^4$ and this extra contribution will
violate self-duality. One can show that these source terms are
singularities as a function of the dual variables $z_\mu$, hence self-duality
will hold almost everywhere. These singularities amount to special boundary conditions
for the dual gauge field.

On the other hand if some of the periods are reduced to zero, that is the original torus is
dimensionally reduced then the dual torus will develop non-compact directions.

From the above it is clear that the Nahm transform relates self-duality on $\R^p\times T^q$
to self-duality on $\R^{4-p-q}\times \hat{T}^q$ for $p+q\leq4$,
hence the dimensionality of the problem goes from $p+q$ to $4-p$, which in our case of the caloron \dash as well as in some other applications \dash
is a considerable simplification. Instead of solving a four dimensional problem directly we can achieve
the same by solving a problem in one dimension.

If all four periods are sent to infinity the dual torus is reduced to a single
point, and self-duality with source terms over this point will give the ADHM equations. In this
way it is possible to derive the whole ADHM construction from Nahm duality and this point of
view may help clarify the mysterious fact that self-duality on $\R^4$ is solved by an algebraic
construction \cite{Atiyah:1978ri, Drinfeld:1978xr}.

If three periods are sent to infinity and one is reduced to zero, then the original setup becomes
the BPS monopole problem on $\R^3$ \cite{Prasad:1975kr,Bogomolny:1975de}. The dual description is then on $\R$, that is Nahm's equation
with specific boundary conditions. This is the original Nahm construction of magnetic monopoles \cite{Nahm:1979yw, Nahm}.

If two periods are decompactified and two are reduced to zero, then we obtain an interesting case where
the duality is between the same type of objects, both descriptions correspond to vortices.

If we only send some of the periods to infinity but keep the remaining finite, then we obtain flat 4-dimensional
spaces. Calorons on $S^1\times \R^3$
will be related to Nahm's equation on the dual circle with periodic boundary conditions with some singularities,
doubly-periodic instantons on $\R^2\times T^2$ will have a description in terms of vortex equations
on $\hat{T}^2$ and instantons in a finite box, that is on $\R\times T^3$ will correspond
to singular monopoles on $\hat{T}^3$ \cite{vanBaal:1995eh, vanBaal:1998hm, Charbonneau:2004ak, Charbonneau}.

It is amusing to note that the manifolds $\R^2$, $\R\times T^2$ and $T^4$ are mapped topologically to themselves and
if the periods are chosen the self-dual value $L=1/\sqrt{2\pi}$ then even the metrics stay the same.

A more detailed presentation of the several variants of Nahm's transform is presented in the rest of
this section.

\subsection{ADHM construction}
\label{ADHM}

Atiyah, Drinfeld, Hitchin and Manin have given a complete recipe to construct all self-dual gauge fields on $\R^4$
with gauge group $SU(n)$ and arbitrary topological charge $k$ \cite{Atiyah:1978ri, Drinfeld:1978xr}.
For comprehensive reviews see \cite{Corrigan:1983sv,AtiyahFermi}. Even though this was the first construction of its kind
we will interpret it in the light of Nahm's duality, which appeared later.

In our notation we will follow the literature and construct the instanton gauge field $A_\mu(x)$ from some
auxiliary data. In our exposition of the Nahm transform on $T^4$ \dash again following the literature \dash we have constructed
an auxiliary instanton out of the physical $A_\mu(x)$. In this sense the ADHM construction is the analog of
the {\em inverse} Nahm transform. This is the reason for shifting the physical gauge field $A_\mu(x)$
by the auxiliary $z_\mu$ variables for Nahm's transform and \dash as we will see \dash shifting the auxiliary
gauge field $B_\mu$ by the physical $x_\mu$ variables for the ADHM construction. We hope this remark
will make it easier to relate the two constructions and clarify the logic behind the notation.

One starts with four $k\times k$ hermitian matrices $B_\mu$
combined into a matrix of quaternions $B=B_\mu \sigma_\mu$ and an $n\times k$ matrix of 2-component spinors
assembled into an $n\times 2k$ matrix $\lambda$. The analog of
the $D_z$ and $D_z^\dagger$ operators is the $(n+2k)\times 2k$ matrix
\bea
\label{delta}
\Delta(x) = \left( \begin{array}{c} \lambda \\ B-x \end{array} \right)
\eea
and its adjoint. We see the appearance of $\lambda$ in $\Delta(x)$ due to the boundary which is absent for the Nahm
transform on $T^4$. For $\R^4$ the $\lambda$-dependent terms will play
the role of a source term as already alluded to in the previous section.

An instanton solution corresponds to the matrices $(B,\lambda)$
if $\Delta^\dagger(x)\Delta(x)$ is a real quaternion, compare with (\ref{sameargument}). This condition
is independent of $x$ and is in fact equivalent to $B^\dagger B + \lambda^\dagger \lambda$ being a real quaternion,
\bea
\label{adhm0}
\im ( B^\dagger B + \lambda^\dagger \lambda ) = 0\,.
\eea
This gives 3 quadratic matrix equations for the algebraic data $(B,\lambda)$,
\bea
\label{adhm}
~[B_0,B_1]-[B_2,B_3]&=&\frac{1}{2i}\left( \lambda_1^\dagger \lambda_2 + \lambda_2^\dagger\lambda_1\right)\nn\\
~[B_0,B_2]-[B_3,B_1]&=&\half\left(\lambda_2^\dagger\lambda_1-\lambda_1^\dagger\lambda_2\right) \\
~[B_0,B_3]-[B_1,B_2]&=&\frac{1}{2i}\left(\lambda_1^\dagger\lambda_1-\lambda_2^\dagger\lambda_2\right)\,.\nn
\eea
An additional constraint is that $B-x=\sigma_\mu\left(B_\mu-x_\mu\right)$ should only be degenerate for $k$ points,
which is an open condition so will generically hold. Once such a data is given, the gauge field, field strength
and a number of other quantities of physical interest can be reconstructed explicitly.

To this end let us look for the kernel of $\Delta^\dagger(x)$ which generically will be $n$ dimensional. Choosing
$n$ normalized basis vectors gives an $(n+2k)\times n$ matrix $v(x)$ of zero-modes for which
\bea
\label{adhmzeromode}
\Delta^\dagger(x)v(x)=0\, , \qquad v^\dagger(x)v(x)=1\,.
\eea
The zero-mode matrix $v(x)$ is the analog of $\psi_z$ as defined by (\ref{zeromodes}).
The gauge field corresponding to the data $(B,\lambda)$ can be written as
\bea
\label{adhmgauge}
A_\mu(x) = v^\dagger(x) \d_\mu v(x)\, .
\eea
It is worth pointing out that the above formula defines a gauge field for any $(B,\lambda)$ matrices, however
it will only be self-dual if $(B,\lambda)$ satisfies the quadratic ADHM equations (\ref{adhm0}).

Once $(B,\lambda)$ does satisfy (\ref{adhm0}) it is clear that the transformation
\bea
\label{adhmnewdata}
B\nyil gBg^{-1}\,,\qquad \lambda\nyil\lambda g^{-1}
\eea
for $g\in U(k)$ leads to a new set of data satisfying (\ref{adhm0}). The form of the gauge field
shows that such a transformation does not change $A_\mu$. Thus it is appropriate to
associate a gauge field to equivalence classes of ADHM data where equivalence is understood with respect to
the transformations (\ref{adhmnewdata}). We conclude that the moduli space of instantons can be explicitly
parametrized by matrices $(B,\lambda)$ satisfying (\ref{adhm0}) modulo the equivalence (\ref{adhmnewdata}).

Obviously if $v(x)$ is a solution to (\ref{adhmzeromode}) then so is $v(x)g(x)^{-1}$ as long as $g(x)$ is unitary
and this will induce a gauge transformation on the gauge field (\ref{adhmgauge}).

It is possible to solve for $v(x)$ explicitly in terms of $(B,\lambda)$. Substituting directly into (\ref{adhmzeromode}) shows
that
\bea
\label{adhmu}
v(x)=\left(\begin{array}{c} -1 \\ u(x) \end{array}\right) \frac{1}{\sqrt{1+u^\dagger(x)u(x)}}\, ,\qquad u(x)=\left(B^\dagger-x^\dagger\right)^{-1} \lambda^\dagger 
\eea
is a normalized solution, where the square root of the positive $n\times n$ matrix,
\bea
\label{fifi}
\phi(x) = 1+u^\dagger(x)u(x)
\eea
is well defined
and $u(x)$ is $2k\times n$. In terms of these new variables the gauge potential becomes
\bea
\label{aphiu}
A_\mu = \phi^{-1/2}\, u^\dagger \d_\mu u \,\phi^{-1/2} + \phi^{1/2}\, \d_\mu \phi^{-1/2}\,.
\eea

The analog of the Green function is the inverse of the real quaternion $\Delta^\dagger(x)\Delta(x)$
which is an ordinary $k\times k$ matrix yielding the definition
\bea
\label{adhmgreen}
f_x &=& (\Delta^\dagger(x) \Delta(x))^{-1}\,.
\eea
Note that on $T^4$ the Green function $f_z(x,y)$ is a bona fide Green function for the second order differential operator $D_z^\dagger D_z$,
whereas in the ADHM construction it is the ordinary inverse of the matrix $\Delta^\dagger(x)\Delta(x)$. We will still
call it the Green function and in terms of this hermitian $k\times k$ matrix we have \cite{Corrigan:1983sv, Corrigan:1978ce, Osborn:1979bx}
\bea
\label{adhmformulae}
\phi^{-1} &=& 1-\lambda f_x \lambda^\dagger\nn\\
A_\mu &=& \half \phi^{1/2}\, \bar{\eta}^j_{\mu\nu}\, \d_\nu \phi_j \, \phi^{1/2} +\half \left[ \phi^{-1/2}, \d_\mu \phi^{1/2} \right]\nn\\
F_{\mu\nu} &=& 2 \phi^{-1/2}\, u^\dagger\, \eta_{\mu\nu} f_x u\, \phi^{-1/2} \nn\\
\tr F^2 &=& - \Box \Box \log \det f_x \\
\psi &=& \frac{1}{2\pi} \phi^{1/2}\, \lambda\, \d_\mu f_x\, \bar{\sigma}_\mu\, \varepsilon\nn\\
\psi^\dagger \psi &=& - \frac{1}{4\pi^2} \Box f_x \, ,\nn
\eea
where $\psi(x)$ is the $2n\times k$ matrix of $k$ normalized fundamental zero-modes of the chiral
Dirac operator in the instanton background, $\bar{\sigma}_\mu D_\mu \psi = 0$, whose existence is guaranteed by the index theorem,
$\Box$ is the four dimensional Laplacian, $\varepsilon =\sigma_2$ is the charge conjugation matrix and we have also introduced
the $n\times n$ matrices
\bea
\label{phij}
\phi_j(x)=\lambda\, \sigma_j\, f_x\, \lambda^\dagger\,.
\eea
We see that $\phi$ and $\phi_j$ completely determine the instanton gauge field.

It is a useful excercise to check the value of the total action and normalization of the fermion zero-modes. Both
the action and zero-mode densities are given as the four dimensional Laplacian of an expression, hence
the integral over 4-space can be evaluated from the asymptotics. The definition (\ref{adhmgreen}) yields for the
Green function $f_x = 1/x^2 + \cdot\cdot\cdot$ which indeed leads to $S = 8\pi^2k$ and $\int d^4x \psi^\dagger\psi = 1$
as it should.

Formulae (\ref{adhmformulae}) show that in order to perform actual calculations the Green function is a useful
instrument. Its analog for the caloron will be used extensively for finding the new exact multi-caloron
solutions.

The above construction is valid for unitary gauge groups. Other classical groups such as $Sp(n)$ or $O(n)$ can
be encorporated by considering their embeddings in higher dimensional unitary groups \cite{AtiyahFermi}. These embeddings are
of the form that the generators should preserve some
additional structure, along with the hermitian metric preserved by the unitary group.
Invoking the ADHM construction and approriately imposing these conditions
one can arrive at explicit instanton solutions for any compact semisimple Lie group. Since $SU(2)=Sp(1)$ one can apply in
this case two forms of the ADHM construction and in practice we will find it convenient to use the $Sp(1)$ realization.

In this variant of the ADHM construction the $B_\mu$ matrices are taken to be real, symmetric. The initially $n\times k$ matrix
of chiral spinors, or equivalently the $n\times 2k$ matrix $\lambda$ is taken to be a
quaternionic $k$-component row vector with real coefficients, $\lambda^a = \lambda^a_\mu \sigma_\mu$.
The symmetries of the ADHM data are modified accordingly, one has the same transformations as in (\ref{adhmnewdata}) but
only $g\in O(k)$ are allowed. The main advantage of using the $Sp(1)$ construction instead of the $SU(2)$ is that
some of the formulae simplify considerably. The quantity $\phi$ in this case is proportional to the identity matrix, hence
a scalar function of $x$. This makes it possible to simplify the formula for the gauge field to
\bea
\label{su2formulae}
A_\mu = \half \phi \,\bar{\eta}^j_{\mu\nu}\,\d_\nu \phi_j\,.
\eea

To summarize, the initial
problem of finding solutions to the self-duality equations,
which are partial differential equations in four variables for the gauge field, is turned into an algebraic
problem of finding roots in a system of quadratic equations and finding the eigenvectors of a matrix
corresponding to zero eigenvalue. In this sense the 4 dimensional problem is reduced to a zero dimensional
one.

\subsection{BPS monopoles}
\label{magnetic}

Assuming that the gauge field is invariant under translations in one of the directions of $\R^4$ one arrives
at the Bogomolny equation for magnetic monopoles. Indeed, if $\phi=A_0$ is introduced as a Higgs field and assuming
that neither $\phi$ nor $A_i$ for $i=1,2,3$ depend on $x_0$ then the self-duality equation will reduce to
\bea
\label{monopoles}
B_i = -D_i \phi \, ,
\eea
where $D_i = \d_i + A_i$ acts in the adjoint representation and $B_i = \half\varepsilon_{ijk}F_{jk}$ is the magnetic field.
The $x_0$-independent gauge transformations descend to the 3 dimensional gauge symmetry of (\ref{monopoles}),
\bea
\label{bpsgaugetr}
\phi &\nyil& g\,\phi\, g^{-1}\nn\\
A_i &\nyil& g\,A_i\,g^{-1} - \d_ig\,g^{-1}\,.
\eea
As BPS monopoles are closely related to our central object \dash the caloron \dash we will summarize some of the well-known
facts; for a review and more details see \cite{MonRing2, Hitchin:1982gh}. All of these facts will be rederived in subsequent
sections from the caloron point of view.

The finiteness of the 3-dimensional action implies that at infinity the Higgs field should
tend to a constant. The appropriate boundary condition at infinity is then
\bea
\label{monopolebc}
\phi_{AB}(r) = i \,\delta_{AB} \left(  \mu_A - \frac{l_A}{2r} + \cdot\cdot\cdot\right)\, ,\qquad \sum \mu_A = \sum l_A = 0\, ,
\eea
or any gauge transform of the above, where the $l_A$ are integers. The magnetic charge of the monopole is
then $(k_1,k_2,\ldots,k_{n-1})$ with $k_A = \sum_{B=1}^A l_B$.
The numbers $i\mu_A$ are the eigenvalues of the Higgs field at infinity.

It turns out that the fields $(\phi,A_i)$ are not the most convenient objects to describe a monopole. One can define so-called spectral
data instead, which are in a one-to-one correspondence with gauge equivalence classes of monopoles \cite{MonRing2, Hurtubise:1985vq}. For gauge
group $SU(2)$ the spectral data consists of a spectral curve which for our present introductory purposes is simply
a complex polynomial $p(\eta)$ of order $k=k_1=-k_2>0$
with leading coefficient 1 and another polynomial $q(\eta)$ of order $k-1$ such that they have no common root. These
two polynomials can be combined into a rational function, $r(\eta)=q(\eta)/p(\eta)$, and
the remarkable fact is that the rational function $r$ completely determines the monopole solution up to gauge transformations.
If it can be written in the form
\bea
\label{largeseparation}
r(\eta)=\sum_{a=1}^k \frac{c_a}{\eta-z_a}
\eea
for non-zero complex numbers $c_a=\exp(v_a + i t_a)$ and arbitrary $z_a$ then such an $r$ represents an approximate
superposition of $k$ charge 1 monopoles with $U(1)$ phases $\exp(i t_a)$ and locations $(-\half v_a,z_a)\in \R\times \C = \R^3$,
provided the separation between the locations is large enough and the choice $\mu_1=-\mu_2=1$ is made \cite{MonRing2}.

If the polynomials have expansions $p(\eta) = \eta^k + \sum_{a=0}^{k-1} p_a \eta^a$ and
$q(\eta) = \sum_{a=0}^{k-1} q_a \eta^a$ then the complex coefficients $p_0, p_1, \ldots, p_{k-1}, q_0, q_1,\ldots,q_{k-1}$
parametrize the $4k$ real dimensional moduli space. The requirement that the two polynomials should not have a
common root is expressible as $\Delta(p,q)\neq 0$, where
\bea
\label{resultant}
\Delta(p,q)=
\left| \begin{array}{ccccccc}
q_0 & q_1 & \ldots & q_{k-1} &         &   & \\
    & q_0 & q_1    & \ldots  & q_{k-1} &   & \\
    &     &  & \ldots&\ldots&\ldots  & \\
    &     &        &      &  q_0     &\ldots& q_{k-1} \\
p_0 & p_1 & \ldots & p_{k-1} & 1 & & \\
    & p_0 & p_1    & \ldots  & p_{k-1} & 1  & \\
    &     & &\ldots&\ldots&\ldots &  \\
    &     &        &   p_0   &  \ldots    &p_{k-1}& 1
\end{array}
\right|
\eea
is a $(2k-1)\times(2k-1)$ determinant, the so-called resultant of the two polynomials. This description of the moduli space
of monopoles as the complement of the algebraic variety $\Delta(p,q)=0$ in $\C^{2k}$
is useful as it provides an explicit holomorphic parametrization and with some extra work
the hyper\kahler metric can also be derived \cite{MonRing2}.

If the rank of the gauge group is larger than one then one has a similar picture with spectral data
for each $SU(2)$ embedding with constraints on the polynomials. We will revisit these questions in chapter \ref{twistors}
in more detail and will also show how the spectral data of monopoles arises from our construction
of caloron solutions and their moduli spaces.

\subsection{Vortices}
\label{vorticessection}

If the gauge field on $\R^4$ is assumed to depend only on $x_0$ and $x_1$, equations relevant for the study of
doubly periodic instantons and vortices are obtained \cite{Jardim:1999mi,Ford:2000zt,Ford:2002pa,Ford:2003vi}. In this case
there are 2 Higgs fields, $\phi_1 = A_2,\; \phi_2 = A_3$
and a 2 dimensional gauge field $A_{0,1}$ both of which can be combined into complexified fields
$\phi = \phi_1 +i\phi_2$ and $A=A_0 - i A_1$. Also, it is convenient to work in complex coordinates $z=x_0+ix_1$.
The self-duality equations reduce to
\bea
\label{vortices}
B&=&-[\phi,\bar{\phi}] \\
\label{vortices2}
D\phi&=&0
\eea
where $B=[D,\bar{D}]=\d \bar{A} - \bar{\d}A + [A,\bar{A}]$ is the curvature of $A$ and $D = \d + A$ is
in the adjoint representation. The symmetry becomes
$\phi\to g\,\phi\, g^{-1}$ and $A\to g\,A\,g^{-1} - \d g\, g^{-1}$ for $SU(n)$ valued gauge transformations
but note that eq.\ (\ref{vortices2}) is invariant under the complexified gauge group $SL(n,\C)$
whereas (\ref{vortices}) only under the original compact group.
This feature is a general phenomenon also occuring in the other dimensionally reduced examples
but perhaps is most transparent in the present case. It will be explained and exploited in chapter \ref{twistors}
in the general context of hyper\kahler geometry.

\subsection{Nahm equation}
\label{nahmequation}

The most important case for our purposes \dash for calorons \dash is the dimensional reduction to 1 dimension. Assuming
that the gauge field only depends on $x_0=t$ self-duality becomes
\bea
\label{nahm}
A_i^\vesszo+[A_0,A_i] = \half \varepsilon_{ijk} [A_j,A_k]\, ,
\eea
the celebrated Nahm equation, where prime denotes differentiation with respect to $t$. It is an ordinary
non-linear differential equation and also plays an important role in the
study of rotating rigid bodies. Gauge transformations only depend on $t$ and act as 
\footnote{An interesting observation is that if the range of $t$ is compact and periodic boundary conditions are imposed \dash
as for the caloron \dash then the
transformation of $A_0$ is the same as the coadjoint action of the centrally extended loop group familiar
from WZNW models \cite{Feher:1992yx}.}
\bea
\label{nahmgaugetr}
A_0&\nyil& g A_0 g^{-1} - g^\vesszo g^{-1} \nn\\
A_i&\nyil& g A_i g^{-1}\, .
\eea
The detailed analysis of Nahm's equation will be done in the next chapter where we use it to
construct new multiply charged caloron solutions.

\subsection{Reduction to zero dimension}
\label{reductiontozerodimension}

There is still a fourth possibility, namely to reduce to zero dimensions and assume that the gauge field does not
depend on any of the coordinates. In this case only the commutator terms in the field strength survive and
we obtain
\bea
\label{zerodim}
~[A_0,A_1]&=&[A_2,A_3] \nn\\
~[A_0,A_2]&=&[A_3,A_1] \\
~[A_0,A_3]&=&[A_1,A_2]\, , \nn
\eea
recognizing immediately the ADHM equations as introduced in section \ref{ADHM}. More precisely, the ADHM
equations are the above equations in the precence of a source given by the $\lambda$-dependent terms. The conclusion
is then clear; the ADHM construction \dash or from our point of view the Nahm transform \dash relates the four
dimensional self-duality equation and its moduli space to self-duality in zero dimensions, hence supplying
an algebraic solution to the former.

\section{Existence and obstruction}
\label{existence}

Nahm's duality transformation states that if a $U(n)$ instanton of charge $k$ exists on $T^4$, so does
a $U(k)$ instanton of charge $n$, if we identify $T^4$ with the toplogically identical $\hat{T}^4$. It follows then
immediately that there can not exist a charge one instanton on the 4-torus \cite{Braam:1988qk}. If it existed, the Nahm transform
would produce a $U(1)$ instanton of charge $n$, which is clearly impossible as $U(1)$ gauge theory is linear.

One can show that no such obstruction exists for higher charge on $T^4$ \cite{Taubes}. It is strongly believed
that on $T^3\times\R$ unit charge instantons exist, although it is not proved. For the remaining cases,
$T^2\times\R^2$, $S^1\times\R^3$ and $\R^4$ it is known that instanton solutions exist with any topological
charge. Those for $S^1\times\R^3$ will be discussed in the next chapter, with an emphasis on
higher topological charge.

\chapter{Multi-caloron solutions}
\label{multicaloronsolutions}

Nahm duality \dash see section \ref{nahmduality} \dash tells us that in order to construct multi-caloron solutions one should study Nahm's equation
on the dual circle \cite{Nahm:1983sv}. This method transforms a solution of a non-linear 4 dimensional partial differential equation
to a solution of an ordinary but still non-linear equation. The precise boundary conditions for the dual gauge
field, formulae for physically interesting quantities and other details of the construction can be obtained
in a number of ways. One could start from the Nahm transform on $T^4$, then carefully perform the limit of 3 periods tending
to infinity and trace what terms arise as sources that violate self-duality, see the comments after eq.\ (\ref{dualF}).
Another possibility is first let $T^4$ tend to $\R^4$ thereby ending up with the ADHM setup and then
compactify one direction \`a la Fourier, resulting in $S^1\times \R^3$. Yet another option is to start from
BPS monopoles on $\R^3$ with corresponding dual discription on $\R$ and compactify this $\R$
in order to have $S^1\times \R^3$ in the original setup \cite{Lee:1997vp}. The compactification will introduce the
time dependence that is absent in the pure monopole situation. These approaches are equivalent and we will
use mainly the second. 

Calorons interpolate between instantons on $\R^4$ and BPS monopoles on $\R^3$ by varying the radius of the circle
corresponding to finite temperature. Because of this it is not surprising that calorons share features with both
objects and for the actual construction one can use a mixture of the ADHM and BPS monopole methods.

We will be seeking solutions of multiple topological charge for which the asymptotic Polyakov loop, or holonomy, defined as
the path ordered exponential at spatial infinity,
\bea
\label{polyakov}
P = \lim_{|\x|\to \infty} {\rm P} \exp \int_0^\beta A_0(t,\x)dt
\eea
is a generic element with all eigenvalues $\exp(i\beta \mu_A)$ distinct. In addition, we assume an ordering
$\mu_1<\mu_2<\cdot\cdot\cdot<\mu_n$. This requirement means that the $SU(n)$ symmetry is maximally
broken to $U(1)^{n-1}$ by the holonomy or equivalently
that all the monopoles inside the caloron will have non-vanishing masses, $8\pi^2\nu_A/\beta$, where $\nu_A=\mu_{A+1}-\mu_A$. The massless limit giving
rise to so-called non-abelian clouds \cite{Lee:1996if, Lu:1998br} can be
taken in a more or less straightforward way but we will not be concerned with it here.

The solutions we obtain
will have one special feature though, they will have vanishing over-all magnetic charge.
Including an arbitrary magnetic charge $(k_1,k_2,\ldots,k_{n-1})$, as mentioned in section \ref{magnetic}, would mean that
the rank of the dual gauge field as defined on different invervals is not a constant but jumps according
to the differences $k_{A+1}-k_A$ \cite{Garland:1988bv}. The boundary conditions for these cases are also known but for the sake
of simplicity we will limit ourselves to vanishing over-all magnetic charge.
We will see that the monopole constituents come in $n$ distinguished types each being associated with a pair of
adjacent eigenvalues $(\mu_A,\mu_{A+1})$. Each of these types has a magnetic charge in the corresponding $U(1)$ subgroup,
but the total magnetic charge of the sum of all constituents will be zero however.

Lattice gauge theory considerations also justify the interest in only zero over-all magnetic charge
as the simulations are performed in a finite box. Clearly, in finite volume with periodic boundary
conditions there can be no net magnetic charge.

Without loss of generality we set the radius of the circle to unity, i.e.\ $\beta = 2\pi$ which results in
the period of the dual circle to be 1.

\section{Dual description of calorons}
\label{dualdescription}

Nahm duality tells us that we should consider the chiral and anti-chiral Dirac operators on the dual
circle parametrized by $z$ in the background of a $U(k)$ dual gauge field $\ahat$,
\bea
\label{chiralantichiralD}
D = \frac{d}{dz} + \sigma_\mu \ahat_\mu(z)\,,\qquad D^\dagger = -\frac{d}{dz} - \bar{\sigma}_\mu \ahat_\mu(z)
\eea
and the requirement of self-duality for $\ahat$ is equivalent to $D^\dagger D$ being a real quaternion. We have
also seen that since $S^1\times \R^3$ is not compact, the dual gauge field is only self-dual up to singularities
and the precise form of the singularities can be obtained by Fourier transforming the
ADHM equations \cite{Kraan:1998kp, Bruckmann:2002vy}. The source term
in the ADHM equations was $\im \lambda^\dagger \lambda$. Adding the Fourier transform of
this term to the self-duality equations for $\ahat$ yields the dual description of $SU(n)$ calorons in terms of
the $U(k)$ dual gauge field $\ahat(z)$ and an $n\times k$ matrix of 2-component spinors or equivalently a $2n\times k$ matrix $\lambda$,
\bea
\label{nahmcaloron}
\ahat^\vesszo_i + [\ahat_0,\ahat_i] - \half \varepsilon_{ijk} [\ahat_j,\ahat_k] = i \sum_A \delta(z-\mu_A) \rho_i^A\,,\qquad
\rho_j^A \sigma_j = i\,\im \lambda^\dagger P_A \lambda\,,
\eea
where the prime stands for the derivative with respect to $z$, the triplet of $k\times k$
hermitian matrices $\rho^A_j$ at each jumping point $z=\mu_A$ are the source terms and $P_A$ is the projection
to the eigenvector corresponding to the eigenvalue $\exp(2\pi i\mu_A)$ of the holonomy. A factor of $i$
in the definition of $\rho_j^A$ is included in order to make them hermitian. It is useful
to fix the gauge such that the asymptotic Polyakov loop is diagonal, in this case $\rho_i^A \sigma_i = i\,\im \bar{\lambda}^A \lambda^A$
with $\lambda^A$ meaning the $A^{th}$ row of $\lambda$, although this will not be always assumed. 

Since (\ref{nahmcaloron}) is a first order equation, the Dirac deltas on the right hand side give rise
to finite jumps,
\bea
\label{jumpppp}
\ahat_j(\mu_A+0)-\ahat_j(\mu_A-0) = i\rho_j^A
\eea
in the dual gauge field at $z=\mu_A$. For this reason the $\rho$-matrices are called the jumps.

Along with the imaginary part, the real part of $\lambda^\dagger P_A \lambda$ will also play a role and for
future use we define the hermitian $k\times k$ matrices
\bea
\label{S}
S_A = \re \lambda^\dagger P_A \lambda\,.
\eea

The similarity between eq.\ (\ref{adhm}) and (\ref{nahmcaloron}) should be clear by now and we recall the essential point once more.
They express the fact that the dual gauge field, $B_\mu$ for instantons and $\ahat_\mu(z)$ for calorons, satisfies the dimensionally reduced
self-duality equation with source terms, or equivalently the fact that the appropriate $D^\dagger D$ operators are real
quaternions. For instantons the reduction leaves only a point and the source is
simply the imaginary part of $\lambda^\dagger \lambda$ whereas for calorons the reduction leaves a circle
and the source is given as the imaginary part of $\lambda^\dagger \sum_A P_A \delta(z-\mu_A) \lambda$. 

It is worth pointing out that the Nahm equation (\ref{nahmcaloron}) also appears in the study of BPS monopoles
on $\R^3$ but in that case the range of $z$ is an open interval \cite{Nahm:1979yw, Nahm, Nahm:1981xg}. This follows from Nahm's duality 
as three periods of $T^4$ have to tend to infinity and one has to shrink to zero in order to have $\R^3$. In the dual description
this means that the dual torus reduces to $\R$, as we have seen in section \ref{compactification}. The finite jumps in the dual gauge field
are the same for BPS monopoles if the rank of the dual gauge group is the same before and after the jumping point.
Having only finite jumps for the caloron corresponds to having no over-all magnetic charge, but for pure monopoles
this is of course impossible, hence in this case the boundary conditions are different and in particular involve poles
for $\ahat_i(z)$. When we discuss in sections \ref{monopolelimit} and \ref{sunmonopolelimit} how BPS monopoles
are embedded in calorons we will show how these poles arise as boundary conditions from the caloron point of view.

We have seen that a simplified variant of the ADHM construction exists for the symplectic series which for $SU(2)=Sp(1)$
makes practical computations swifter. The requirement of $B_\mu$ being real and symmetric translates into
\bea
\label{spseries}
\ahat_\mu(-z) = \ahat_\mu(z)^T\,.
\eea
The $\lambda$ matrix in this case is a $k$-vector of quaternions, thus can be written $\lambda^a = \lambda^a_\mu \sigma_\mu$
with real coefficients $\lambda^a_\mu$. There are only 2 jumping points and we have the following restriction on
$\rho_i^A$ and $S_A$,
\bea
\label{spjumps}
{\rho_i^1}^T &=& - \rho_i^2 \,,\qquad S_1^T = S_2\,.
\eea
The condition on the jumps is easily seen to be consistent with (\ref{spseries}).

The first step in constructing the caloron solution is to solve eq.\ (\ref{nahmcaloron}). The solutions
in the bulk of the intervals $(\mu_A,\mu_{A+1})$ are solutions to the homogenous Nahm equation (\ref{nahm}) and the
Dirac deltas give finite jumps across $z=\mu_A$. First we will investigate the general structure of Nahm's equation
in the bulk of a fixed interval and then the structure of the matching conditions.

\subsection{Nahm's equation}
\label{solvingnahm}

On each interval $(\mu_A,\mu_{A+1})$ the dual $U(k)$ gauge field satisfies the homogenous Nahm equation \cite{Nahm}
\bea
\label{nahmagain}
\ahat_i^\vesszo+[\ahat_0,\ahat_i] = \half \varepsilon_{ijk} [\ahat_j,\ahat_k]\,,
\eea
which is the dimensional reduction of the self-duality equations to 1 dimension as we have seen in chapter \ref{selfdual}.
It follows that $\tr \ahat_i$ is a constant. Furthermore, taking derivatives explicitly it is easy to see that
$\tr \ahat_i \ahat_j - \delta_{ij}\tr \ahat_k \ahat_k/3$ is also
constant. More generally, $\tr \ahat_{i_1}\ldots \ahat_{i_m}$ is conserved as
long as it is made totally symmetric and traceless in the indices $i_1,\ldots,i_m$, giving rise
to constant tensors
\bea
\label{conservedCs}
C_{i_1\ldots i_m} = \frac{(-i)^m}{m!}\tr \ahat_{i_1}\ldots \ahat_{i_m} + ({\rm permutations}) - ({\rm traces})\,,
\eea
where the factor of $(-i)^m$ was introduced to make them real. Note that only the first $k$ of them are independent.
Since they are totally symmetric and traceless, $C_{i_1\ldots i_m}$ is in the spin $m$ irreducible representation
of $SO(3)$ and has $2m+1$ independent components.

Totally symmetric and traceless combinations can be simply encoded using a complex null vector $y_i$ for which $\y^2=0$.
The conservation laws are then equivalent to the statement that
$y_{i_1}\ldots y_{i_m} \tr \ahat_{i_1}\ldots \ahat_{i_m}$ is conserved
for any null vector $\y$ because the monomials $y_{i_1}\ldots y_{i_m}$ project on precisely the totally
symmetric and traceless part. Obviously scaling $\y$ with an arbitrary non-zero complex number is irrelevant
so it is best to think of $\y$ as living in a conic of $\cp$ defined by $y_1^2+y_2^2+y_3^2=0$.
Such a complex submanifold is necessarily a $\cpone$. The simplest way to see this is by the
explicit parametrization of complex null vectors \dash up to an over-all factor \dash as
\bea
\label{zetaparam}
\y = \left(\zeta,\,\frac{1-\zeta^2}{2},\,\frac{i(1+\zeta^2)}{2}\right)\,.
\eea
Here $\zeta$ is a complex number which naturally lives on the Riemann sphere.

The conservation laws thus can be expressed in a very compact form by saying that $\det (\eta+\y\bahat)$
is conserved for any complex $\eta$ and null vector $\y$
since expanding the determinant in $\eta$ will reproduce the trace of any power of $\y\bahat$ up to $k$. As
a result the algebraic curve in the variables ($\eta,\zeta$)
\bea
\label{curve}
\det\left(\eta + \frac{\ahat_2 + i\ahat_3}{2} + \zeta \ahat_1 - \zeta^2 \frac{\ahat_2-i\ahat_3}{2}\right)=0
\eea
is independent of $z$. This is called the spectral curve and has genus $(k-1)^2$.

Let us count the number of degrees of freedom. Locally $\ahat_0$ can always be gauged away. There are $3k^2$
remaining real variables in the 3 anti-hermitian $k\times k$ matrices $\ahat_i$. The conserved quantities with $m$ indices we have
found have $2m+1$ independent components and summing them from 1 to $k$ gives
a total of $k^2+2k$ conservation laws. After gauging away $\ahat_0$, constant gauge transformations are still allowed,
giving $k^2-1$ gauge parameters. Thus we are left with $3k^2-(k^2+2k)-(k^2-1) = (k-1)^2$
gauge invariant degrees of freedom to be determined. They are related to the $(k-1)^2$ globally defined holomorphic 1-forms
of the spectral curve \cite{Nahm:1982jt}.

We will see that the conserved tensors (\ref{conservedCs}) determine the long-range or abelian properties of the caloron
and the remaining constants of integration are responsible for the short-range or non-abelian behaviour.

The case $k=1$ is special because $U(1)$ is abelian. There are no commutator terms and 
the Nahm equation simply says that $\ahat_i$ is (covariantly) constant. Based on constant Nahm data the complete construction
of the most general charge 1 caloron for unitary gauge group has been derived in \cite{Kraan:1998kp,Kraan:1998pm,Kraan:1998sn,Lee:1998bb}.
For an extension to arbitrary simple groups, see \cite{Lee:1998vu}.

Even if the topological charge is greater than unity one can look for special solutions which have constant
Nahm data. Such an ansatz requires the constant $\ahat_i$ to mutually commute leading to axially symmetric solutions
for any charge \cite{Bruckmann:2002vy}. Here, however, we will be concerned with {\em generic} solutions with topological charge $k>1$. 

\subsection{Structure of the jumps}
\label{structureofthejumps}

There are $n$ jumps in the dual gauge field for gauge group $SU(n)$ and these will be related to locations of
the constituent monopoles. More precisely, the jump in $\tr \ahat_i$ at $z=\mu_A$ is $i\tr \rho_i^A$ and since
the $\x$-dependence always enters as $\ahat_i - ix_i$ on both intervals, $\tr \rho_i^A$ can be interpreted
as the shift between the center of masses of monopoles of type $A-1$ and $A$. For this reason these traces
carry a clear physical interpretation.

Solution of eq.\ (\ref{nahmcaloron}) proceeds with solving the homogenous Nahm equation in the bulk of the intervals
giving independent Nahm data on each. Then the differences at each jumping point, $\ahat_i(\mu_A+0)-\ahat_i(\mu_A-0)$,
should equal $i\rho_i^A$. However, the $\rho_i^A$ are not arbitrary matrices, but are of a special form 
\bea
\label{jumpingform}
\rho_i^A \sigma_i = i\,\im \lambda^\dagger P_A \lambda\,,
\eea
and any 3 differences $\ahat_i(\mu_A+0)-\ahat_i(\mu_A-0)$ in general can not be written in this form.
This fact is in contrast
with the situation for unit topological charge in which case arbitrary \dash and necessarily constant \dash Nahm solutions on
the two sides of the jump can be matched.

In this section we will
analyse in detail what the constraints are on the jumps $\rho_i^A$ for arbitrary topological charge \dash
and consequently on the differences $\ahat_i(\mu_A\!+0)\!-\!\ahat_i(\mu_A\!-0)$ \dash
and what their most general form is
for fixed $\tr \rho_i^A$. This will be of great practical help for the actual construction of the
caloron because with prescribed center of mass locations we will be in a position to constrain the various moduli present
in the most general solution of the homogeneous Nahm equation on each interval.

We will be concerned with a fixed jumping point $\mu_A$ only, and the $A$ index will be dropped in the remainder of this section.
A diagonal holonomy will be assumed and the $A^{th}$ row of $\lambda$ will be denoted by $\xi$. So $\xi$
is a $k$-vector of chiral spinors.

There are $4k$ real parameters in $\xi$ and $3k^2$ real parameters in the three $k\times k$ hermitian matrices $\rho_i$.
There is a $U(1)$ symmetry $\xi \to e^{ic} \xi$ that rotates the $\xi$ but does not change the jumps $\rho_i$, hence there are $4k-1$
moduli entering a jump at each jumping point. We know already that 3 of these moduli, $\tr \rho_i$, are associated to constituent locations
and our analysis will reveal the physical interpretation of the remaining $4k-4$ parameters.

Before doing so, let us mention a constraint that holds for the $\rho_i$ matrices as a direct consequence of
their definition. For a complex null vector (\ref{zetaparam}) we have $\det (y_i \tau_i) = 0$. This means
that $y_i \tau_i$ is of rank 1, thus can be written as the product of two chiral spinors. Concretely,
\bea
\label{tauy}
y_i \tau_i = \frac{i}{2} \lmatrix{r} \zeta-i \\ -\zeta-i \rmatrix \lmatrix{cc} \zeta+i, & \zeta-i \rmatrix\,,
\eea
where the fact that the two spinors are linear in $\zeta$ will become important in chapter \ref{twistors}. Now
we only wish to point out that as a result, $y_i \rho_i$ has also rank 1 and can be written as the
product of two $k$-vectors. This constrains the jumps and they must satisfy the following quadratic
equation,
\bea
\label{yrhorank1}
(y_i \rho_i )^2 = y_i y_j\, a_i \,\rho_j\,,
\eea
where we have introduced $a_i=\tr \rho_i$. To keep $a_i$ fixed, we seek the most general form allowed by the constraints for the
traceless part $\rho_i-a_i/k$ and in particular its dependence on $a_i$.

Spelling out all the indices in (\ref{jumpingform}) for diagonal holonomy yields
\bea
\label{giveshejho}
S^{ab} \delta^{\alpha\beta} - \rho_i^{ab} \tau_i^{\alpha\beta} = \bar{\xi}^a_\alpha \xi^b_\beta\,,
\eea
where the real part $S$ was introduced in (\ref{S}). It is clear that if the above equation is viewed as
an identity for $2k\times 2k$ matrices, then the requirement is that the left hand side should
be a hermitian rank 1 projector. Hermiticity is guaranteed by $S$ and $\rho_i$ being hermitian as $\tau_i$ is hermitian.
A necessary and sufficient condition for a hermitian matrix $M_{pq}$ to be of rank 1 is that
$M_{pq}M_{rs} = M_{ps} M_{rq}$. Imposing this condition on the left hand side of $(\ref{giveshejho})$ and then
decomposing the answer into its real and imaginary quaternion parts gives
\bea
\label{decomp1}
2S^{ab} S^{cd} &=& S^{cb} S^{ad} + \rho_i^{cb} \rho_i^{ad}\\
S^{ad} \rho_k^{cb} + \rho_k^{ad} S^{cb} - 2 \rho_k^{cd} S^{ab} &=& i\varepsilon_{ijk} \rho_i^{cb} \rho_j^{ad}\,.\nn
\eea
Further decomposition of the second equation with respect to symmetric and antisymmetric $(cb) - (ad)$ indexpairs
leads to
\bea
\label{decomp2}
S^{ad}\rho_i^{cb}+S^{cb}\rho_i^{ad}&=&S^{ab}\rho_i^{cd}+S^{cd}\rho_i^{ab}\\
\label{decomp3}
S^{cd}\rho_i^{ab}-S^{ab}\rho_i^{cd}&=& i \varepsilon_{ijk} \rho_i^{cb} \rho_j^{ad}\,.
\eea
Equations (\ref{decomp1}-\ref{decomp3}) are the necessary and sufficient conditions for the $\rho_i$ and $S$ matrices
to have the special form (\ref{giveshejho}) with some $\xi$. Now taking the trace in (\ref{decomp3}) with respect to
indices $ad$ and $ab$ gives two conditions that resemble a $u(2)$ algebra,
\bea
\label{resemble}
[S,\rho_k]=i \varepsilon_{ijk} \rho_i a_j\,,\qquad [\rho_i, \rho_j]=i\varepsilon_{ijk}\left(s \rho_k - a_k S\right)\,,
\eea
where $s=\tr S$. In order to really see this introduce the normalized traceless part of $\rho_i$ and $S$ as well as their
normalized traces,
\bea
\label{hulyejeloles}
E_i &=& \frac{\rho_i}{a}-\frac{e_i}{k}\,,\qquad e_i = \frac{a_i}{a}\nn\\
E_0 &=& \frac{S}{a}-\frac{e_0}{k}\,,\qquad  e_0 = \frac{s}{a}\,,
\eea
where $a$ is the length of $a_i$. Then changing variables from $E_0, E_i$ to $C_0, C_i$ by
\bea
\label{Cs}
C_i &=& \frac{1}{(e_0^2-1)^{1/2}}\left(\delta_{ij}-e_i e_j\right) E_j + \frac{e_0}{e_0^2-1} e_i e_j E_j - \frac{e_i}{e_0^2-1} E_0\,,\nn\\
C_0 &=& \frac{e_0}{e_0^2-1} E_0 - \frac{1}{e_0^2-1} e_i E_i
\eea
translates (\ref{resemble}) to
\bea
\label{u2}
[C_i,C_j]=i\varepsilon_{ijk}C_k\,,\qquad [C_0,C_i]=0\,,
\eea
which indeed shows that the $C_0, C_i$ matrices constitute a $k$ dimensional representation of $u(2)$.
Contracting the indices $ab$ in (\ref{giveshejho}) shows that $e_0\geq1$. We have assumed
for a moment that $e_0>1$, but we will see that the limit $e_0\to1$ is smooth and is relevant for the axially
symmetric solutions \cite{Bruckmann:2002vy}.

So far we have imposed condition (\ref{decomp3}) but not yet (\ref{decomp1}) and (\ref{decomp2}). A lengthy but straightforward
calculation of what these two conditions mean for the new $C_i$ and $C_0$ matrices leads to the simple result for the Casimir operator
\bea
\label{casimir}
\tr C_iC_i = \frac{3}{2}\,,
\eea
which means that the $k$ dimensional representation is the sum of a spin $\half$ and $k-2$ trivial representations, hence there exists
a unitary matrix $U$ such that
\bea
\label{unitaryanyad}
C_i = \half U \tilde{\tau}_i U^{-1}\,,
\eea
where $\tilde{\tau}_i$ is a $k\times k$ matrix with the usual Pauli matrices $\tau_i$ in the upper left $2\times 2$ block and zero elsewhere.
Because it is so sparse only the first two rows of $U$ will contribute to $C_i$. Let us denote these
two rows by the two $k$-vectors $u_\alpha$ for $\alpha=1,2$
and assemble them into a $k\times 2$ matrix $u$. The index $\alpha$ can be thought of as a chiral spinor index.
Since $U$ is unitary the $u_\alpha$ vectors are orthonormal, $u^\dagger u = 1$.
The lengthy but straightforward calculation we have referred to above also determines $C_0$ and after putting everything together
for the original variables $\rho_i$ and $S$ we have
\bea
\label{everything}
\rho_i&=& \frac{a}{2}\left(\sinh H (\delta_{ij}-e_i e_j)+ \cosh H e_i e_j \right) u \tau_i u^\dagger + a_i\, \frac{u\, u^\dagger}{2}\nn \\
S &=& \frac{1}{2} a_i u \tau_i u^\dagger + a\, \cosh H  \frac{u\, u^\dagger}{2}\,,
\eea
where we have introduced the new variable $H\geq0$ through $\cosh H = e_0$. These are our final formulae for the jumps $\rho_i$
and $S$, they must have these special forms. One can easily show by direct substitution that once $\rho_i$ is given, $S$ is completely fixed as well,
\bea
\label{skiralysag}
S=\frac{2\rho_i\rho_i+a_i\rho_i}{\sqrt{3}\sqrt{2\tr\rho_i\rho_i+a^2}}\,.
\eea
Let us count the parameters. We have $a_i$ and $H$ as 4 real moduli, in the $u_\alpha$ orthonormal vectors $2k-2$ complex moduli
but multiplication by a phase does not change $\rho_i$ nor $S$, leaving $4k-5$ real parameters, all together $4k-1$ moduli as it should be.

Let us elaborate on our result. The $u$ matrix is clearly a dual gauge moduli and apart from this and
the already familiar $a_i$ there is only one more parameter, $H$. This means that once $a_i$ is fixed by putting the
center of mass of the given type of monopoles to a prescribed location there is only the $H$ parameter to be determined
up to $U(k)$ dual gauge transformations.

Setting $H=0$ we obtain \dash after gauging away $u$ \dash the simple expressions
\bea
\label{aximaxi}
\rho_i&=&\frac{a_i}{2}\left(e_j\tilde{\tau}_j+\tilde{\tau}_0\right)\nn\\
S&=&\frac{a}{2}\left(e_j\tilde{\tau}_j+\tilde{\tau}_0\right)\,,
\eea
where $\tilde{\tau}_0$ is the $2\times 2$ identity matrix in the upper left block and zero elsewhere, similarly to $\tilde{\tau}_i$.
This shows that the matrix structure of $\rho_i$ is the same for all $i$, namely $e_j\tilde{\tau}_j+\tilde{\tau}_0$, and in particular they
mutually commute. This is precisly
the situation of the axially symmetric solutions found in \cite{Bruckmann:2002vy}. However, mutually commuting
jumps do not necessarily imply axial symmetry.

We also see that the jump in the dual gauge field $\ahat$ is limited essentially to a $u(2)$ subgroup. Most of
the components do not change across a jumping point if $k$ is large. Another way of seeing this is by noticing
that the number of parameters in the jumps grows linearly in $k$ whereas the number of components in $\ahat$ grows
quadratically.

Above we have assumed that $a\neq0$. The case of vanishing relative separation that corresponds to $a=0$ can be obtained
by taking the limit $a\to0$ and $H\to\infty$ while keeping $\half a \exp H = s$ fixed. In this case we have
\dash again after gauging away $u$ \dash the simple forms
\bea
\label{aHlimit}
\rho_i &=& \frac{s}{2}\tilde{\tau}_i\nn\\
S &=& \frac{s}{2} \tilde{\tau}_0\,.
\eea
These are the most general forms for the jumps and $S$ up to dual gauge transformations for $\tr\rho_i=0$.

Once $\ahat$ is found over the dual circle, the $n$ normalizable zero-modes of the anti-chiral Dirac operator in the background of
$\ahat_\mu(z)-ix_\mu$ have to be determined, from which $A_\mu(x)$ can be computed according to the
general prescription of the Nahm transform. Alternatively, as we have seen
for the ADHM construction, the Green function can be used and in the next section we proceed with its construction. 

\section{Green function}
\label{greenfunction}

The Green function is the inverse of $\Delta^\dagger(x)\Delta(x)$ for the ADHM construction.
Compactification from $\R^4$ to $S^1\times \R^3$ by the Fourier transform
turns eq.\ (\ref{adhmgreen}) into \cite{Nahm,Kraan:1998kp, Bruckmann:2002vy,Lee:1998bb}
\bea
\label{green}
\left[ -\left( \frac{d}{dz} + \ahat_0-it \right)^2 - \left(\bahat-i\x\right)^2 + \sum_A \delta(z-\mu_A) S_A \right] f_x(z,z^\vesszo) = \delta(z-z^\vesszo)\,,
\eea
where the $k\times k$ matrices $S_A$ are defined by (\ref{S}) for each jumping point $z=\mu_A$.
The above form of the Green function equation should not come as a surprise following our extensive discussion of Nahm duality.
On the 4-torus eq.\ (\ref{inversegreendef}) defined the Green function and
we recognize the operator $D_x^\dagger D_x$ of the (inverse) Nahm transform in the first two terms of eq.\ (\ref{green}). 
The $S$-dependent \dash and consequently $\lambda$-dependent \dash third term is the real quaternionic part of the
source $\lambda^\dagger \sum_A P_A \delta(z-\mu_A) \lambda$ and comes from the Fourier transformation of $\re \lambda^\dagger \lambda$. 
Naturally, $f_x(z,z^\vesszo)$ is periodic in $z,z^\vesszo$ just as the Dirac delta on the right
hand side in eq. (\ref{green}).

By applying a dual gauge transformation if necessary, $\ahat_0(z)$ can be assumed to be a constant without loss of generality.
Using subsequently the non-periodic gauge transformation $g(z)=\exp{(\ahat_0-it)z}$ it is possible to gauge away $\ahat_0-it$ completely and we define
\bea
\label{ftwidle}
\tilde{f}_x(z,z^\vesszo) = g(z) f_x(z,z^\vesszo) g(z^\vesszo)^{-1}\,,
\eea
however doing so introduces periodicity only up to gauge transformation by the dual holonomy $h = g(1) = \exp{(\ahat_0-it)}$,
\bea
\label{ftwidleperiod}
\tilde{f}_x(z+1,z^\vesszo) = h \tilde{f}_x(z,z^\vesszo)\,,\qquad \tilde{f}_x(z,z^\vesszo+1) = \tilde{f}_x(z,z^\vesszo) h^{-1}\,.
\eea
Along with $f_x$ one has to gauge transform accordingly all other quantities in the Green function equation (\ref{green})
and we introduce
\bea
\label{tildequantities}
\atilde_i(z) &=& g(z) \ahat_i(z) g(z)^{-1} = \exp(\ahat_0z)\,\ahat_i(z)\,\exp(-\ahat_0z)\nn\\
\tilde{S}_A &=& g(\mu_A) S_A \,g(\mu_A)^{-1} = \exp(\ahat_0\mu_A)\,S_A\,\exp(-\ahat_0\mu_A) 
\eea
thus $\atilde_i$ is only periodic up to gauge transformation by $h$ similarly to $\tilde{f}_x$. Note that
the above adjoint transformation introduces time dependence neither into $\atilde_i$ nor $S_A$. Clearly, 
since $\ahat_i$ satisfies Nahm's equation (\ref{nahmagain}) with an $\ahat_0$ term, $\atilde_i$ will satisfy
one without it,
\bea
\label{gaugefixednahm}
\atilde_i^\vesszo = \half \varepsilon_{ijk} [\atilde_j,\atilde_k]\,,
\eea
which we will call the gauge fixed Nahm equation.

The construction of $\tilde{f}_x$ follows the same methodology as the solution of Nahm's equation on the full dual
cirle. First we construct solutions on the bulk of each interval and then
take into account the boundary conditions given by the matching conditions.

\subsection{Bulk}
\label{bulk}

After gauging away $\ahat_0-it$ every column $w$ of $\tilde{f}_x$ satisfies the equation
\bea
\label{greenbulk}
\left[ \frac{d^2}{dz^2}+\left(\batilde-i\x\right)^2 \right] w = 0
\eea
in the bulk of each interval. As noted earlier, the operator on the left hand side \dash the covariant
Laplacian in one dimension \dash can be written
also as $-D_x^\dagger D_x$ with
\bea
\label{ttutt}
D_x &=& \;\;\;\frac{d}{dz}+\sigma_j\left(\atilde_j-ix_j\right) =\;\;\;\frac{d}{dz} - \tau_j\left(i\atilde_j+x_j\right)\nn\\
D_x^\dagger &=& -\frac{d}{dz}-\bar{\sigma}_j\left(\atilde_j-ix_j\right)=-\frac{d}{dz} - \tau_j\left(i\atilde_j+x_j\right)\,,
\eea
again invoking our
general argument (\ref{selfdualityfermions}). Thus it is natural
to look for solutions of $D_x \psi =0$. To avoid confusion we note that the {\em global} operator $D_x$ including
the jumping data on the whole circle as boundary conditions has no normalizable zero-modes as stated in section \ref{nahmduality}.
However, now we are seeking
{\em local} solutions to eq.\ (\ref{greenbulk}) and these are the same as {\em local} solutions to $D_x \psi =0$.
Although it may seem to imply the existence of zero-modes with the wrong chirality, these {\em local} solutions do not
give rise to {\em global}, normalizable zero-modes.

To solve $D_x \psi = 0$ we use the ansatz $\psi(z) = (1+u_i\tau_i) w(z)$ where $u_i$ is a $z$-independent unit vector and
both $u$ and $w(z)$ may depend on $\x$ \cite{Nahm:1982jt}. In fact the spinor part $1+u_i\tau_i$ is a $2\times 2$ matrix
and hence represents 2 chiral spinors but one may take any column of it as they are linearly dependent as a result of $|\u|=1$.
It is straightforward to show that once $\psi(z)$ is annihilated by $D_x$ then the corresponding $w(z)$
in the ansatz will satisfy the bulk Green function equation (\ref{greenbulk}).

The $D_x \psi = 0$ equation for the ansatz leads to
\bea
\label{ansatzw}
\left[ \frac{d}{dz} - (i\batilde+\x)\u \right] w &=& 0 \\
\label{ansatzy}
\y(i\batilde+\x) w &=& 0\,,
\eea
where we have introduced the complex vector $\y=\n^{(1)} + i \n^{(2)}$ with $\n^{(1)}, \n^{(2)}$ and $\n^{(3)}=\u$ forming a right-handed
orthonormal basis of $\R^3$, in other words $\n^{(i)} \times \n^{(j)} = \varepsilon_{ijk} \n^{(k)}$ and $\n^{(i)} \n^{(j)} = \delta_{ij}$.

Obviously there is a $U(1)$ ambiguity in defining $\y$ from $\u$ but this ambiguity $\y\to e^{ic} \y$ leaves (\ref{ansatzy})
invariant. Scaling $\y$ with an arbitrary non-zero complex number also leaves (\ref{ansatzy}) invariant, so it is best to
think of $\y$ as an element in $\cp$. Once $\y$ is given, $\u$ can be reconstructed and the fact that
$\n^{(1)}$, $\n^{(2)}$ and $\u$ are orthonormal is translated into the properties
\bea
\label{yu}
\u=i\,\frac{\y\times\bar{\y}}{\y\bar{\y}}\,,\qquad \u\times \y = -i\y\,,\qquad \u\y=0\,,\qquad \y^2 = 0\,,
\eea
so $\y$ is null.  It already appeared in the context of the spectral curve (\ref{curve}). Its reappearance here is not an accident,
eq.\ (\ref{ansatzy}) is an algebraic equation for $w$ and implies
\bea
\label{detstuff}
\det \y(i\batilde+\x) = 0
\eea
for non-trivial solutions, which is exactly the equation of the spectral curve (\ref{curve}) with the identification $\eta = -i \y\x$.
It shows that the ansatz is self-consistent,
the assumed $z$-independence of $\u$ and consequently of $\y$ is justified, since we have shown that the spectral curve
is $z$-independent.

The algebraic constraint (\ref{detstuff}) is a polynomial equation of order $2k$ for $\zeta$
if we use the parametrization (\ref{zetaparam}) and will generically have $2k$ roots. Taking its
complex conjugate we see immediately that if $\y$ is a solution, so is $\bar{\y}$.
This means that the $2k$ roots for $\zeta$ naturally come in pairs. The transformation $\y\to\bar{\y}$ induces the anti-podal map
$\zeta\to-1/\bar{\zeta}$ on $\cpone$ and the transformation $\u\to -\u$ for $\u$. We note in passing that the anti-podal map will be an important
ingredient in the twistor construction of the moduli space in chapter \ref{twistors}. 

The polynomial (\ref{detstuff}) depends on $\x$ and this will specify the $\x$-dependence of the roots $\zeta$ and
the corresponding $\y$ and $\u$. Let us label the roots as $\zeta^{(a)}$, $\zeta^{(a+k)}=-1/\bar{\zeta}^{(a)}$ and accordingly
$\y^{(a)}$, $\y^{(a+k)}=\bar{\y}^{(a)}$ as well as $\u^{(a)}$, $\u^{(a+k)}=-\u^{(a)}$. Restricting $\u$ and $\y$ to these
special values makes eq.\ (\ref{ansatzy}) solvable for non-zero $w$ which as a result has to lie in the kernel of $\y(i\batilde+\x)$.
The matrix
\bea
\label{M}
M(z)=\adj \y(i\batilde(z)+\x)
\eea
projects exactly to the kernel, where $\adj$ stands for the adjoint \dash or matrix
of subdeterminants \dash defined by $M\adj M = \det M$. Thus the solution for $w$ is
proportional to any column of $M$,
\bea
\label{solutionw}
w_a(z)=\varphi(z) M(z)_{ad} \,,
\eea
for arbitrary $d$ and with $\varphi(z)$ a scalar function that is to be determined. This 
completely solves eq.\ (\ref{ansatzy}) and now we turn to eq.\ (\ref{ansatzw}).

As a result of $\atilde$ satisfying the gauge fixed Nahm equation, it follows
that\footnote{In fact for any holomorphic function $h$, the matrix $H(z)=h(\y\batilde(z))$ satisfies the differential equation
$H^\vesszo(z) = [\u\batilde(z),H(z)]$.}
\bea
\label{MMMM}
M^\vesszo(z) = [\u(i\batilde(z)+\x),M(z)]\,.
\eea
Using this result in the substitution of (\ref{solutionw}) into (\ref{ansatzw})
gives for the function $\varphi(z)$ the differential equation
\bea
\label{phieqfinal}
\varphi^\vesszo\,  M_{cd}= \varphi\, \left(M \u(i\batilde+\x)\right)_{cd}\,,
\eea
where on both sides arbitrary $cd$ components can be taken as $M$ has rank 1. Now using
\bea
\label{mtrick}
M \u(i\batilde+\x) &=& \half [M,\u(i\batilde+\x)] + \half \{M,\u(i\batilde+\x)\} =\nn\\ 
&=& -\half M^\vesszo + \half \{M,i\u\batilde\} + \u\x M
\eea
where $\{ , \}$ is the anti-commutator, we see that $\varphi(z)$ can be factorized as
\bea
\label{chi}
\varphi(z) = \frac{\exp(z\u\x+\chi(z))}{\sqrt{M(z)_{cd}}} \,.
\eea
This defines the function $\chi(z)$ which satisfies the rather simple equation
\bea
\label{chieq}
\chi^\vesszo = \frac{\{M,i\u\batilde\}_{cd}}{2M_{cd}}\,,
\eea
which encodes all the information about the non-abelian cores as we will see shortly.

The parity transformation $\zeta\to-1/\bar{\zeta}$ or equivalently $\y\to\bar{\y}$ or $\u\to-\u$ can be
characterized by inspecting the large-$\x$ behaviour of the roots. For $|\x|\to\infty$ the polynomial equation (\ref{detstuff})
becomes $(\y\x)^k=0$ which has a $k$-fold degeneracy and has roots $\y$ that are orthogonal to $\x$. The vector $\u$
is orthogonal to $\y$ by definition thus we conclude that the solution for $\u$ in this limit is
\bea
\label{usolution}
\u = \pm \frac{\x}{|\x|}\,,
\eea
and due to the parity $\u\to -\u$ exactly $k$ roots are $\u=\x/|\x|$ and the $k$ others are $\u=-\x/|\x|$. This
analysis of the asymptotics gives us a natural characterization of the grouping of the $2k$ roots into two sets of $k$ elements, one
corresponding to $\u\x > 0$, the other to $\u\x < 0$, at least for large enough $|\x|$.
We will arrange the $2k$ roots in such a way that $\u^{(b)}\x > 0$ and $\u^{(b+k)}\x < 0$ for large enough $|\x|$.

This asymptotic region has a transparent
meaning also on the level of the bulk Green function equation (\ref{greenbulk}). For large $|\x|$ the equation
and its solutions tend to
\bea
\label{asymptgreen}
\left(-\frac{d^2}{dz^2}+\x^2\right) w = 0\,,\qquad w(z) = e^{\pm |\x|z} w_0\,,
\eea
whereas eq.\ (\ref{ansatzw}) and its solution tend to
\bea
\label{asymptw}
\left(\frac{d}{dz}-\u\x\right) w = 0\,,\qquad w(z) = e^{\x \u z} w_0\,,
\eea
for some constant $k$-vector $w_0$. This serves as a cross check that the roots $u=\pm \x/|\x|$ are
indeed correct for large $|\x|$. Also, we have
established that the roots $\u=\x/|\x|$ correspond to exponentially growing and the roots $\u=-\x/|\x|$
to exponentially decaying solutions of the bulk Green function equation in the large $|\x|$ limit.

To package all the solutions, introduce the two $k\times k$ matrices $f^\pm(z)$ with $f^-$ containing
as columns the $k$ exponentially decaying and $f^+$ containing as columns the exponentially growing
solutions $w$. Explicitly,
\bea
\label{fpfm}
f^+_{ab}(z) &=& \varphi^{(b)}(z)\, M^{(b)}_{ac}(z)\nn\\
f^-_{ab}(z) &=& \varphi^{(b+k)}(z)\, M^{(b+k)}_{ac}(z)\,,
\eea
where the column $c$ can be arbitrary and the superscripts $(b)$ and $(b+k)$ mean that the
corresponding roots should be plugged into $\varphi(z)$ and $M(z)$. 

Let us summarize what has been done. We started off looking for solutions of the bulk Green function
equation (\ref{greenbulk}) which is a second order differential equation for $k$ variables so we expect
$2k$ solutions. We have produced exactly $2k$ solutions packaged in $f^\pm(z)$, where $\varphi$ satisfies
(\ref{phieqfinal}), $M$ is defined by (\ref{M}) and the number $2k$ comes from the $2k$ possible choices for $\y$ and $\u$. In
particular the $z$-dependence of the solutions are quite similar but one should resist the temptation to
think that they are \dash as solutions \dash the same or linearly dependent. Formulae (\ref{fpfm})
represent $2k$ linearly independent, and as such the full set of solutions. The problem of solving (\ref{greenbulk})
in full generality is thus reduced to finding roots of a polynomial and solving the first order differential equation (\ref{chieq}) in
a single variable.

For the subsequent discussion on various asymptotic regions it will be useful to introduce \cite{Bruckmann:2002vy}
\bea
\label{rprm}
R^\pm(z) = \pm {f^\pm}^\vesszo(z) {f^\pm(z)}^{-1}\,,
\eea
which is easily seen to be independent of the initial condition for $f^\pm$. As a result of $f^\pm$
satisfying (\ref{greenbulk}) a Riccati type of equation holds for $R^\pm$,
\bea
\label{riccati}
\pm{R^{\,\pm}}^\vesszo(z)+ R^\pm(z)^2 = (i\batilde(z)+\x)^2\,.
\eea
From the large $|\x|$ behaviour of the columns of $f^\pm$ given in (\ref{asymptgreen})
one can deduce that $R^\pm(z)\to |\x|$ for large $|\x|$, which is indeed consistent with the above
Riccati equation.

The reason the quantities $R^\pm$ are useful is that they only contain algebraic dependence on $\x$ and when
isolating exponential terms from algebraic ones they will naturally arise. The roots $\y$ are certainly
algebraic in $\x$, so is the matrix $M=\adj \y(i\batilde+\x)$ and all the exponential dependence is contained in $\varphi$.

Let us introduce yet two more matrices $F^\pm(z)$ that take this into account by leaving out
the $\varphi$ factors from $f^\pm$, thus $F^\pm$ are simply columns of $M^{(b)}(z)$,
\bea
\label{FpFm}
F^+_{ab}(z) &=& M^{(b)}_{ac}(z)\nn\\
F^-_{ab}(z) &=& M^{(b+k)}_{ac}(z)\,,
\eea
which are clearly algebraic in $\x$ \cite{Bruckmann:2004nu}. Again, the index $c$ is arbitrary. The definitions (\ref{rprm}),
the fact that the $\varphi$ factors drop out and the labelling convention $\u^{(b+k)}=-\u^{(b)}$ lead to
\bea
\label{algebraicR}
R^\pm_{ad} = \u^{(b)}(i\batilde+\x)_{ae} F^\pm_{eb} {{F_{bd}^\pm}^{-1}}\,,
\eea
(note that the index $b$ appears 3 times and is summed over from 1 to $k$) which
shows directly that $R^\pm$ has only an algebraic dependence on $\x$ as well as on $\batilde$. This means that once $\batilde$ is known no further
integration is needed to determine $R^\pm$. It is also clear that as a result of $\u^{(b)}\x \to |\x|$ for all $b$
in the $|\x|\to\infty$ limit the above formula leads to $R^\pm\to |\x|$ as it should.

After this short intermezzo on $R^\pm$, to be used later in section \ref{asymptoticregions}, we
continue with the construction of the Green function.

\subsection{Matching at the jumping points}
\label{jumpingpoints}

The boundary conditions \dash periodicity in $z$, jumps at $z=\mu_A$ and the behaviour of $f_x(z,z^\vesszo)$
as $z\to z^\vesszo$ dictated by $\delta(z-z^\vesszo)$ in (\ref{green}) \dash have not been taken into
account yet.

In order to incorporate them properly note first that the terms $\delta(z-\mu_A) S_A$
give jumps in the derivative of $f_x(z,z^\vesszo)$ as the equation is second order. It will prove useful
to switch to a first order equation in the standard way by assembling
$\tilde{f}_x(z,z^\vesszo)$ and $\frac{d}{dz} \tilde{f}_x(z,z^\vesszo)$ into a $2k\times k$ matrix. It satisfies the equation
\bea
\label{firstorder}
\left[ \frac{d}{dz} - \lmatrix{cc} 0 & 1 \\ -(\atilde_j-i\x_j)^2+\sum_A \delta(z-\mu_A)\tilde{S}_A & 0 \rmatrix \right] \lmatrix{c} \tilde{f}_x(z,z^\vesszo) \\ \frac{d}{dz} \tilde{f}_x(z,z^\vesszo) \rmatrix = 
-\lmatrix{c} 0 \\ \delta(z-z^\vesszo) \rmatrix
\eea
and as a result has jumps in its second component. As this equation is first order, the 
solution is a path ordered exponential with known evolution between the jumps, this we have 
computed in the previous section giving matrices $f_A^\pm(z)$ for each interval.
This evolution for $\mu_A < z,z^\vesszo < \mu_{A+1}$ is given by the $2k\times 2k$ matrix
\bea
\label{kutya}
W(z,z^\vesszo)=\lmatrix{cc} f_A^+(z) & f_A^-(z) \\ {f_A^+}^\vesszo(z) & {f_A^-}^\vesszo(z) \rmatrix
\lmatrix{cc} f_A^+(z^\vesszo) & f_A^-(z^\vesszo) \\ {f_A^+}^\vesszo(z^\vesszo) & {f_A^-}^\vesszo(z^\vesszo) \rmatrix^{-1}\,.
\eea
For the full evolution from $\mu_A$ to $\mu_{A+1}$ we write $W_A = W(\mu_{A+1},\mu_A)$. Due to $\tilde{f}_x$ being
only periodic up to a dual gauge transformation, we have $W_{A+n}=h W_A h^{-1}$. Here and henceforth the products
of the type $h W_A$ where a $k\times k$ matrix multiplies a $2k\times 2k$ matrix is understood $k\times k$ blockwise.

As $z$ reaches any of the jumping points $\mu_A$ one has to insert into the evolution the $2k\times 2k$ matrix
\bea
\label{Jjump}
J_A = \exp \lmatrix{cc} 0 & 0 \\ \tilde{S}_A & 0 \rmatrix = \left(\begin{array}{cc} 1 & 0 \\ \tilde{S}_A & 1 \end{array}\right)\,,
\eea
responsible for the jump in $\frac{d}{dz}\tilde{f}_x(z,z^\vesszo)$. Thus the full evolution
from $\mu_B < z^\vesszo < \mu_{B+1}$ to $\mu_A < z < \mu_{A+1}$,
which range may now contain any number of jumping points, is given by
\bea
\label{homsol}
W(z,z^\vesszo) = W(z,\mu_A)\,J_A W_{A-1}\,J_{A-1} \cdot\cdot\cdot W_{B+1} J_{B+1} W(\mu_{B+1},z^\vesszo)\,.
\eea
Incorporating the inhomogenous right hand side proportional to $\delta(z-z^\vesszo)$ is easy, it gives
\bea
\label{krokodil}
\lmatrix{c} \tilde{f}_x(z,z^\vesszo) \\ \frac{d}{dz} \tilde{f}_x(z,z^\vesszo) \rmatrix
= W(z,z_0)\, C(z_0,z^\vesszo) - W(z,z^\vesszo) \lmatrix{c} 0 \\ \theta(z-z^\vesszo) \rmatrix\,, 
\eea
with an arbitrary $2k\times k$ matrix $C(z_0,z^\vesszo)$, arbitrary initial point $z_0$ and the step function $\theta(z-z^\vesszo)$
whose derivative is $\delta(z-z^\vesszo)$. 

All what is left is taking
into account the periodicity of $f_x(z,z^\vesszo)$ under $z\to z+1$ and $z^\vesszo\to z^\vesszo + 1$. This requires
a little care as we have gauged away $\ahat_0-it$ completely which introduced into $\tilde{f}_x(z,z^\vesszo)$
periodicity only up to gauge transformation
by the dual holonomy $h$, see (\ref{ftwidleperiod}). This requirement determines $C(z_0,z^\vesszo)$ to be
\bea
\label{C}
C(z_0,z^\vesszo) = \frac{-1}{1-h^{-1}W(z_0+1,z_0)} h^{-1}W(z_0+1,z^\vesszo) \lmatrix{c} 0 \\ 1 \rmatrix\,.
\eea
Substitution into (\ref{ftwidle}) and (\ref{krokodil})
leads to the final solution for the full, periodic and continous Green function
\bea
\label{finalgreen}
f_x(z,z^\vesszo) = g(z)^{-1}\left[-{\curly W}(z,z^\vesszo) + \theta(z^\vesszo-z) W(z,z^\vesszo) \right]_{12} g(z^\vesszo)\,,
\eea
where $g(z) = \exp{(\ahat_0-it)z}$, the subscript 12 means that from the $2k\times 2k$ matrix the upper right $k\times k$ block should be taken
and we have defined the important contribution,
\bea
\label{curlyw}
{\curly W}(z,z^\vesszo) = W(z,z_0) \left( \frac{1}{1-h^{-1}W(z_0+1,z_0)}\right) W(z_0,z^\vesszo)\,.
\eea
The choice of $z_0$ is arbitrary, different choices are convenient for different applications of formula (\ref{finalgreen}). The
factor $W(z_0+1,z_0)$ is the evolution operator over a full period and, just as any other $W$ factor, is known explicitly
in terms of $f^\pm(z)$'s  which we constructed in the previous section for each interval. All the $t$-dependence
of the Green function is in the dual holonomy $h=\exp(\ahat_0-it)$ and the factors $g(z)^{-1}, g(z^\vesszo)$.

Generally the Green function is hermitian, however, if we use the variant of the construction
specific to the symplectic series for $SU(2)=Sp(1)$ the restriction (\ref{spseries}) implies in addition that
\bea
\label{spgreen}
f_x(z,z^\vesszo)^T = f_x(-z^\vesszo,-z)\,,
\eea
which, since $\mu_2=-\mu_1$, means
\bea
\label{spgreenmu}
f_x(\mu_2,\mu_2)^T = f_x(\mu_1,\mu_1)\,,\quad f_x(\mu_1,\mu_2)^T = f_x(\mu_1,\mu_2)\,,\quad f_x(\mu_2,\mu_1)^T=f_x(\mu_2,\mu_1)\,.
\eea

\section{Gauge field}
\label{gaugefield}

One of the advantages of calculating the Green function is that the gauge field and its field strength can
be expressed entirely in terms of $f_x$ and $\lambda$. The general $SU(n)$ ADHM formulae (\ref{adhmformulae} - \ref{phij})
can be directly used for the caloron and the Fourier transform gives
\bea
\label{caloronphiphij}
\phi(x)^{-1}=
1-\sum_{A,B} P_A \lambda\, f_x(\mu_A,\mu_B)\lambda^\dagger\, P_B\,,\qquad
\phi_j(x) = \sum_{A,B} P_A \lambda\, \sigma_j\, f_x(\mu_A,\mu_B) \lambda^\dagger\, P_B\,,
\eea
which shows that we only need to know the Green function evaluated at the jumping points.
Fourier transforming the ADHM formula (\ref{adhmformulae}) of the action density yields \cite{Bruckmann:2002vy}
\bea
\label{actiondensity}
-\half \tr F_{\mu\nu}^2 = - \half \Box\Box \log \det {\curly W}(z_0,z_0)^{-1}\,,\quad {\curly W}(z_0,z_0)^{-1} = 1-h^{-1} W(z_0+1,z_0)\,,
\eea
which expression is easily seen to be independent of $z_0$ due to the determinant.

We have noted that for $Sp(1)$ the matrix $\phi$ is proportional to the identity hence a scalar function of $x$ and
that $\lambda$ in this case is a $k$-vector of quaternions, thus can be written $\lambda^a = \lambda^a_\mu \sigma_\mu$ with
real coefficients $\lambda^a_\mu$. The quantities $\phi_j$ are arbitrary quaternions. Furthermore, we have properties (\ref{spgreen} - \ref{spgreenmu}).
Now the following combinations appear in $\phi$ and $\phi_j$,
\bea
\label{su2lambdacombi}
\lambda^a \bar{\lambda}^b &=& \lambda^a_\mu \lambda^b_\mu + ( \lambda^a_\mu \lambda^b_\nu ) \eta_{\mu\nu}\nn\\
\lambda^a \sigma_j \bar{\lambda}^b &=& \lambda^a_\mu \lambda^b_\nu \bar{\eta}^j_{\mu\nu}+
\half (\lambda^a \sigma_j \bar{\lambda}^b + \lambda^b \sigma_j \bar{\lambda}^a)\,,
\eea
where the identity (\ref{sigmaidentities}) has been used. The decomposition
into symmetric and anti-symmetric parts with respect to $ab$ is useful
because $f_x(\mu_1,\mu_2)$ and $f_x(\mu_2,\mu_1)$ are symmetric and thus give zero when contracted
with the anti-symmetric parts. If the asymptotic holonomy is parametrized by $\omega_i$ as $P=\exp(2\pi i \omega_i\tau_i)$,
then we have $\mu_2=-\mu_1=|\bomega|=\omega$ and we obtain
\bea
\label{su2phi}
\phi^{-1} = 1 - \sum_A \tr f_x(\mu_A,\mu_A) S_A = 1 - 2 \tr f_x(\omega,\omega)S_2\,,
\eea
where we have used the definition (\ref{S}) of $S_A$ and the fact that $S_1^T = S_2$ as well as (\ref{spgreenmu}).

Similarly we can calculate $\phi_j$ using the decompositions (\ref{su2lambdacombi}) together with (\ref{spjumps}) and
the orthogonal projections
\bea
\label{projections}
P_1 = \half\left(1-\frac{\omega_i\tau_i}{\omega}\right)\,,\qquad P_2 = \half\left(1+\frac{\omega_i\tau_i}{\omega}\right)\,,
\eea
to obtain
\bea
\label{su2phij}
\phi_j = \half f_x^{ab}(-\omega,\omega) P_1 (\lambda^a \sigma_j \bar{\lambda}^b + \lambda^b \sigma_j \bar{\lambda}^a) P_2 - {\rm h.c.}
+ 2 \frac{\omega_i\sigma_i}{\omega} \tr f_x(\omega,\omega)\rho^2_j\,.
\eea
The usefulness of these quantities is that the gauge field assumes the simple form,
\bea
\label{gaugefieldcaloron}
A_\mu = \half \phi\, \bar{\eta}^j_{\mu\nu}\, \d_\nu \phi_j\,,
\eea
in exactly the same way as (\ref{su2formulae}) for the ADHM construction.

\section{Asymptotic regions}
\label{asymptoticregions}

In order to demonstrate the physical content of formula (\ref{finalgreen}) we define
4 sensible limits. First, the zero-mode limit in
which all but one type of monopoles are far away from $\x$. In this
limit we drop all exponential corrections coming from monopoles not of the preferred type but still keep algebraic
corrections. All contributions from the preferred type of monopoles will be kept fully. Second, we define
the $SU(2)$ monopole limit in which case also the algebraic corrections from the non-preferred types will be dropped. This
leads to quantities being sensitive exclusively to the preferred type, thus we should recover the known results for
$SU(2)$ BPS monopoles. Third, the $SU(n)$ monopole limit, when only one type of monopoles is assumed to be far and
all contributions \dash both algebraic and exponential \dash are dropped, but everything is kept from the remaining
$n-1$ types. This setting is the same as for $SU(n)$ BPS monopoles. The forth kind of limit we consider
is the abelian \dash or far field \dash limit in which case $\x$ is assumed to be far from every monopole. Exponential corrections coming
from any of the non-abelian cores will be dropped and only algebraic tails will be kept.

We emphasize that maximal symmetry breaking leads to all monopoles being massive and as a result to exponentially
decaying cores thus justifying our procedures regarding these contributions.

The structure of the Nahm formalism tells us that there is a notion of a nearest neighbour
for the various types of monopoles. This is given by the ordering in group space originating from the ordering of the
eigenvalues $\mu_1 < \mu_2 < \cdot\cdot\cdot < \mu_n$. Both in the Nahm and Green function equation only
quantities from neighbouring intervals are directly interacting through the jumping conditions. Hence one
may suspect that monopoles of type $A$ are interacting with types $A-1$ and $A+1$ differently than with all
other types. Since in our analysis we carefully separate algebraic and exponential contributions coming from the
various types of monopoles, we will be able to determine in a precise way how the interation between the
monopole types varies.

\subsection{Zero-mode limit}
\label{zeromodelimit}

We have chosen the name `zero-mode limit' because this case will be relevant for the behaviour of fermion zero-modes
which are only sensitive to one type of monopoles, see chapter \ref{diracoperator}. In this limit all types but one
are assumed to be far away and we neglect all exponentially
small contributions coming from these, but still keep all algebraic contributions \cite{Bruckmann:2003ag}. Let the preferred type be associated
to the interval $(\mu_A,\mu_{A+1})$.

The basic ingredient is the evolution operator $W(z,z^\vesszo)$ defined by (\ref{kutya}).
It can be written
\bea
\label{wtilde}
W(z,z^\vesszo) =
\lmatrix{cc} 1 & 1 \\ R_A^+(z) & - R_A^-(z) \rmatrix
\lmatrix{cc} U_A^+(z,z^\vesszo)  & 0 \\ 0 & U_A^-(z,z^\vesszo) \rmatrix
\lmatrix{cc} 1 & 1 \\ R_A^+(z^\vesszo) & - R_A^-(z^\vesszo) \rmatrix^{-1}
\eea
for $\mu_A\leq z,z^\vesszo\leq\mu_{A+1}$, where we have introduced
\bea
\label{wtildedef}
U_A^\pm(z,z^\vesszo) = f_A^\pm(z) f_A^\pm(z^\vesszo)^{-1}\,.
\eea
Since we have seen in (\ref{algebraicR}) that $R^\pm$ is algebraic, it is clear that only $U_A^\pm(z,z^\vesszo)$ contains
exponential terms in $W(z,z^\vesszo)$. In particular $U_A^\pm(z,z^\vesszo) \sim \exp\pm|\x|(z-z^\vesszo)$ for large $|\x|$.
Let us introduce furthermore $U_A^\pm$ without arguments to mean $U_A^\pm = U_A^\pm(\mu_{A+1},\mu_A)$ and
also $R_A(z)=\half (R_A^+(z)+R_A^-(z))$. For the Green function (\ref{finalgreen}) we will
need the full evolution operator $W(z_0+1,z_0)$ where we are free to make the choice $z_0=\mu_A+0$. In this
case it becomes
\bea
\label{fullwanyad}
W(\mu_A+1,\mu_A)= J_{A+n} W_{A+n-1} J_{A+n-1} \cdot\cdot\cdot J_{A+2} W_{A+1}J_{A+1}W_A\,,
\eea
where $J_{A+n} = h J_A h^{-1}$. The crucial simplification comes from the fact that for
$B\neq A$ we drop all $U_B^-$ terms as they are exponentially small, which leads to
\bea
\label{wanyad2}
W_B &=& 
\half\lmatrix{c} 1 \\ R_B^+(\mu_{B+1}) \rmatrix U_B^+ R_B(\mu_B)^{-1} 
\lmatrix{cc} R_B^-(\mu_B), & 1 \rmatrix\,, \\
W_B J_B W_{B-1} &=&
\frac{1}{4}\lmatrix{c} 1 \\ R_B^+(\mu_{B+1}) \rmatrix
U_B^+ R_B(\mu_B)^{-1}\Sigma_BU_{B-1}^+ R_{B-1}(\mu_{B-1})^{-1}
\lmatrix{cc}R_{B-1}^-(\mu_{B-1}), & 1 \rmatrix,\nn
\eea
where we have introduced $\Sigma_B = \tilde{S}_B +R_B^-(\mu_B)+R_{B-1}^+(\mu_B)$. Note that while $R_A(z)$ only carries
information about the interval $(\mu_A,\mu_{A+1})$, the term $\Sigma_B$ involves quantities from both $(\mu_{B-1},\mu_B)$
and $(\mu_B,\mu_{B+1})$.

Upon multiplying more and more terms in order to form (\ref{fullwanyad}) the pattern in (\ref{wanyad2}) continues with more
and more $U^+$'s in the middle. Eventually for the Green function (\ref{finalgreen}) the term
$(1-h^{-1}W(\mu_A+1,\mu_A))^{-1}$ is needed in the limit of large $U_B^+$ matrices for $B\neq A$. Evaluation
of the limit is done using the following formula which can be checked directly,
\bea
\label{Xab}
\lim_{K\to\infty} \left(1-\lmatrix{c}a\\ b \rmatrix K \lmatrix{cc} c & d \rmatrix \right)^{-1}= 
1-\lmatrix{c} a \\ b \rmatrix (ca+db)^{-1} \lmatrix{cc}c&d\rmatrix
\eea
for $k\times k$ matrices $a,b,c,d$ in the limit of a large and generic $k\times k$ matrix $K$. Applying this formula
for the evaluation of $(1-h^{-1}W(\mu_A +1,\mu_A))^{-1}$ with
all $U_B^+$ for $B\neq A$ contained in $K$ and
\bea
\label{apply}
\lmatrix{c} a \\ b \rmatrix = \lmatrix{c} 1 \\ \tilde{S}_A + R_{A-1}^+(\mu_A) \rmatrix\,,\quad
\lmatrix{cc} c & d \rmatrix = \lmatrix{cc} \tilde{S}_{A+1} + R_{A+1}^-(\mu_{A+1})\,, & 1 \rmatrix W_A
\eea
leads to
\bea
\label{wz0}
{\curly W}(z,z^\vesszo)=
\eea
\vskip -1cm
\bea
=W(z,\mu_A) \left(W_A^{-1}-  \lmatrix{c} 1 \\ b_{A-1} \rmatrix \left[ \lmatrix{cc} c_{A+1}&1 \rmatrix W_A \lmatrix{c} 1\\ b_{A-1}\rmatrix \right]^{-1} \lmatrix{cc} c_{A+1}&1\rmatrix \right) W(\mu_{A+1},z^\vesszo)\nn\,.
\eea
Here we introduced $b_{A-1} = \tilde{S}_A + R^+_{A-1}(\mu_A)$ and $c_{A+1} = \tilde{S}_{A+1}+ R^-_{A+1}(\mu_{A+1})$ which contain data
from the neighbouring intervals.
Using another identity that can be checked directly to hold for any matrices involved,
\bea
\label{bigtrick}
W_A^{-1}-\lmatrix{c} 1 \\ b_{A-1} \rmatrix \left[ \lmatrix{cc} c_{A+1}&1 \rmatrix W_A \lmatrix{c}1\\ b_{A-1}\rmatrix\right]^{-1}\lmatrix{cc}c_{A+1}&1\rmatrix=\nn\\
=W_A^{-1} \lmatrix{c} -1 \\ c_{A+1} \rmatrix \left[ \lmatrix{cc} b_{A-1} & -1\rmatrix W_A^{-1} \lmatrix{c}-1\\ c_{A+1}\rmatrix \right]^{-1}
\lmatrix{cc}b_{A-1} & -1\rmatrix W_A^{-1}\,,
\eea
we arrive at the final expressions for $\mu_A\leq z^\vesszo\leq z\leq \mu_{A+1}$
\bea
\label{finalex}
\tilde{f}_x(z,z^\vesszo)&=&
\lmatrix{cc} 1, & 0 \rmatrix W(z,\mu_{A+1})\lmatrix{c}-1 \\c_{A+1}\rmatrix\times\nn\\
&\times&\left[ \lmatrix{cc}b_{A-1},&-1\rmatrix W_A^{-1} \lmatrix{c}-1\\c_{A+1}\rmatrix\right]^{-1}\times\\
&\times&\lmatrix{cc}b_{A-1},&-1\rmatrix W(\mu_A,z^\vesszo)\lmatrix{c}0\\1\rmatrix\nn
\eea
or writing out explicitly,
\bea
\label{finalexplicit}
\tilde{f}_x(z,z^\vesszo)&=&\left( U_A^-(z,\mu_{A+1})-U_A^+(z,\mu_{A+1}) Z_{A+1}^r \right)\times\nn\\
&\times&\left( U_A^-(\mu_A,\mu_{A+1}) - Z_{A-1}^l U_A^+(\mu_A,\mu_{A+1}) Z_{A+1}^r \right)^{-1}\times\\
&\times&\left(U_A^-(\mu_A,z^\vesszo) - Z_{A-1}^l U_A^+(\mu_A,z^\vesszo) \right) \frac{1}{2R_A(z^\vesszo)}\,,\nn
\eea
where
\bea
\label{zfactors}
Z_{A-1}^l=1-2\Sigma_A^{-1} R_A(\mu_A)\,,\qquad Z_{A+1}^r=1-2\Sigma_{A+1}^{-1}R_A(\mu_A)\,.
\eea

Let us analyse the final formulae (\ref{finalex}-\ref{finalexplicit}) in more detail. We have started off with the approximation that
any exponentially small contribution from monopoles not of type $A$ are neglected. This means that the non-abelian
cores of all monopoles except the preferred type are pushed far away and may only contribute through their algebraic \dash or abelian \dash tails.
The non-abelian core of the preferred type should of course be visible as no approximation was made regarding that.
Hence the expectation could have been that the monopoles of type $A$ will be present fully in the final formula
whereas all other monopoles contribute algebraically. The terms $W(z,\mu_{A+1})$,
$W(\mu_A,\mu_{A+1})$ and $W(\mu_A,z^\vesszo)$
in (\ref{finalex}) contain data only from the monopoles of type $A$ and the interaction with neighbouring monopoles of type
$A-1$ and $A+1$ enter only through $b_{A-1}$ and $c_A$ in (\ref{finalex}) or through $Z_{A-1}^l$ and $Z_{A+1}^r$ in (\ref{finalexplicit})
and are algebraic. Thus monopoles not of type $A-1$, $A$ or $A+1$ contribute nothing at all in this limit, their effect
is purely exponential, once these are dropped not even algebraic tails survive.

\subsection{SU(2) monopole limit}
\label{monopolelimit}

In the $SU(2)$ monopole limit we neglect all contributions coming from monopoles not of type $A$. In the zero-mode limit
we have neglected the exponential corrections, now we have to drop the algebraic ones as well. These \dash as we have seen \dash
only come from monopoles of type $A-1$ and $A+1$. We have also seen that
these contributions are controlled by $b_{A-1} = \tilde{S}_A + R_{A-1}^+(\mu_A)$ and $c_{A+1} = \tilde{S}_{A+1} + R_{A+1}^-(\mu_{A+1})$
in (\ref{finalex}) and $Z_{A-1}^l, Z_{A+1}^r$ in (\ref{finalexplicit}). We have also established that $R_A^\pm(z)\to|\x|$
for large $|\x|$ and thus $R_{A-1}^\pm(z)\to|\bcy_{A-1}|$ for large center of mass location $|\bcy_{A-1}|$ of monopoles of type $A-1$,
and similarly $R_{A+1}^\pm(z)\to|\bcy_{A+1}|$ for monopoles of type $A+1$. Hence pushing the two types $A-1$ and $A+1$
to infinity means that
\bea
\label{bcz}
b_{A-1} \nyil |{\rm\bf y}_{A-1}|\nyil\infty\,,&&\qquad Z_{A-1}^l \nyil 1\nn\\
c_{A+1} \nyil |{\rm\bf y}_{A+1}|\nyil\infty\,,&&\qquad Z_{A+1}^r \nyil 1\,.
\eea
Inspection of (\ref{finalex}) or (\ref{finalexplicit}) gives in this limit for
$\mu_A\leq z^\vesszo\leq z\leq \mu_{A+1}$
\bea
\label{fmonopole}
\tilde{f}_x(z,z^\vesszo) = -U_A(z,\mu_{A+1})\, U_A(\mu_A,\mu_{A+1})^{-1}\, U_A(\mu_A,z^\vesszo) \frac{1}{2R_A(z^\vesszo)}\,, 
\eea
where we have introduced $U_A(z,z^\vesszo)=U_A^+(z,z^\vesszo)-U_A^-(z,z^\vesszo)$. This Green function can be computed
from data on the interval $(\mu_A,\mu_{A+1})$ only as appropriate for an $SU(2)$ BPS monopole.

We note that the quantity $U_A(z,z^\vesszo)$ in terms of which the Green function is expressed
may also be defined as the solution to
the bulk Green function equation (\ref{greenbulk}) with the boundary conditions
\bea
\label{ubc}
U_A(z,z)=0\,,\qquad\left. \frac{d}{dz}U_A(z,z^\vesszo)\right|_{z^\vesszo\to z} =
- \left. \frac{d}{dz^\vesszo} U(z,z^\vesszo)\right|_{z^\vesszo\to z} = R_A(z)\,.
\eea

The boundary conditions for the monopole Green function can easily be obtained as well. Since
$U_A(z,z)=0$ for coinciding arguments, formula (\ref{fmonopole}) implies
\bea
\label{trallala}
\tilde{f}_x(\mu_{A+1},z^\vesszo)=\tilde{f}_x(z,\mu_A)=0\,,
\eea
that is, the Green function vanishes at both ends of the interval $(\mu_A,\mu_{A+1})$,
which are precisely the right conditions for BPS monopoles \cite{Nahm:1979yw}.

Let us now derive the energy density, $\epsilon_A(\x)$, of monopoles. This will be calculated from the
caloron action density (\ref{actiondensity}) in a similar manner as the Green function above. In
approximating $\det(1-h^{-1}W(\mu_A+1,\mu_A))$ we first keep exponentially large terms $U_B^+$ to highest order only
and drop exponentially small $U_B^-$ terms for $B\neq A$. Similarly to (\ref{Xab})
the following formula may be checked directly to hold to highest order in a large matrix $K$,
\bea
\label{fasz2}
\det\left( 1 - \lmatrix{c} a \\ b \rmatrix K \lmatrix{cc} c & d \rmatrix \right) \nyil \det\left(-K(ac+bd)\right)\,,
\eea
which we will use with the same substitution as in (\ref{apply}). This leads to
\bea
\label{mmmm}
\det\curly W(\mu_A,\mu_A)^{-1} \nyil \det(-K)\,\det \left( \lmatrix{cc} c_{A+1} & 1 \rmatrix W_A \lmatrix{c} 1\\b_{A-1} \rmatrix \right)\,,
\eea
which, upon applying the logarithm, shows that the action density is a sum of two terms, the energy density
of monopoles of type $A$ and another term that tends to zero algebraically as the monopoles not of type $A$ are
pushed to infinity. Thus we arrive at
\bea
\label{monopoleE}
\epsilon_A(\x) = -\half \Delta\Delta \log \det [W_A]_{12} = -\half\Delta\Delta\log\det\left( U_A(\mu_{A+1},\mu_A) R_A(\mu_A)^{-1} \right)\,,
\eea
where again the subscript 12 indicates that the upper right $k\times k$ block should be taken from the $2k\times 2k$ matrix $W_A$.
Since no approximation is made with respect to the monopoles of type $A$, formula (\ref{monopoleE}) is the exact
energy density of $SU(2)$ BPS monopoles of charge $k$. It is a generalization of the well-known formula
$\epsilon(\x) = -1/2\,\Delta\Delta\log\left(\sinh(\nu |\x|)/|\x|\right)$ for a charge 1 monopole of mass $4\pi\nu$, located at the origin.

The asymptotic behaviour
\bea
\label{mm}
U_A(\mu_{A+1},\mu_A)=U^+_A(\mu_{A+1},\mu_A)-U^-_A(\mu_{A+1},\mu_A) \sim \exp\nu_A|\x|\,,
\eea
for large $|\x|$ assures the mass formula $\int d^3x\,\epsilon_A(\x) = 4\pi k\, \nu_A$, which, when summed over all monopoles
and integrated over the $S^1$ of unit radius gives the correct action for the caloron, $S=8\pi^2k\sum \nu_A = 8\pi^2k$.

We note in passing that it follows from our discussion of the general structure of the jumps in section \ref{structureofthejumps}. 
that once $|{\rm\bf y}_{A\pm1}|$ tends to infinity, the corresponding $\rho^A_i, \rho^{A+1}_i$ matrices also tend to infinity.
And since they determine the boundary conditions for the dual gauge field on the interval $(\mu_A,\mu_{A+1})$ a diverging
jump will ultimately mean a pole for $\ahat_i(z)$ at $z=\mu_A$ and $z=\mu_{A+1}$. Dimensional analysis
shows that since $\rho^A_i$ and $\rho^{A+1}_i$ diverge linearly in $|{\rm\bf y}_{A\pm1}|$ the divergence
in $\ahat_i(z)$ will be linear in $(z-\mu_A)^{-1}$ and $(z-\mu_{A+1})^{-1}$ thus it must develop simple poles at both
ends of the interval \cite{Nahm:1979yw}.
This is how the appropriate boundary conditions for Nahm's equation describing BPS monopoles are recovered from the caloron.
In this sense a full caloron configuration acts as a regulator for monopoles because with every type 
at a finite distance no poles are present anywhere in the computations, only in the limit of pushing some of them to infinity.

\subsection{SU(n) monopole limit}
\label{sunmonopolelimit}

In the previous section we made a detour to $SU(2)$ monopoles because they were swiftly obtained from the zero-mode
limit which was actually our primary interest. Making a similar detour to general $SU(n)$ monopoles is also possible
by moving the $n^{th}$ type to infinity resulting in rather compact and explicit expressions for the energy
density.

A caloron has net vanishing magnetic charge and removing the $n^{th}$ type \dash which, essentially, was neutralizing
the caloron \dash leaves a configuration with magnetic charge $(k,k,\ldots,k)$. The moduli space of such BPS monopoles
and its metric for $k=1$ has been fully described in \cite{Kraan:1998pn}.

Since the $n^{th}$ type of monopoles to be pushed to infinity are associated to the interval $(\mu_n,\mu_1 + 1)$ it is convenient to
choose $z_0 = \mu_1$ for the evaluation of the Green function (\ref{finalgreen}). Exponentially small $U^-_n$ terms
are dropped as usual, as well as algebraic ones in order to have no effect at all from the removed $n^{th}$ type.

Going through the same steps as before \dash using the key formula (\ref{Xab}) to perform the limit of dropping
exponentially small term with algebraic corrections still kept, then rewriting $\tilde{f}_x(z,z^\vesszo)$ analogously to (\ref{finalex})
and finally dropping the algebraic $b_n^{-1}$ and $c_n^{-1}$ terms as well \dash we conclude that the Green
function for $SU(n)$ BPS monopoles of arbitrary charge is
\bea
\label{sungreen}
\tilde{f}_x(z,z^\vesszo) = W(z,\mu_1)_{12} \left[W(\mu_1,\mu_n)_{12}\right]^{-1} W(\mu_n,z^\vesszo)_{12}\,,
\eea
where $\mu_1 \leq z^\vesszo \leq z \leq \mu_n$ and again the 12 subscript means that the upper right $k\times k$ block should be taken from the $2k\times 2k$
matrices $W$. The term $W(\mu_1,\mu_n) = W(\mu_n,\mu_1)^{-1}$ is the inverse of the full evolution operator
over the full range of $z$ from $\mu_1$ to $\mu_n$. In this way we see precisely how the dual circle of calorons
opens up into $\R$ appropriate to BPS monopoles. In the original setup this of course means that $S^1\times \R^3$
reduces to $\R^3$ and our Green function (\ref{sungreen}) is indeed static.

Regarding the boundary conditions for Nahm's equation and the Green function the same comment applies as for
the $SU(2)$ case. The jumps $\rho^1_i$ and $\rho^n_i$ at $\mu_1$ and $\mu_n$ diverge as a result of
$b_n = \tilde{S}_1 + R^+_n(\mu_1)\to|{\rm\bf y}_n|\to\infty$ and $c_n = \tilde{S}_n + R^-_n(\mu_n)\to|{\rm\bf y}_n|\to\infty$,
where ${\rm\bf y}_n$ is the center of mass location associated to the pushed away monopoles of type $n$ leading
to a simple pole for the dual gauge field, $\ahat_i(z) \sim (z-\mu_1)^{-1}$ and $\ahat_i(z) \sim (z-\mu_n)^{-1}$
as $z\to\mu_1$ and $z\to\mu_n$. The jumps at $\mu_2,\ldots,\mu_{n-1}$ remain finite.

Since $W(\mu_1,\mu_1) = W(\mu_n,\mu_n)=1$, the resulting Green function
(\ref{sungreen}) has the property $\tilde{f}_x(\mu_1,z^\vesszo) = \tilde{f}_x(z,\mu_n) = 0$
as appropriate for magnetic monopoles. Now this Green function is essentially
given in terms of the $f_A^\pm(z)$ matrices computed in section \ref{bulk} with the matching at $z=\mu_2,\ldots,\mu_{n-1}$
being taken care of by our general formalism.

The energy density of a general $SU(n)$ monopole of charge $k$ can be obtained analogously to the $SU(2)$ case
in the previous section. We arrive at
\bea
\label{mmmmmmm}
\epsilon(\x) = -\half\Delta\Delta\log\det W(\mu_n,\mu_1)_{12}\,.
\eea

This concludes our detour to BPS monopoles and we continue with the abelian limit of the caloron Green function that
will be used in the context of fermion zero-modes in chapter \ref{diracoperator}.

\subsection{Abelian limit}
\label{abelianlimit}

In this limit \dash also called far field limit \dash all exponential contributions are neglected and only algebraic tails are kept.
Since in the zero-mode limit we have dropped
already all but one type of exponential contributions, only one type is left corresponding to the interval $(\mu_A,\mu_{A+1})$.

For $\mu_A < z^\vesszo\leq z < \mu_{A+1}$ the terms  $U_A^+(z,\mu_{A+1})$ and
$U_A^+(\mu_A,z^\vesszo)$ are both exponentially decaying in (\ref{finalexplicit}), leading to
\bea
\label{abelianlimitf}
\tilde{f}_x(z,z^\vesszo) = U_A^-(z,z^\vesszo) \frac{1}{2R_A(z^\vesszo)}\,.
\eea
If $z^\vesszo < z$ this result itself is exponentially decaying so we obtain a non-vanishing Green function only if $z=z^\vesszo$ in which case it is
\bea
\label{abelianlimitff}
\tilde{f}_x(z,z) = \frac{1}{2R_A(z)}\,.
\eea
Recall that $R_A(z) = \half (R_A^+(z)+R_A^-(z))$ is purely algebraic in terms
of solutions of Nahm's equation. Once $\ahat$ is known so is the Green function, no further integration is needed.

It is also possible to determine the Green function for $z=z^\vesszo=\mu_A$ or $z=z^\vesszo=\mu_{A+1}$ in the abelian limit.
Following the same logic as before, $U_A^+(\mu_A,\mu_{A+1})$ is exponentially decaying hence can be dropped,
$U_A^\pm(\mu_A,\mu_A)=U_A^\pm(\mu_{A+1},\mu_{A+1})=1$, and the remaining growing terms cancel identically
in (\ref{finalexplicit}), which leads to the simple results
\bea
\label{bazevaze}
\tilde{f}_x(\mu_A,\mu_A) = (1-Z_{A-1}^l)\frac{1}{2R_A(\mu_A)} = \frac{1}{\Sigma_A}\,,\qquad\tilde{f}_x(\mu_{A+1},\mu_{A+1}) = \frac{1}{\Sigma_{A+1}}\,,
\eea
with $\Sigma_A = \tilde{S}_A + R^-_A(\mu_A)+R^+_{A-1}(\mu_A)$.
Note that the Green function is not continuous at the jumping points. We have seen that $\tilde{f}_x(\mu_A-0,\mu_A-0)=(2R_{A-1}(\mu_A-0))^{-1}$
and $\tilde{f}_x(\mu_A,\mu_A) = \Sigma_A^{-1}$ whereas $\tilde{f}_x(\mu_A+0,\mu_A+0)=(2R_{A+1}(\mu_A+0))^{-1}$. Naturally,
the full Green function is continuous but in the limit we are taking \dash which can be considered to be the high temperature
limit as well \dash it develops a discontinuity. The width of the transition from one side of $\mu_A$ to the other
is inversely proportional to the temperature and in the limit of infinite temperature it becomes a finite discontinuity.

In order to construct the gauge field the Green function evaluated at different jumping points is also needed.
For $z^\vesszo=\mu_A$ and $z=\mu_{A+1}$, however, the only remaining exponential term is $U_A^-(\mu_A,\mu_{A+1})^{-1}$
which is exponentially small, hence
\bea
\label{bazevaze2}
\tilde{f}_x(\mu_{A+1},\mu_A) = 0\,.
\eea
As a result, the matrices $\phi,\,\phi_j$ in (\ref{caloronphiphij}) simplify considerably in the abelian limit,
\bea
\label{abelianphiphij}
\phi^{-1}=1-\sum_A P_A\, \lambda\, \Sigma_A^{-1}\, \lambda^\dagger P_A\,,\quad
\phi_j = \sum_A P_A\, \lambda\, \sigma_j\, \Sigma_A^{-1}\, \lambda^\dagger P_A\,.
\eea
These can be used directly to compute the gauge field as in (\ref{adhmformulae}).

The action density (\ref{actiondensity})
simplifies as well. For this the choice $z_0=\mu_1$ is a convenient one and then we need to approximate $\det \curly W(\mu_1,\mu_1)^{-1}$.
Using the expressions (\ref{fullwanyad} - \ref{wanyad2}) leads to
\bea
\label{fasz}
&\curly W(\mu_1,\mu_1)^{-1}=& \\ 
&= \frac{1}{2^n} \lmatrix{c}1\\\tilde{S}_1+R_n^+(\mu_{n-1})\rmatrix U_n^+ R_n(\mu_n)^{-1}\Sigma_n \cdot\cdot\cdot 
U_1^+R_1(\mu_1)^{-1} \lmatrix{cc} R_1^-(\mu_1),\, 1 \rmatrix\,,&\nn
\eea
where all $U_A^+$ terms are exponentially growing. Approximation of the determinant 
is done in the same way as in formula (\ref{fasz2}) using the substitutions,
\bea
\label{fasz3}
\lmatrix{c} a \\ b \rmatrix = \lmatrix{c}1\\\tilde{S}_1+R_n^+(\mu_{n-1})\rmatrix\,,\quad \lmatrix{cc} c & d \rmatrix = \lmatrix{cc} R_1^-(\mu_1),\, 1 \rmatrix\,,
\eea
yielding
\bea
\label{fasz4}
\det \curly W(\mu_1,\mu_1)^{-1} = \prod_A \det U_A^+ \frac{\det \Sigma_A}{\det 2R_A(\mu_A)}\,,
\eea
finally leading to the following formula for the action density in the abelian limit,
\bea
\label{fasz5}
-\half\tr F^2_{\mu\nu} = -\half \Box\Box \log \prod_A \det U_A^+ \frac{\det \Sigma_A}{\det R_A(\mu_A)}\,.
\eea

\section{Solutions for SU(2) and charge 2}
\label{charge2}

For $k=2$ the spectral curve is a torus and as a result elliptic functions appear in the solution of Nahm's equation, which are quite manageable
for practical calculations \cite{Bruckmann:2004nu}. The complication for $k>2$ is partly due to the fact that the genus of the spectral
curve is greater than 4 and the generalization of elliptic functions to higher genus are not so elementary \cite{Nahm:1982jt}.
It should be stressed that conceptually \dash and qualitatively \dash there is no major difference between $k=2$ and $k>2$,
except for writing explicit formulae. Thus our detailed analysis of double topological charge is illustrative of the
general multi-charge case and henceforth we set $k=2$.

The general strategy has been outlined before and we will use the symplectic formulation.
First we solve Nahm's equation on a fixed interval, then deal
with the matching at the jumping points and finally solve for the Green function.
As the gauge group in this section is $SU(2)$, the asymptotic Polyakov loop will
be parametrized as $P = \exp (2\pi i \omega_i \tau_i)$. There will be two intervals
and we have $\mu_1 = -\omega$ and $\mu_2 = \omega$, where $\omega = |\bomega|$.

\subsection{Bulk}
\label{charge2bulk}

We have seen that the traces $\tr \atilde_i$ are conserved and essentially decouple from the $su(2)$ traceless part.
They parametrize the center of mass and henceforth $\tr \atilde_i = 0$ will be assumed without loss of generality.
We have also seen that the traceless part (in the $i,j$ indices) of $\tr \atilde_i \atilde_j$ is conserved, so
introducing $F(z)=-\frac{1}{6} \tr \atilde_i(z)\atilde_i(z)$ for the trace part we can write
\bea
\label{Cij}
-\half \tr \atilde_i(z) \atilde_j(z) = C_{ij} + F(z)\delta_{ij},
\eea
with the constant, symmetric and traceless $3\times 3$ matrix $C$. Using
the basis $\sigma_j$ for $su(2)$ the dual gauge field can be written as $\atilde_i(z) = - X_{ji}(z) \sigma_j$ and
the above relation translates into $X^T X = C + F$.
It follows from the gauge fixed Nahm equation (\ref{gaugefixednahm}) that
\bea
\label{dP}
F^\vesszo = \frac{1}{3} \varepsilon_{ijk} \tr [\atilde_i,\atilde_j]\atilde_k =
\frac{2}{3} \varepsilon_{ijk}\,\varepsilon_{mnp}\,X_{mi}X_{nj}X_{pk} = 4\,\det X,
\eea
which together with $X^T X = C + F$ leads to
\bea
\label{dF}
{F^\vesszo}^{\,2} = 16(F+c_1)(F+c_2)(F+c_3)\,,
\eea
where the $c_i$ are the eigenvalues of $C$ ordered as $c_2 \leq c_1 \leq c_3$. In addition we have $c_1+c_2+c_3 = 0$.
Now recall that the Weierstrass elliptic function ${\curly P}(z;g_2,g_3)$ satisfies the differential
equation \cite{AbSt}
\bea
\label{weier}
{{\curly P}^\vesszo}^{\,2} = 4({\curly P} - e_1)({\curly P} -e_2)({\curly P}-e_3)\,,
\eea
where the elliptic invariants are $g_2 = -4 (e_1 e_2 +e_1 e_3+e_2e_3)$ and $g_3 = 4e_1e_2e_3$.
Thus the most general solution for $F(z)$ is
\bea
\label{solF}
F(z) = \frac{1}{4} {\curly P}(z-z_0;g_2,g_3)\,,\quad g_2 = -64(c_1 c_2 + c_1 c_3 + c_2c_3)\,,\quad g_3 = -256c_1c_2c_3\,,
\eea
with an arbitrary $z_0$. This completely determines the $z$-dependence of the dual
gauge field which we now reconstruct. 

It is possible to diagonalize $C$ by an orthogonal matrix $R$ and then it follows from $X^T X = C +F$ that
\bea
\label{Xz}
X(z) = \tilde{R}\;\diag\left( \sqrt{F(z)+c_1} , \sqrt{F(z)+c_2}, \sqrt{F(z)+c_3} \right) R^T \,,
\eea
with another orthogonal matrix $\tilde{R}$. Clearly,
$R$ corresponds to spatial rotations while $\tilde{R}$ is coming from a global $SU(2)$ dual gauge rotation $U$.
Let us parametrize the eigenvalues of the traceless $C$ as
\bea
\label{c123}
c_1 = D^2 \frac{1-2\k^2}{12}\,,\qquad c_2 = D^2 \frac{\k^2-2}{12}\,,\qquad c_3 = D^2\frac{\k^2+1}{12}\,,
\eea
with a so-called shape parameter $0\leq \k\leq 1$ (not to be confused with the topological charge $k$ which is 2
in this section) and scale parameter $D>0$. In terms of these the elliptic
invariants are
\bea
\label{ellipticinvkD}
g_2=\frac{4}{3}D^4(1-\k^2+\k^4)\,,\qquad g_3 = \frac{4}{27} D^6 (2\k^2-1)(\k^2-2)(\k^2+1)\,.
\eea
It is useful to express the result in terms of Jacobi elliptic functions,
\bea
\label{jacobi}
f_1(z) = \frac{\k^\vesszo}{\cn_\k(z)}\,,\qquad f_2(z) = \k^\vesszo\, \frac{\sn_\k(z)}{\cn_\k(z)}\,,\qquad f_3(z)=\frac{\dn_\k(z)}{\cn_\k(z)}\,,
\eea
which were used in \cite{Dancer:1992kn} in the BPS monopole context. Note that $f_1$ and $f_3$ are symmetric
functions of $z$ whereas $f_2$ is anti-symmetric in accordance with the symmetry properties of $\sigma_i$, hence
the restriction (\ref{spseries}) is fulfilled. We refer to standard definitions \cite{AbSt},
\bea
\label{jacobidef}
z = \int_0^{\phi(z)} \frac{dt}{\sqrt{1-\k^2\sin^2 t}}\,,\qquad K(\k)=\int_0^{\pi/2} \frac{dt}{\sqrt{1-\k^2 \sin^2t}}\,,\qquad \k^\vesszo = \sqrt{1-\k^2}\nn\\
\sn_\k(z) = \sin \phi(z)\,,\quad \cn_\k(z) = \cos \phi(z)\,,\quad \dn_\k(z) = \sqrt{1-\k^2\sn_\k^2(z)}\,.
\eea
A useful identity is $f_1^\vesszo = f_2f_3$ and its cyclic permutations.
Then \dash using the relations $F(z)+c_i = D^2 f_i^2(Dz)/4$, which follow from an identity between the Jacobi
and Weierstrass elliptic functions \cite{AbSt} \dash we obtain
\bea
\label{nahmsolution}
\atilde_i(z) = -i a_i + \frac{D}{2} R_{ij} f_j(D(z-z_0)) U \sigma_j U^\dagger\,,
\eea
where we have added the arbitrary trace part $a_i$. The arbitrary constant $z_0$ will be fixed
by the condition (\ref{spseries}). We see that in order to fully specify a solution to Nahm's equation
on a fixed interval we have to specify its location $a$, spatial orientation $R$, dual gauge orientation $U$, scale $D$ and shape $\k$,
as well as $z_0$.
The condition (\ref{spseries}) further restricts $U$ to be generated by the symmetric $\sigma_2$.

A special case is of interest. Taking $\k=1$ we see from (\ref{jacobi}) that $f_1(z)=0, f_2(z)=0$ and $f_3(z)=1$ which
gives, up to spatial and gauge rotations,
\bea
\label{constantahat}
\atilde_i(z)= \lmatrix{r} -ia_1 \\ -ia_2 \\ \frac{D}{2}\sigma_3 -ia_3 \rmatrix\,,
\eea
i.e.\ constant Nahm data used in \cite{Bruckmann:2002vy} to construct axially symmetric solutions. Thus we
see that the shape parameter $\k$ controls how much the configuration deviates from the axially symmetric
arrangement. 

\subsection{Matching at the jumps}
\label{charge2matching}

Having found the solutions for Nahm's equation in the bulk of both intervals one has to impose the matching conditions
at the jumping points $z=\pm\omega$. For definiteness we take the choice of equal mass constituents and set $\omega=1/4$
as most appropriate for the confined phase where on average the trace of the Polyakov loop vanishes and we have $1/2\,\tr P = \cos{2\pi\omega}=0$.
The jumping conditions will restrict the pair of moduli $(a_i,R,U,D,\k,z_0)_{1,2}$ found in the previous section
for each interval. For charge 2 and gauge group $SU(2)$ we expect 4 constituent monopoles with arbitrary locations. 

We have found two sets of non-trivial solutions that interpolate between overlapping and well-separated constituents. It
should be stressed that solving the matching conditions for the various moduli is, from a numerical point of view, a
not very difficult task. One has to solve a finite number of equations \dash although transcendental \dash for a
finite number of variables. In the present section
we present exact solutions in order to illustrate in a controlled way how typical configurations behave. This investigation
makes it clear how the jumping conditions restrict the moduli, which parameter controls which behaviour of the solution, etc.
After all this is understood one can generate any number of numerical solutions essentially in a straightforward way.

We will see in section \ref{zeromodeabelianlimit}
when we discuss the behaviour of fermionic zero-modes that in the abelian limit they localize to ellipses. For large separation
between the constituents these 2 ellipses collapse to 2 finite segments and the support for the zero-modes become 4 points at the
ends of these 2 segments. Since in this case the singularity structure given by the 4 Dirac deltas clearly signals 4 point-like objects
with well-defined locations we will introduce the extremal points of the major axis of the ellipses as approximate constituent
locations for arbitrary finite separation.

For the first particularly simple parametrization \dash which we call 'rectangular' \dash the two disks are parallel and separated in height
by a distance $d$. The asymptotic Polyakov loop is chosen not to be diagonal but rather $P=\exp{(2\pi i \omega \tau_2)}$.

The parameters entering the Nahm solutions on the two intervals are $D_1=D_2=D$, $\k_1=\k_2=\k$,
$R_1=R_2=1$, $U_1=U_2=1$, $\Delta a_i = a^2_i - a^1_i = ( 0\,,  -d\,,  0)$ and $z_0^1=0$, $z_0^2=1/2$. We take $\ahat_0=0$,
thus $\ahat_i(z) = \atilde_i(z)$. Then the Nahm data on the first interval $(-1/4,1/4)$ is
\bea
\label{nahmdata1}
\ahat_i(z) =\half
\lmatrix{l} D f_1\left(\frac{D}{4}z\right) \sigma_1 \\D f_2\left(\frac{D}{4}z\right)\sigma_2-id \\ D f_3\left(\frac{D}{4}z\right)\sigma_3 \rmatrix\,,
\eea
whereas on the second interval $(1/4,3/4)$ we have
\bea
\label{nahmdata2}
\ahat_i(z) = \half
\lmatrix{l} Df_1\left(\frac{D}{4}\left(z-\half\right)\right)\sigma_1 \\ Df_2\left(\frac{D}{4}\left(z-\half\right)\right)\sigma_2 +id\\ D f_3\left(\frac{D}{4}\left(z-\half\right)\right)\sigma_3\rmatrix\,.
\eea
Now the matching condition at $z=1/4$ gives
the following jumps (remembering that for $SU(2)$ the requirements for $z=-\omega$ follow from $z=\omega$, see (\ref{spjumps}))
\bea
\label{followingjumpsxx}
\rho^2_i = d \lmatrix{c} 0 \\ 1+\tau_2 \\ 0 \rmatrix\,,\qquad \rho^1_i = d \lmatrix{c} 0 \\ -1+\tau_2 \\ 0 \rmatrix\,,
\eea
provided we impose $d=Df_2(D/4)$. This is an example of how the matching conditions constrain the various Nahm moduli
entering the two intervals. The jumps above are seen to be of the allowed general form derived in
section \ref{structureofthejumps}. In fact the constraints
discussed in that section were used to find these \dash and subsequent \dash solutions. All parameters,
and in particular the separations between constituents, are determined by $D$ and $\k$. 

The configuration is characterized by the 4 approximate constituent locations
\bea
\label{approxconsti}
{\rm y}^{(a)}_A = \lmatrix{c} 0 \\ \half(-1)^A d \\ \half(-1)^a D \rmatrix\,.
\eea

In order to see how the point-like constituents of the axially symmetric solutions are deformed by the generic non-constant
dual gauge fields, we have considered a second family that interpolate between axially symmetric and
generic solutions. These we called 'crossed' configurations because
in the abelian limit the two ellipses are in the same plane and form a shape of an X as shown in figure \ref{diskss}.
The interpolation is indicated by the arrows from the configuration of 4 constituents forming a cross
to all being alined along the $z$ axis.

The moduli corresponding to this family of solutions is $D_2=D_1=D$, $\k_2=\k_1=\k$,
$\Delta a_i = a^2_i-a^1_i=(0,0,-d\cos\alpha)$, $z_0^1=0$, $z_0^2=1/2$, $\ahat_0=0$ and
\bea
\label{crossedparam}
U_2=U_1^{-1} =\exp\frac{i\theta\tau_2}{2}\,,\qquad
R_2=R_1^{-1} = \lmatrix{ccc} \cos\phi & 0 & \sin\phi \\ 0 & 1 & 0 \\ -\sin\phi & 0 & \cos\phi \rmatrix\,.
\eea
The jumps are taken to be
\bea
\label{crossjumps}
\rho^2 = d\lmatrix{l} - \tau_3\sin\alpha \\ - \tau_2\sin\alpha \\ \tau_1+\cos\alpha \rmatrix\,,\qquad
\rho^1 = d\lmatrix{l} \;\;\; \tau_3\sin\alpha \\ - \tau_2 \sin\alpha \\ -\tau_1-\cos\alpha \rmatrix\,, 
\eea
again in accordance with our results in section \ref{structureofthejumps} for the most general allowed form.
These data are subject to the matching condition $\ahat_i(1/4+0)-\ahat_i(1/4-0) = i \rho^2_i$ which
means that the following constraints should hold,
\bea
\label{crossconstraints}
D\sin(\theta-\phi)\left(f_3\left(\frac{D}{4}\right)-f_1\left(\frac{D}{4}\right)\right)&=&d(1-\sin\alpha)\nn\\
D\sin(\theta+\phi)\left(f_3\left(\frac{D}{4}\right)+f_1\left(\frac{D}{4}\right)\right)&=&d(1+\sin\alpha)\\
Df_2\left(\frac{D}{4}\right) &=& d \sin\alpha\,.\nn
\eea
The 4 constituent monopoles for this family of solutions are located at the approximate positions
\bea
\label{approxloccross}
{\rm y}^{(a)}_A = \half \lmatrix{c} (-1)^a D \sin\phi \\ 0 \\ (-1)^{A+a}D\cos\phi- (-1)^A d \cos\alpha \rmatrix\,.
\eea
In order to have exactly point-like constituents in the abelian limit we need to impose $\k=1$, implying $\sin\alpha=0$
and $\cos\theta \sin\phi = 0$. The first possibility is $\cos\theta =0$ for which $|\cos\phi| D = d$. Such parameter
values lead to two constituents of opposite charge to coincide as can be read off from (\ref{approxloccross}). A solution
of this type describes a singular (zero-size) instanton on top of a smooth caloron. Excluding this singular case
we are left with the choice $\sin\phi=0$ for which $|\sin\theta| D = d$, implying $D > d$. We now find axially symmetric solutions
with constituent locations at
\bea
\label{approxlocaxial}
{\rm y}^{(a)}_A = \pm \half \left((-1)^Ad + (-1)^a D\right) \lmatrix{c} 0 \\ 0 \\ 1 \rmatrix\,,
\eea
where the overall sign comes from the fact that $\cos\alpha = \pm1$. For $\cos\theta \neq 0$ all constituents
are now separated from each other, giving a regular solution. The jumps (\ref{crossjumps}) become in this
case one dimensional and it can be shown that for $SU(2)$ exactly point-like constituents, that is $\k_1 = \k_2 = 1$,
forces the jumps to be one dimensional for any choice of the remaining parameters.

\begin{figure}[htb]
\vspace{6.5cm}
\includegraphics{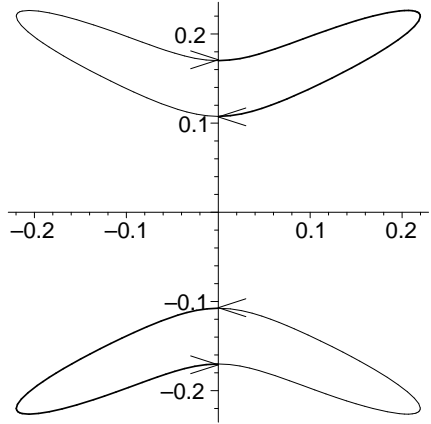}
\caption{\label{butter}\footnotesize The locations of the monopoles and antimonopoles (fat vs. thin curves) in the 1-3 plane for
what we called 'crossed' configuration. The arrows indicate that by varying $\alpha$ from $\pi$ to $0$ the configuration interpolates
between the 4 constituents forming a cross and all being alined along the $z$ axis.}
\end{figure}

Nevertheless for $\k\neq 1$, or $\sin\alpha \neq 0$, insisting as before that equal charge constituents are well
separated while keeping the centers of mass of these pairs at a fixed distance $d \cos\alpha$, forces $\k\to 1$
for increasing $D$. Thus this setting also tends to point-like constituents. We will illustrate this behaviour for
$\theta = \pi/4$. In figure \ref{butter} we plot for a typical value of $d$ the constituent locations as given by (\ref{approxloccross}),
varying $\alpha$ from $\pi$ to $0$. Note that for a given $d,\alpha$ and $\theta$ the constraints (\ref{crossconstraints})
can be used to solve for $\phi, D$ and $\k$. The asymptotic behaviour for $\alpha=\pi/2$ is determined by
\bea
\label{kdblabla}
\k^\vesszo = \frac{4\exp(-\frac{D}{4})}{3+2\sqrt{2}} \left(1 + O({\k^\vesszo}^2) \right)\,,\qquad D=2\sqrt{2} d \left(1+O({\k^\vesszo}^2)\right)\,.
\eea

\begin{figure}[htb]
\vspace{6cm}
\includegraphics{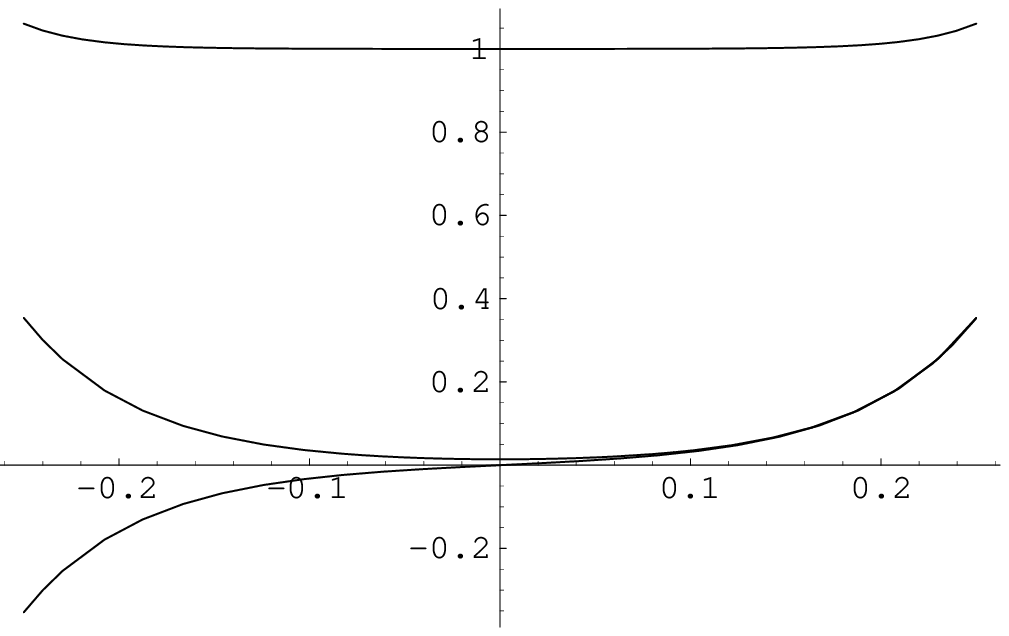}
\includegraphics{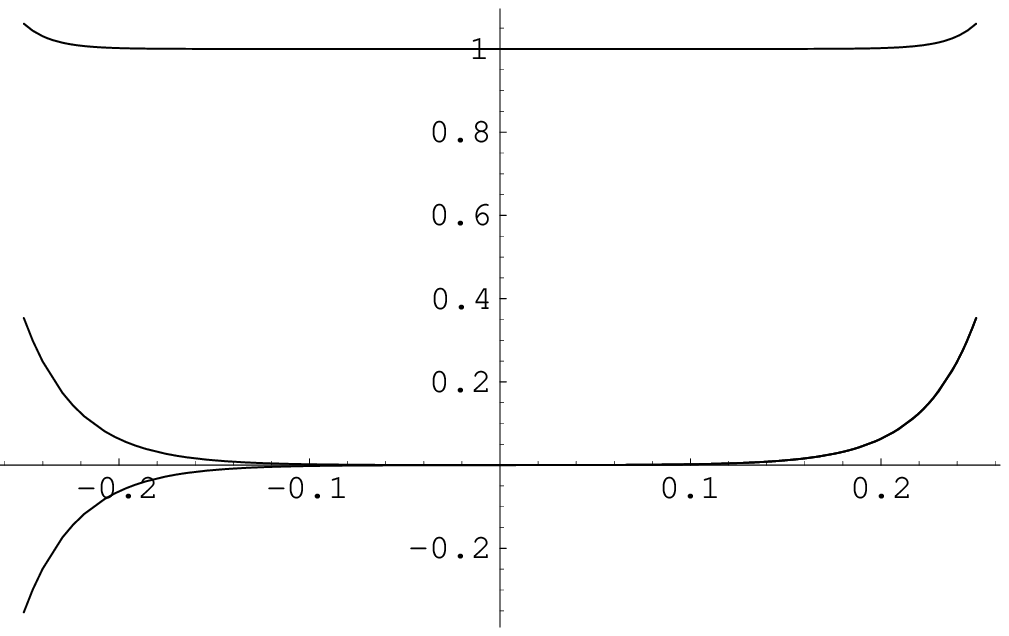}
\caption{\label{fi}\footnotesize The three functions $f_j(D(\k)z)$ for $-1/4\leq z\leq 1/4$, at $\k=1-10^{-4}$
(left) and $\k=1-10^{-8}$ (right), illustrating the approach to the point-like limit $\k\to1$.}
\end{figure}
It is also interesting to inspect $\ahat_i(z)$ in the limit $\k\to 1$ (or $D\to\infty$) in order to understand to which
extent we retrieve the piecewise constant solutions of Nahm's equation as these give rise to exactly point-like
constituents. To this end we plot $f_i(D(\k)z)$ in figure \ref{fi}, which up to an overall scale and rotation represent
the constituent locations. The plotted cases are for $\k=1-10^{-4}$ and $\k=1-10^{-8}$, clearly showing how in the bulk
of the interval $f_{1,2}\to 0$ and $f_3\to 1$ which is the characteristic behaviour of the axially symmetric and hence
exactly point-like constituents. We also see, however, that at the jumping points $z=\pm1/4$ all three functions
deviate from their near constant bulk value and develop a pole that scales as $D^{-1}$. This behaviour is in fact
in perfect agreement with the general construction of BPS monopoles in which case poles are imposed on the Nahm data
as opposed to finite discontinuities.

\subsection{Green function}
\label{greenfunctionsu2}

Having found complete solutions for Nahm's equation including the jumping conditions, the next step in constructing
$SU(2)$ calorons for topological charge 2 is to obtain the Green function. The general method was described in section
\ref{greenfunction}. and now we will apply the results to the present case.

One of the ingredients were the zeros of the spectral curve. In terms of the parametrization (\ref{zetaparam}) one
has to solve a $4^{th}$ order polynomial equation for $\zeta$. However \dash using a simple trick that is specific
to charge 2 \dash one can solve directly for $\y$. The spectral curve
condition $\det \y(i\bahat+\x)=0$ and the fact that $\y$ is null yields
\bea
\label{curvenull}
y_iy_i = 0\,,\qquad (x_ix_j-\frac{\delta_{ij}}{3}x^2-C_{ij})y_iy_j=0\,,
\eea
where the symmetric, traceless, conserved tensor $C_{ij} = -\half \tr \ahat_i\ahat_j + \frac{\delta_{ij}}{6} \tr \ahat_k\ahat_k$
surfaced again.

Let us denote the eigenvalues of the symmetric, traceless matrix $x_ix_j-\delta_{ij}\,\x^2/3-C_{ij}$ by $\lambda_i=\lambda_i(\x)$
ordered as $\lambda_1\leq\lambda_3\leq\lambda_2$
and the orthogonal transformation that brings it into diagonal form by ${\cal O}={\cal O}(\x)$.
Introducing $\tilde{\y}={\cal O}^T \y$ and $\tilde{\u}={\cal O}^T \u$ for the corresponding quantities in the new
frame, the equations (\ref{curvenull}) reduce to
\bea
\label{reducedcurvenull}
\tilde{y}_1^2+\tilde{y}_2^2+ \tilde{y}_3^2= 0\,,\qquad \lambda_1 \tilde{y}_1^2+\lambda_2\tilde{y}_2^2+\lambda_3\tilde{y}_3^2= 0\,.
\eea
These are easily solved and we obtain
\bea
\label{solutiony}
\tilde{\y}^{(a)} = \lmatrix{r} \sqrt{\lambda_2-\lambda_3} \\ (-1)^{a+1} \sqrt{\lambda_3-\lambda_1} \\ i \sqrt{\lambda_2-\lambda_1} \rmatrix\,,\qquad
\tilde{\y}^{(a+2)} = \overline{\tilde{\y}^{(a)}}\,,
\eea
in accordance with our general discussion on the parity transformation $\y\to\bar{\y}$ in section \ref{bulk}. For the subsequent
formulae we set $z_0=0, a_i=0, R=1, U=1$ and $D=1$ all of which can be reinstated at the end by appropriate scaling,
spatial rotation and gauge rotation. In this case we have $M(z) = \adj \y(i\batilde(z) +\x) = \y(\x-i\batilde(z))$, from which
the explicit expressions for the matrices $F^\pm$ in (\ref{FpFm}) are
\bea
\label{fpfmsu2}
F^+(z)&=&-\half \lmatrix{cc} y^{(1)}_1 f_1(z) - i y^{(1)}_2f_2(z) & y^{(2)}_1f_1(z)-iy^{(2)}_2f_2(z) \\
2\x\y^{(1)}-y^{(1)}_3f_3(z) & 2\x\y^{(2)}-y^{(2)}_3f_3(z) \rmatrix \nn\\
F^-(z)&=&-\half \lmatrix{cc} y^{(3)}_1 f_1(z) - i y^{(3)}_2f_2(z) & y^{(4)}_1f_1(z)-iy^{(4)}_2f_2(z) \\
2\x\y^{(3)}-y^{(3)}_3f_3(z) & 2\x\y^{(4)}-y^{(4)}_3f_3(z) \rmatrix\,,
\eea
with the choice $c=2$ for the arbitrary index.
Next we address solving (\ref{chieq}), which for charge 2 can be written as
\bea
\label{chieqsu2pre}
\chi^\vesszo = - \frac{2y_iu_j\,C_{ij}\,\delta_{cd}-i(\y\x)(\u\batilde)_{cd}}{\y\x\delta_{cd}-i\y\batilde_{cd}}\,,
\eea
reducing with the choice $c=1, d=2$ to
\bea
\label{chieqsu2}
\chi^\vesszo = - \x\y \frac{u_1 f_1-i u_2 f_2}{y_1f_1-iy_2f_2}\,.
\eea
This equation was actually considered and solved in terms of $\vartheta$-functions in the BPS 2-monopole context in \cite{Panagopoulos:1983yx}
but we found it more convenient to express the solution in terms of elliptic integrals.
One only has to use the properties (\ref{yu}), the facts
that $f_1^\vesszo = f_2 f_3$ and cyclically as well as $f_i^2(z) - f_j^2(z) = 4(c_i-c_j)$ to show
by direct substitution that the solution of (\ref{chieqsu2}) with the
initial condition $\chi(0)=0$ is
\bea
\label{chisolution}
\chi(z)&=&-\x\y\frac{u_3}{y_3}z + \frac{1}{4}\log\left(\frac{2\x\y-f_3(z)y_3}{2\x\y+f_3(z)y_3}\right)-\\
&&i\sign(z)\frac{{\k^\vesszo}^2 y_1y_2}{4 \x\y\, y_3} \left[ \Pi_\k(f_3^{-1}(z),m)-\Pi_\k(1,m)+|z| \right]\nn\,,
\eea
where the elliptic integral of the third kind is defined as
\bea
\label{ellipticpi}
\Pi_\k(s,m) = \int_0^s \frac{dt}{(1-mt^2)\sqrt{(1-\k^2t^2)(1-t^2)}}\,,\qquad m=4\frac{(\x\y)^2}{y_3^2}\,.
\eea
This solution for $\chi$ determines $\varphi$ through the definition (\ref{chi}), which in turn specifies
the solution for the bulk Green function equation, see (\ref{solutionw}), and we obtain,
\bea
\label{heeee}
w_a(z) = \varphi(z) M(z)_{a2} = \exp(z\u\x +\chi(z)) \frac{M(z)_{a2}}{\sqrt{M(z)_{12}}}\,.
\eea
Finally combining our results we have the following exact solution for the bulk
Green function equation,
\bea
\label{finalkiralysag}
w_a(z) &=&
\exp\left(iz\frac{(\x\times\y)_3}{y_3}-i\sign(z)\frac{{\k^\vesszo}^2 y_1y_2}{4 \x\y\, y_3}\left[\Pi_\k(f_3^{-1}(z),m)-\Pi_\k(1,m)+|z| \right]\right)\times\nn\\
&&\times\left(\frac{-y_1f_1(z)-(-1)^aiy_2f_2(z)}{2}\right)^{1/2}\left(\frac{2\x\y+(-1)^ay_3f_3(z)}{2\x\y-(-1)^ay_3f_3(z)}\right)^{1/4}\,.
\eea
These can be packaged into $f_{ab}^+(z)=w^{(b)}_a$ and $f_{ab}^-(z)=w^{(b+2)}_a$ where, again, the superscript $(b)$ means that
the explicit solutions (\ref{solutiony}) for $\y={\cal O}\tilde{\y}$ should be plugged into $w$.

\subsection{Gauge field}
\label{su2gaugefield}

Now that we have the exact Green function for topological charge 2 we are in a position to construct the gauge
field, field strength and action density in a straightforward manner using the general formulae of section \ref{gaugefield}.

To illustrate our exact solutions we have plotted the action density for two sets of parameters for what was called
the 'crossed' configuration. The plots are in the 1-3 plane at $t=0$ with holonomy $\omega=1/4$.

\begin{figure}[htb]
\begin{center}
\includegraphics[width=11cm,height=11cm]{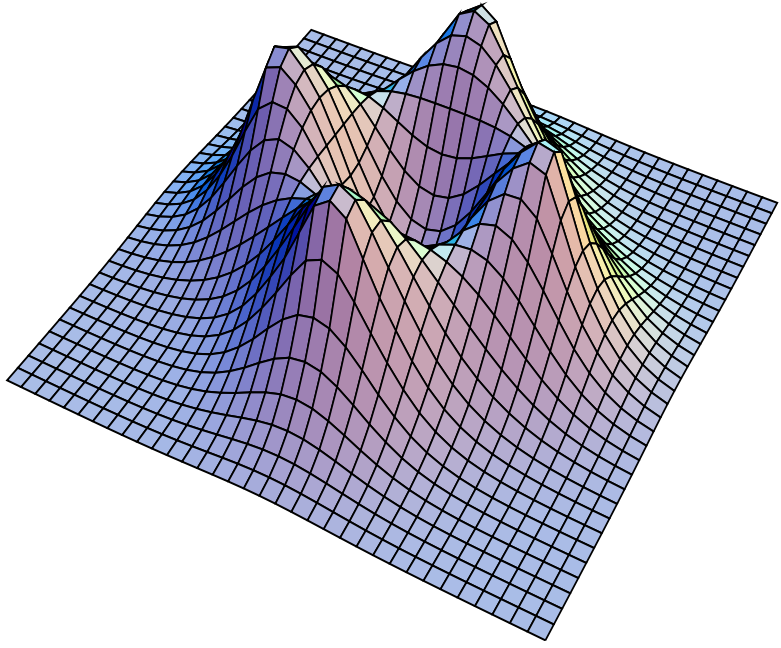}
\caption{\label{crossedad}\footnotesize Field strength squared of an $SU(2)$ caloron of charge 2.}
\end{center}
\end{figure}

In figure \ref{crossedad} the parameters determining the solution are
$\k=0.997,\;$ $D=8.753,$ $\alpha=\pi/2$, $\theta=\pi/4,$ and $\phi=-\pi/4$, see section \ref{charge2matching}. As $\k$
is sufficiently close the unity, corresponding to the point-like limit, the constituents are clearly visible.

\begin{figure}[htb]
\begin{center}
\includegraphics[width=11cm,height=11cm]{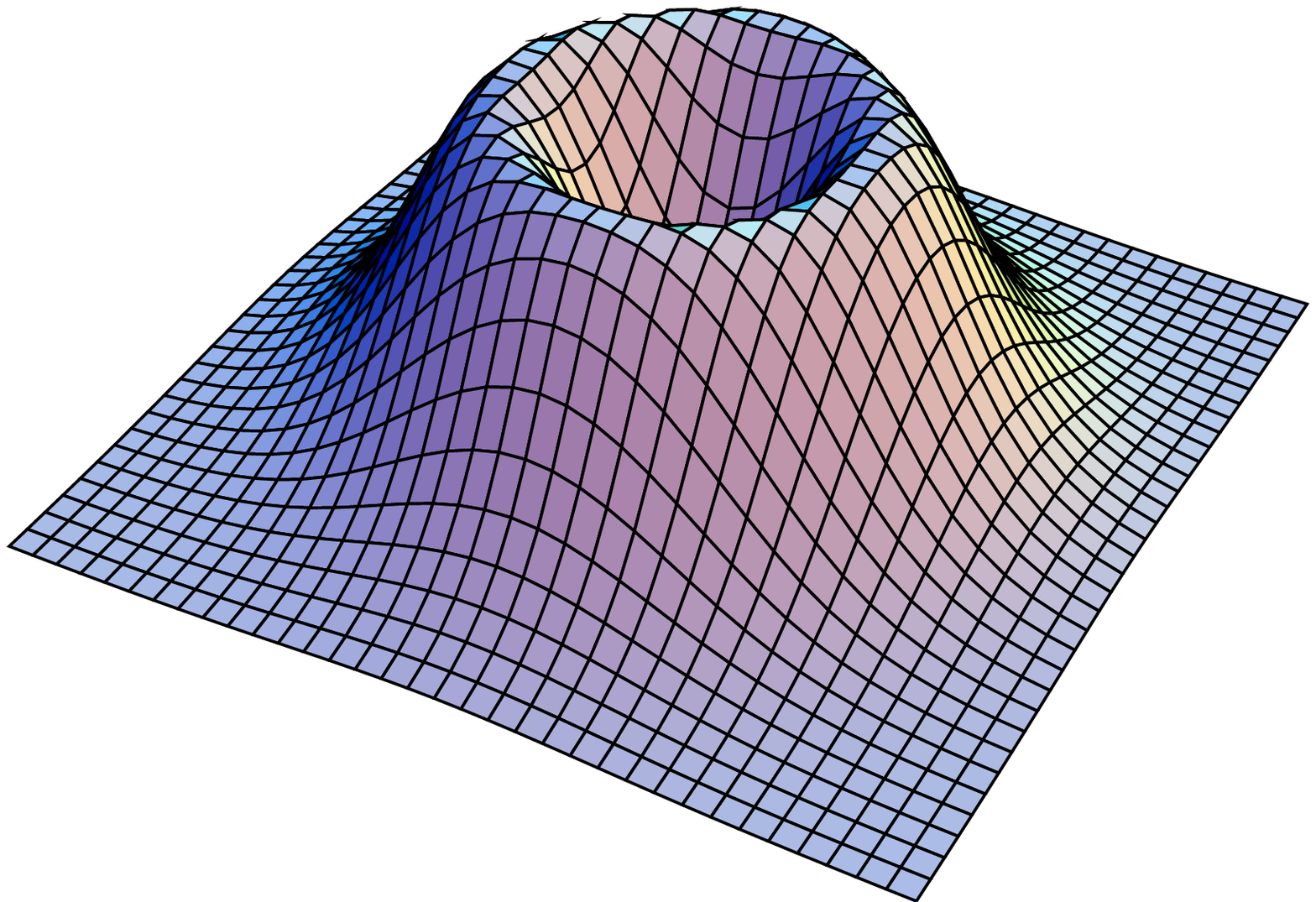}
\caption{\label{crossedad2}\footnotesize Field strength squared of an $SU(2)$ caloron of charge 2 with considerably overlapping constituents.}
\end{center}
\end{figure}

For an example of strongly overlapping constituents, see figure \ref{crossedad2}. This configuration has
parameters $\k=0.962,\,D=3.894$ and for the rest the same as in the previous case. Since monopole constituents
of opposite charge are closer to each other then in the previous case, the time dependence is also stronger. They
are overlapping so much, that the action density does not even show 4 separate peaks. The plot is similar
to the well-known doughnut shape \cite{Forgacs:1981ve} of coinciding static charge 2 monopoles, however,
here we have a combination of 2 pairs of monopoles, and the caloron configuration is time dependent.

We will see in section \ref{exactresults}
that using the zero-modes, it is possible to isolate the constituents, even though the action density blurs them.

\begin{figure}[htb]
\begin{center}
\includegraphics[width=11cm,height=11cm]{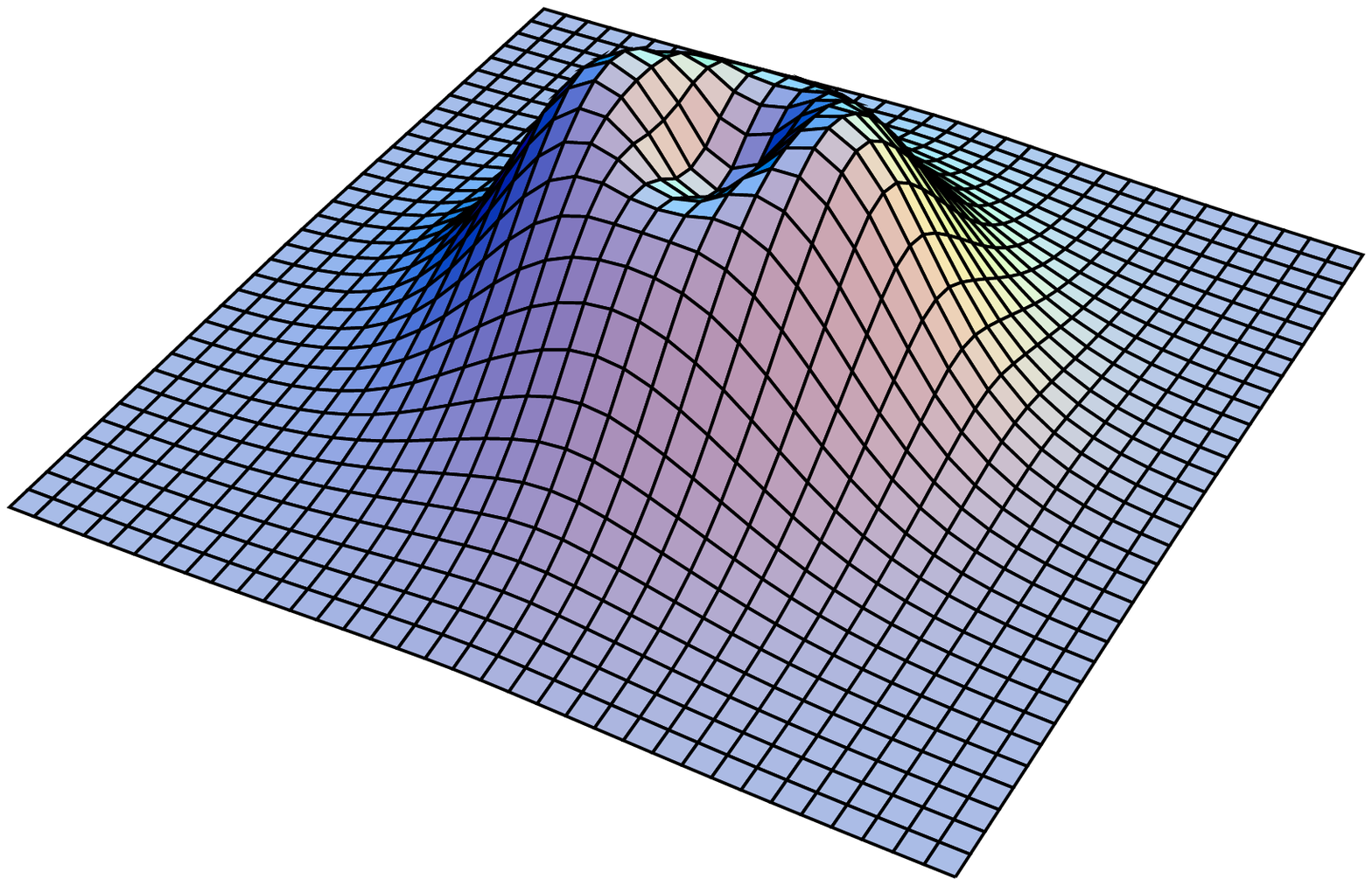}
\caption{\label{monopolead}\footnotesize Energy density of an $SU(2)$ monopole of charge 2.}
\end{center}
\end{figure}

The exact results from the previous section can be used for BPS monopoles too, this was shown in
section \ref{monopolelimit} where we derived the monopole limit of calorons. In figure \ref{monopolead} we plotted
the energy density (\ref{monopoleE}) of a typical
charge 2 configuration with parameter values $\k=0.57$ and $D=6.915$, somewhere between the known doughnut shape
\cite{Forgacs:1981ve} corresponding to $\k=0$ and well-separeted monopoles at $\k=1$.

\subsection{Abelian limit}
\label{su2abelianlimit}

In this section we make some comments on the gauge field in the abelian limit. We have seen in section \ref{abelianlimit}
that upon dropping all exponential contributions coming from the non-abelian cores of the constituent monopoles
\dash essentially shrinking them to zero \dash the gauge field becomes abelian. Let us focus on the temporal
component. It can be written in this limit as
\bea
\label{higgs}
A_0(\x) = \frac{\omega_i\sigma_i}{2\omega} \Phi(\x) \,,
\eea
which is clearly abelian and defines the field $\Phi(\x)$ (remember that in the present gauge the asymptotic
value $i\omega_i\tau_i$ is gauged away). Using our results on the Green function we can compute $\Phi$,
which we know in any case to depend algebraically on $\x$. Let us recall
the Green function evaluated at the jumping points in the abelian limit (\ref{bazevaze}),
\bea
\label{su2abeliangreen}
\tilde{f}_x(\omega,\omega) = \frac{1}{\Sigma_2}=\frac{1}{\tilde{S}_2 + R_2^-(\omega)+R_1^+(\omega)}\,,\qquad \tilde{f}_x(-\omega,\omega) = 0\,,
\eea
in terms of which we can calculate the quantities $\phi$ and $\phi_j$ (\ref{su2phi}-\ref{su2phij}) to obtain,
\bea
\label{su2abelianphi}
\phi = \frac{1}{1-2\tr \tilde{S}_2 \Sigma_2^{-1}}\,,\qquad \phi_j = 2\frac{\omega_i\sigma_i}{\omega} \tr\tilde{\rho}^2_j \Sigma_2^{-1}\,,
\eea
where $\tilde\rho^2_j = \exp (\ahat_0\,\omega)\,\rho^2_j\,\exp(-\ahat_0\,\omega)$, see (\ref{tildequantities}). 
Now using (\ref{gaugefieldcaloron}) to express the gauge field by $\phi$ and $\phi_j$ we arrive at
\bea
\label{asympthiggs}
\Phi = \frac{2\,\tr\tilde\rho_i^2\,\d_i\Sigma_2^{-1}}{1-2\,\tr \tilde{S}_2\Sigma_2^{-1}}\,.
\eea
Let us emphasize that in this formula for the asymptotic Higgs field of the caloron we know every term explicitly.
Since for abelian fields linear superposition preserves self-duality we expect that $\Phi$ can be written
as a difference $\Phi = \Phi_1 - \Phi_2$, with each contribution $\Phi_A$ coming from monopoles of type $A$.
The minus sign is due to the sign change in magnetic charge. Such a factorization is not at all obvious from (\ref{asympthiggs}).
In \cite{Hurtubise:1985af} it was shown, using the twistor correspondence, that for BPS monopoles the algebraic
tail of the Higgs field \dash in other words in our abelian limit \dash is harmonic almost everywhere with 
peculiar support on extended disks. If we indeed wish to identify the two parts $\Phi_A$ with contributions
of the two types of monopoles, then we should recover this result.
We will come back to this question in section \ref{zeromodeabelianlimit} where we
discuss the abelian limit of the fermion zero-mode densities.

\chapter{Dirac operator and zero-modes}
\label{diracoperator}

The index theorem states that the anti-chiral Dirac operator $D^\dagger=-\bar{\sigma}_\mu (\d_\mu+A_\mu)$ has
$k$ zero-modes for charge $k$ calorons and $D=\sigma_\mu(\d_\mu+A_\mu)$ has none. 
The zero-modes of the former are used in this chapter as probes for constituent monopoles. 

The circle corresponding to finite temperature necessitates a choice
of boundary condition. We seek normalizable solutions of $D^\dagger \psi_z = 0$ which are perodic up to an arbitrary phase factor,
\bea
\label{psiperiod}
\psi_z(t+2\pi,x) = \exp(-2\pi iz)\psi_z(t,x)\,,\qquad \int d^4x\,\psi_z^\dagger \psi_z = 1\,,
\eea
in the gauge where the gauge field is periodic. The choice $z=0$ corresponds to periodic zero-modes most relevant
for supersymmetric gauge theory, whereas $z=1/2$ is the canonical choice for the anti-periodic fermions of finite temperature field theory. 

It is well-known for instantons on $\R^4$ that if the $k$ lumps making up the instanton are sufficiently well separated
then a zero-mode density $\psi^\dagger\psi$ will peak roughly at each lump. It is also well-known that for instantons
both the gauge field and the zero-modes have algebraic dependence on the coordinate whereas for monopoles \dash
due to their mass \dash the dependence in the vicinity of the non-abelian core is exponential. The
same behaviour applies to calorons in the vicinity of each constituent monopole. Zero-modes localize exponentially
for $z\neq\mu_A$ as stated by the Callias index theorem \cite{Callias:1977kg}
and they delocalize \dash or spread \dash for $z=\mu_A$ with only algebraic dependence.
Localization of the zero-modes for $z\neq\mu_A$ will be used as a tool to identify the constituents in a full caloron solution.
In particular we will show to what extent can the constituents be identified as point-like objects.

The $k$ zero-modes may be packaged into a $2n\times k$ matrix $\psi_z$ and the straightforward generalization
of the corresponding ADHM formula in (\ref{adhmformulae}) states
that the densities can be computed from the Green function,
\bea
\label{greendensity}
\psi_z^\dagger(x) \psi_z(x) = -\frac{1}{4\pi^2} \Box f_x(z,z)\,,
\eea
with $f_x(z,z)$ given by formula (\ref{finalgreen}). Here $\Box$ is the 4-dimensional Laplacian. We have computed the Green
function in several asymptotic regions \dash including the one relevant for exact BPS monopoles \dash in section \ref{asymptoticregions}
and now will use these results for the zero-modes. Exact solutions will be presented for $SU(2)$ and charge 2.

\section{Abelian limit}
\label{zeromodeabelianlimit}

We have seen in section \ref{abelianlimit} that in the abelian limit the Green function becomes static and accordingly the zero-modes will
be static too. The abelian limit can be thought of as the high temperature limit as well. Since the masses of the
monopoles are proportional to the temperature, the non-abelian cores shrink to zero size in this limit. This
gives rise to an infinite localization for the zero-modes, hence they can be used to trace the cores \cite{Bruckmann:2003ag}.

Quite remarkably
the full zero-mode density \dash that is the sum over all $k$ zero-modes \dash in the abelian limit will
not depend on $z$ as long as it is in the bulk of an interval to allow for exponential localization. It can then be expressed
by the conserved quantities of the Nahm equation. For $z=\mu_A$ the zero-modes do not localize exponentially,
but spread algebraically over both $A-1$ and $A$ types of monopoles to allow for the hopping between the two types.

Let us recall the result for the Green function in the abelian limit for $\mu_A < z < \mu_{A+1}$,
\bea
\label{recallabgreen}
\tilde{f}_x(z,z) = \frac{1}{2R_A(z)}\,,
\eea
where $R_A(z)=\left(R_A^+(z) + R_A^-(z)\right)/2$ and the quantities $R_A^\pm(z)$ have been computed in section
\ref{bulk} and have been shown to have only algebraic dependence on $\x$ and $\ahat_\mu$, the solution for Nahm's
equation on $(\mu_A,\mu_{A+1})$. Thus for the zero-mode density we have
\bea
\label{zeromodeswehave}
\psi_z^\dagger(\x)\psi_z(\x) = -\frac{1}{8\pi^2} \Delta \frac{1}{R_A(z)}\,,
\eea
where the 3 dimensional Laplacian $\Delta$ was used as we have shown that the Green function is static in this limit.
A $z$-dependent dual gauge transformation may rotate the individual zero-modes, thus in order not to have this
ambiguity let us sum over the $k$ zero-modes and define ${\curly V}_A(\x) = \tr R_A(z)^{-1}$. Now we may
interpret $\rho_A(\x) = 2\pi \,\tr \psi_z^\dagger(\x) \psi_z(\x)$ as an electrostatic charge distribution 
with potential ${\curly V}_A(\x)$,
\bea
\label{vlaplace}
\Delta {\curly V}_A(\x) = -4\pi \rho_A(\x)\,.
\eea
We have seen for $|\x|\to\infty$ that $R_A(z)\to|\x|$ which gives ${\curly V}_A(\x)\to k/|\x|$ ($k$ coming from
the trace of the identity matrix) implying that the total integral of $\rho_A(\x)$ over 3-space is the topological charge $k$.
This can also be seen from the normalization of the zero-modes, $\int d^4x \psi_z^\dagger \psi_z = 1$,
which when summed over all $k$ of them and remembering that $\beta=2\pi$ also gives $\int d^3x\, \rho_A(\x) = k$.

Following our discussion on the infinite localization of the zero-modes we expect ${\curly V}_A(\x)$ to be harmonic as a
function of $\x$ almost everywhere. In addition we expect it not to depend on $z$ since the cores on which $\rho_A(\x)$ is
supported are determined by the gauge field, which clearly does not depend on $z$.

The zero-modes of a class of axially symmetric solutions with arbitrary charge \cite{Bruckmann:2002vy} and the most general charge 1 solution
\cite{Kraan:1998pm,Kraan:1998sn,Lee:1998bb} can be obtained easily and will illustrate most of what we have said above.
These cases are special in the sense that they can be described by a {\em constant} dual gauge field on each interval.
Consequently the constant matrices $\atilde_i$ must mutually commute on each interval and hence can be
diagonalized simultaneously, yielding on $(\mu_A,\mu_{A+1})$,
\bea
\label{constantnahmdata}
\batilde = i\, \diag \left({\rm\bf y}^{\,1}_A,\ldots,{\rm\bf y}^{\,k}_A\right)\,,
\eea
defining all together $nk$ vectors ${\rm\bf y}_A^{\,a}$. From these, $k$ are associated to each each interval $(\mu_A,\mu_{A+1})$,
representing the locations of the $k$ monopoles of type $A$. In this case
the matrices $f_A^\pm(z)$ solving the bulk Green function equation (\ref{greenbulk}) are particularly simple,
\bea
\label{constantfpfm}
f_A^\pm(z) = \diag \left( \exp\left(\pm z\left|\x-{\rm\bf y}_A^{\,1}\right|\right),\ldots,\exp\left(\pm z\left|\x-{\rm\bf y}_A^{\,k}\right|\right) \right)\,.
\eea
resulting in similarly simple expressions for $R_A^\pm(z)=\pm {f_A^\pm}^\vesszo(z){f_A^\pm}^{-1}$,
\bea
\label{constantrprm}
R_A^\pm(z) = \diag \left( \left|\x-{\rm\bf y}^{\,1}_A\right|,\ldots, \left|\x-{\rm\bf y}^{\,k}_A\right| \right)\,,
\eea
which are also constant, but nevertheless satisfy the Riccati equation (\ref{riccati}) as they should.
Now taking $R_A(z)=(R_A^+(z) + R_A^-(z))/2$ yields,
\bea
\label{constantV}
{\curly V}_A(\x) = \sum_a \,\frac{1}{\left|\x-{\rm\bf y}^{\,a}_A\right|}\,,\qquad \rho_A(\x) = \sum_a \,\delta\left(\x-{\rm\bf y}_A^{\,a}\right)\,.
\eea
We see that ${\curly V}_A$ is independent of $z$, also that it is
harmonic almost everywhere and the localization of the zero-modes for $A$ fixed is on the $k$
monopoles of type $A$. This is a clear signal that the constituents can be described as point-like objects for arbitrary separations
between them in the present case.

For generic higher charge solutions it is not directly obvious that the zero-modes
follow a similar point-like behaviour. Due to their infinite localization we expect in any case a singular
behaviour that is almost everywhere zero with a distributional support on the cores. In general we will find an extended disk-like
structure, which, however becomes point-like for large separation between the constituents. The singularity is of course
a byproduct of the abelian limit, in a full non-abelian solution everything is smooth. We have calculated
the exact Green function in section \ref{charge2} for charge 2 so we will see in this case
how the singularity structure is resolved by the exact solution.

\subsection{Multipole expansion}
\label{multipoleexpansion}

In this section we will study
in detail \dash for generic caloron solutions \dash the potential ${\curly V}_A(\x)$ and the charge distribution $\rho_A(\x)$ it gives rise to.
Since we will be focusing on one given type, the index $A$ will be dropped.

In principle our formula (\ref{algebraicR}) can be used to calculate $R(z)=(R^+(z)+R^-(z))/2$ exactly and hence $\curly V$ as well. However,
this way of analysing $\curly V$ is a bit awkward as the position $\x$ is hidden in the roots $\zeta$ or null vector $\y$ of the algebraic curve
in a rather complicated fashion. It is much more practical to use the Riccati equation (\ref{riccati}) and
the asymptotic conditions $R{\,^\pm}(z)\to r=|\x|$ for large $r$ we have found earier,
\bea
\label{riccatianyad}
\pm{R^{\,\pm}}^\vesszo(z)+ R^\pm(z)^2 = (i\batilde(z)+\x)^2=-\batilde(z)^2+2ir\, \e\batilde(z) + r^2\,,
\eea
where we have introduced the unit vector $e_i=x_i/r$. One can assume that $\tr\atilde_i = 0$ since any trace part
can be incorporated by a shift in $x_i$.
Solving (\ref{riccatianyad}) for a general solution of Nahm's equation on the right hand side is very difficult but an efficient
expansion is possible in powers of $r$, furthermore one does not need to solve any differential equation in doing so. The reason is
that in terms of the variable $Y^{\,\pm}(z)=R^{\,\pm}(z)/r$ the Riccati equation (\ref{riccatianyad}) becomes
\bea
\label{yric}
Y^{\,\pm}(z)^2 \pm  {Y^{\,\pm}}^\vesszo(z) \frac{1}{r}= 1 + 2i\batilde \e \frac{1}{r} -  \batilde(z)^2\frac{1}{r^2}\,,
\eea
and the term containing the derivative is only subleading in $1/r$, resulting in purely algebraic recursion relations.
The asymptotic condition for the new variable is $Y^{\,\pm}\to1$ for large $r$.

It is a simple excercise to extract the first few terms of $Y^\pm$ from (\ref{yric}). Let us
introduce the useful combinations $U=-i\batilde \e$ and $V=-\batilde^2$, in terms of which the expansion of $Y^{\,\pm}$ becomes
\bea
\label{yexpansion}
Y^{\,\pm}\!=\!1-U \frac{1}{r} + \half (-U^2\!+V\!\pm\! U^\vesszo)\frac{1}{r^2} + \frac{1}{4} (\!-\!2U^3\!-\!U^{\vesszo\vesszo}\!+\!\{U,V\}\!\pm2{U^2}^\vesszo\!\mp V^\vesszo)\frac{1}{r^3}+\ldots
\eea
where $\{,\}$ stands for the anti-commutator. Wherever the derivatives $U^\vesszo$ or $V^\vesszo$ appear one
has to use the Nahm equation to substitute some polynomial of $\atilde_i$ into the expressions. From the above
expansion one can compute $Y(z)=\half (Y^+(z)+Y^-(z))$ and the expansion of its inverse, the trace of which finally
becomes,
\bea
\label{vexp}
\tr \frac{1}{Y} = k + \half \tr (3U^2-V) \frac{1}{r^2} + \half \tr(5U^3-3UV)\frac{1}{r^3}+\ldots\,.
\eea
Two important observations are in order. First, all the potentially $z$-dependent terms are actually constants
as a result of $\atilde_i$ satisfying Nahm's equation. For example,
\bea
\label{consts}
\tr(3U^2-V)&=&(-i)^2\,\tr(\atilde_i\atilde_j)(3e_ie_j-\delta_{ij})\nn\\
\tr(5U^3-3UV)&=&(-i)^3\,\tr(\atilde_i\atilde_j\atilde_k)(5e_ie_je_k-e_i\delta_{jk}-e_j\delta_{ik}-e_k\delta_{jk})\,,
\eea
with tensors built out of $e_i$ in such a way that they are totally symmetric and traceless. We have seen in section
\ref{solvingnahm} that if $\tr\atilde_{i_1}\ldots\atilde_{i_m}$ is contracted with such a tensor then the result
is conserved. Thus, with the help of the constant, totally symmetric and traceless $C_{i_1\ldots i_m}$ tensors
defined in eq.\ (\ref{conservedCs}) the expansion for ${\curly V} = r^{-1} \tr Y^{-1}$ can be written as,
\bea
\label{vexpander}
{\curly V}(\x) = \frac{k}{r} + 3\,\frac{C_{ij} e_ie_j}{r^3} + 15\,\frac{C_{ijk}e_ie_je_k}{r^4}+\ldots\,.
\eea
The dipole term is absent because we have chosen $\tr\atilde_i =0$ by shifting $x_i$ appropriately.
The second very important observation we wish to make is that $\curly V$
is harmonic. It is well-known that if $C_{i_1\ldots i_m}$ is a totally symmetric and traceless
constant tensor then the monomials $C_{i_1\ldots i_m} e_{i_1}\ldots e_{i_m}$ defined on $S^2$ are spherical harmonics with
eigenvalues $-m(m+1)$ for the Laplacian. Therefore each term $C_{i_1\ldots i_m} e_{i_1}\ldots e_{i_m}/ r^{m+1}$ is harmonic
and so is ${\curly V}(\x)$, at least to this order of the multipole expansion.

In fact one can continue the expansion to any given order, compute first $Y^{\,\pm}(z)$ from the Riccati
equation (this is quite tedious, but only amounts to algebraic recursion relations),
then $Y(z)$ as well as its inverse (also a very tedious expansion) and finally its trace
at which point everything simplifies considerably. The pattern observed above will continue to hold and for definiteness
we give the term with coefficient of $r^{-5}$ in $\curly V$,
\bea
\label{4order}
{\curly V}_4 = \frac{1}{8}\tr \left( 35 U^4+10UU^{\vesszo\vesszo}-30VU^2+3V^2-V^{\vesszo\vesszo}+5{U^\vesszo}^2\right)=
105\,C_{ijkl} e_ie_je_ke_l\,.
\eea
As mentioned above, the pattern continues to hold for higher orders as well and the coefficient ${\curly V}_m$ of $r^{-m-1}$
will be proportional to $(-i)^m\,\tr(\atilde_{i_1}\ldots\atilde_{i_m})$, contracted with a tensor with first term
$e_{i_1}\ldots e_{i_m}$ and remaining terms that are exactly such that they make it totally symmetric and traceless. Hence
instead of $(-i)^m\,\tr(\atilde_{i_1}\ldots\atilde_{i_m})$ one can use the conserved tensors $C_{i_1\ldots i_m}$, leading to
${\curly V}_m \sim C_{i_1\ldots i_m}e_{i_1}\ldots e_{i_m}$. The constant of proportionality is a trivial symmetry factor,
the number of ways $2m$ indices can be contracted pairwise, yielding the remarkable formula
\bea
\label{finalV}
{\curly V}(\x) = \frac{k}{r} + \sum_{m=1}^\infty \frac{(2m)!}{2^mm!}\, C_{i_1\ldots i_m} e_{i_1}\ldots e_{i_m}
\frac{1}{r^{m+1}}\,.
\eea
Here we have included an arbitrary dipole term $C_ie_i/r^2$ that was assumed zero by an appropriate choice of the
coordinate system. Thus we conclude that the multipole moments of the total zero-mode density in the abelian limit 
are the conserved quantities of the Nahm equation.

Let us summarize what has been done. We are not even close to solving Nahm's equation for arbitrary
topological charge, but it turned out that only the conserved tensors are needed in the abelian limit. The more complicated
remaining $(k-1)^2$ constants of integration we have described in section \ref{solvingnahm} are solely responsible for the non-abelian cores.
We are not able to solve the Riccati equation (\ref{riccatianyad}) either,
yet we were able to derive an exact multipole series for $\curly V(\x)$ in terms of the conserved tensors.
This series is manifestly
harmonic almost everywhere and independent of $z$.

Let us repeat that the relevance of ${\curly V}(\x)$ is
that it determines the total zero-mode density in the abelian limit through
\bea
\label{psipsiV}
\tr \psi_z^\dagger(\x)\psi_z(\x) = -\frac{1}{8\pi^2} \Delta {\curly V}(\x)\,.
\eea

Now let us check our formula against the special case of constant Nahm data where we did not need to resort to the
multipole expansion. In this case, the diagonal $\atilde_i$ given in (\ref{constantnahmdata}) leads to
\bea
\label{blabla}
\frac{(-i)^m}{m!}\,\tr\atilde_{i_1}\ldots\atilde_{i_m} = \frac{1}{m!} \sum_a \cy^{\,a}_{i_1}\ldots \cy^{\,a}_{i_m}\,,
\eea
which means that the conserved tensors are just the above monomials made totally symmetric and traceless.
Using one of the standard definitions of the Legendre polynomials, $P_m$, leads to
\bea
\label{constantCe}
\frac{(2m)!}{2^mm!} C_{i_1\ldots i_m}e_{i_1}\ldots e_{i_m} =
\sum_a \frac{1}{m!} P_m\left(\cos\theta^a\right) \left|\bcy^{\,a}\right|^m\,,
\eea
where $\theta^{\,a}$ is the angle between $\bcy^{\,a}$ and $\e$. Substitution into formula (\ref{finalV}) indeed gives the well-known
expansion of $\sum_a |\x-\bcy^{\,a}|^{-1}$ in terms of spherical harmonics.

Naturally, one would wish to sum the series (\ref{finalV}) into a closed expression in the general case of non-constant
Nahm data as well. One would then analyse its singularity structure in order to determine what set the zero-modes
are supported on. We will perform this analysis for charge 2, although by a different method, rather then by summing the multipole series.

A final note about formula (\ref{finalV}). For general charge $k$ only the conserved tensors with at most $k$ indices
are independent, the others can be expressed in terms of these in a polynomial way. This relationship between
the multipole coefficients of $\curly V$ results in a polynomial equation,
\bea
\label{polyV}
\sum_m a_m(\x) {\curly V}^m =0\,,
\eea
where the coefficients $a_m(\x)$ are polynomials in $x_i$ and the independent conserved quantities.
For example, in the particularly simple charge 1 case the only invariant is $C_i = \cy_i$, all further
tensors are polynomials of $\cy_i$ and we have
\bea
\label{charge1poly}
(\x-\bcy)^2 {\curly V}^2 - 1 = 0\,.
\eea
For charge 2 the only invariant is $C_{ij}$ if we assume $C_i = 0$ and all higher rank $C$ tensors
can be expressed by $C_{ij}$. In the next section we will derive the polynomial equation satisfied by
$\curly V$ in this case.

\subsection{Charge 2}
\label{su2abelianzeromodes}

First note that the shape parameter $\k=1$ corresponds to constant Nahm data on the two intervals, resulting in
\bea
\label{pointlike}
{\curly V}(\x) = \frac{1}{|\x+{\rm\bf y}|} + \frac{1}{|\x-{\rm\bf y}|}\,,\qquad \rho(\x) = \delta(\x+{\rm\bf y})+\delta(\x-{\rm\bf y})\,,
\eea
with ${\rm\bf y}=(0,0,D/2)$. Now observe that
the interchange $c_2 \leftrightarrow c_3$ induces $D\to iD$ and $\k\to \k^\vesszo$ as can be seen from their
definition (\ref{c123}). Thus if we interchange the axis $x_2$ and $x_3$ and at the same time transform
both $D$ and $\k$ then ${\curly V}$ will stay invariant. Therefore, from eq. (\ref{pointlike}) we can immediately
obtain the corresponding result for $\k=0$. Strangely, the constituents moved to complex locations, something
that has been observed already in the BPS monopole context \cite{Nahm:1981xg}.
Nevertheless a manifestly real expression for the analytic continuation of (\ref{pointlike}) is
\bea
\label{realformV}
{\curly V}(\x) = \sqrt{8} \frac{\sqrt{4r^2-D^2+\sqrt{(4r^2-D^2)^2+16D^2x_2^2}}}{\sqrt{(4r^2-D^2)^2+16D^2x_2^2}}\,.
\eea
This function is singular on the circle of radius $D/2$ in the 1-3 plane, furthermore the $x_2$ derivative
is discontinous on the disk bounded by the circle. This can be seen by expanding around small $x_2$, but
for $4(x_1^2+x_3^2)<D^2$,
\bea
\label{expandring}
{\curly V}(\x) = \frac{8D|x_2|}{(D^2-4(x_1^2+x_3^2))^{3/2}} + O(x_2^3)\,,
\eea
which indeed reveals a discontinuity in the $x_2$ derivative. We conclude that the charge
distribution $\rho(\x)$ is singular on the whole disk, elsewhere it is smooth. This
behaviour is very far from the exactly point-like situation and in order to understand how that is approached
as $\k\to 1$ we now analyse the intermediate cases $0<\k<1$.

We know that $R^\pm(z)$ satisfies the Riccati equation (\ref{riccati}). However, not every solution of (\ref{riccati})
will have the property that $\tr R(z)^{-1}$ is independent of $z$. This means that $R^\pm(z)$ are very special
solutions and since the equation is first order the whole solution is determined by the initial condition, which
in turn has to be very special. Expanding $R^\pm(z)$ and $\atilde_i(z)$ in Taylor series around $z=0$, eq.\ (\ref{riccati})
will determine all Taylor coefficients of $R^\pm(z)$ as a polynomial function of $R^\pm(0)$ and the Taylor coefficients of $\atilde_i(z)$,
which are polynomials in $D$ and $\k$. Now one can compute the Taylor series of $\tr R(z)^{-1}$ and impose
the constraints that all coefficients should vanish except for the constant term. This gives an infinite
system of equations for $R^\pm(0)$ \dash of coure it is infinitely redundant \dash
which actually fixes it as an algebraic function of $\x,\, D$ and $\k$. We immediately
obtain ${\curly V}(\x)$ since, by construction, ${\curly V}(\x)=\tr R(z)^{-1} = \tr R(0)^{-1}$. Note that the choice $z=0$
corresponds to the interval $(-\omega,\omega)$, hence to monopoles of type $A=1$, but by shifting $z$ appropriately
one can do the same for the interval $(\omega,1-\omega)$, that is for type $A=2$.

Let us illustrate this method for unit topological charge. In this case $\batilde(z) = i \bcy$, a constant.
The Taylor expansion of $R^\pm(z)$ \dash which is now a scalar, not a matrix \dash is taken to be
\bea
\label{taylorrr}
R^\pm(z) = a_\pm + b_\pm z + c_\pm z^2 + \ldots.
\eea
Substitution into the Riccati equation (\ref{riccati}) determines the coefficients $b_\pm,\,c_\pm,\ldots$
as functions of $R^\pm(0)=a_\pm$ and we obtain,
\bea
\label{taylorrrrr}
b_\pm = \pm \left((\x-\bcy)^2-a_\pm^2\right)\,,\qquad c_\pm = a_\pm \left(a_\pm^2-(\x-\bcy)^2\right)\,,\qquad\ldots
\eea
giving rise to the following expansion of $R(z)^{-1}$,
\bea
\label{tayll}
\frac{2}{R^+(z) + R^-(z)} = \frac{2}{a_++a_-} + 2\,\frac{a_+ - a_-}{a_++a_-}z + 2\,\frac{(\x-\bcy)^2-a_+a_-}{a_++a_-}z^2 + \ldots.
\eea
The requirement of $z$-independence implies $a_+ = a_-$ and $(\x-\bcy)^2 = a_+ a_-$, resulting in all
further coefficients to vanish as well. Clearly, this leads to the correct result as it should, $\curly V(\x) = |\x-\bcy|^{-1}$.

This rather cumbersome way of obtaining $\curly V$ should in principle work for any topological charge, however,
in practice we could only make use of it for charge 2 (apart from the trivial charge 1 case above). Even in this
case it is very complicated as the first 11 Taylor
coefficients in $z$ are needed to constrain $R^\pm(0)$ unambigously. Leaving technicalities aside we only present the
result of this procedure. Let us parametrize the initial conditions for the Riccati equation (\ref{riccati}) by
dimensionless quantities $X_\mu$ and $Y_\mu$,
\bea
\label{initialcond}
R^+(0)+R^-(0) = D(X_0+X_i\tau_i)\,,\qquad R^+(0)-R^-(0)=D(Y_0+Y_i\tau_i)\,,
\eea
and measure $x_i$ in units of $D/2$. Then we have
\bea
\label{vX}
{\curly V} = \frac{2X_0}{X_0^2-X_i^2}\,.
\eea
The constraints from the first 11 terms in the Taylor expansion imply that $Y_0 = Y_1 =Y_3 = X_2 = 0$ and that the
variable $Q = X_0^2-X_1^2-X_3^2-Y_2^2$ satisfies the polynomial equation
\bea
\label{Q}
&&Q^3+\left(\k^2-2+r^2\right)Q^2-\left(\k^4+2\k^2(x_1^2-x_2^2-x_3^2)+4x_2^2+r^4\right)Q-r^6+(2-\k^2)\k^4+\nn \\
&&+4\k^2(x_1^2-x_3^2)-\k^4(3x_1^2-x_2^2-x_3^2)+\left((2-\k^2)(3x_1^2-x_2^2+x_3^2)-4x_1^2\right)r^2=0\,,
\eea
whereas $V_1 = X_1/X_0, V_3 = X_3/X_0$ and $Y_2$ can be solved for in terms of $Q$,
\bea
\label{kutyulek}
V_1&=&\frac{r^4-Q^2+2({\k^\vesszo}^2(Q-2x_1^2)+x_1^2+x_2^2-x_3^2)+(2-\k^2)\k^2}{4x_1\k^\vesszo \k^2}\nn\\
V_3&=&\frac{r^4-Q^2+2(Q-2x_3^2-{\k^\vesszo}^2(x_1^2-x_2^2-x_3^2))-(2-\k^2)\k^2}{4x_3\k^2} \\
Y_2&=&\frac{r^4-Q^2+2\k^2(x_1^2-x_2^2-x_3^2)+4x_2^2+\k^4}{4x_2\k^\vesszo}\nn\,.
\eea
Thus we obtain from (\ref{vX}) the explicit expression for ${\curly V}$ as an algebraic function of $\x$,
\bea
\label{vQ}
\curly V = \frac{1}{2\sqrt{(Q+Y_2^2)(1-V_1^2-V_3^2)}}\,,
\eea
and the proper root of the cubic equation for $Q$ is fixed by the asymptotic condition $\curly V \to 2/r$, singling out the one with
\bea
\label{Qasympt}
Q = r^2 + \frac{2x_2^2+\k^2(x_1^2-x_2^2-x_3^2)}{r^2} + O(1)\,.
\eea
This completes our derivation of the electrostatic potential $\curly V$ for the most general charge 2 solution. We have
checked that the first 21 orders in the multipole expansion (\ref{finalV}) agree with the closed form given above. In
addition, we have checked numerically that the evaluation of $\curly V$ using formulae (\ref{FpFm}-\ref{algebraicR}) also agrees
with (\ref{vQ}).

An alternative form for $\curly V$ can be given as the root of a 6$^{th}$ order polynomial with coefficients themselves
polynomials in $\x$ and $\k$. This is because $Q$ is a root of a cubic polynomial and ${\curly V}^2$ is a rational
function of $Q$, hence also fulfills a cubic equation. The coefficients of this polynomial are rather
lengthy and are given in an appendix.

Yet another way of writing $\curly V$, which turns out to be the most useful, is the following. Expanding it
for small $x_2$ yields,
\bea
\label{smallx2ellipse}
\curly V(\x) = \frac{8\k^\vesszo D|x_2|}{\left(D^2{\k^\vesszo}^2-4\left(x_1^2+{\k^\vesszo}^2x_3^2\right)\right)^{3/2}} + O(x_2^3)\,,
\eea
for $D^2{\k^\vesszo}^2-4\left(x_1^2+{\k^\vesszo}^2x_3^2\right) > 0$, generalizing the corresponding expansion
for $\k=0$ in (\ref{expandring}). In particular the cirlce of diameter $D$ is deformed into an ellipse with major axis $D$ and
minor axis $\k^\vesszo D$. The singularity structure is similar to the $\k=0$ case, the $x_2$ derivative jumps at the disk bounded
by the ellipse. It is useful to introduce polar coordinates $(p,\varphi)$ suited for the ellipse
in the 1-3 plane by $(x_1,x_3)=(\k^\vesszo p \cos\varphi, p \sin\varphi)$. Taking the Laplacian then gives
the following distribution,
\bea
\label{laplacedisk}
\rho(\x) = -\frac{1}{4\pi} \Delta \curly V(\x) = -\delta(x_2) \frac{D}{\pi\k^\vesszo p}\, \frac{\d}{\d p} \frac{\theta(D-2p)}{\sqrt{D^2-4p^2}}\,.
\eea
We see that due to the step function, $\curly V$ is harmonic almost everywhere with support on the disk, including
its boundary ellipse. We also see that there is a radially symmetric \dash in an elliptical sense \dash 
charge distribution on the disk. Further inspection of (\ref{laplacedisk}) shows that the total charge on
the disk excluding its boundary is $-\infty$, the total charge on the boundary ellipse is $+\infty$ such
that the overall total charge is 2 as it should be.

\begin{figure}[htb]
\vspace{6.5cm}
\includegraphics{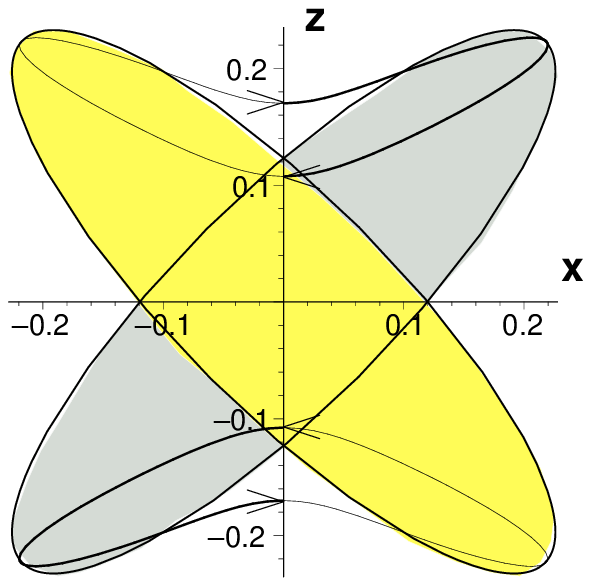}   
\includegraphics{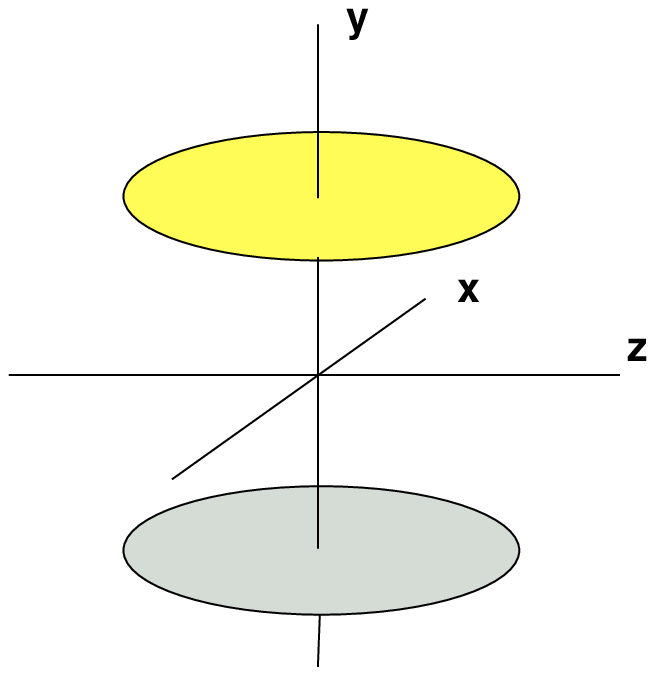}
\begin{center}
\caption{\label{diskss}\footnotesize Two examples to illustrate the location of the disk singularities of the zero-modes for $SU(2)$ and charge 2. What
we called 'rectangular' configuration is shown on the left, the 'crossed' configuration on the right.}
\end{center}
\end{figure}

In section \ref{charge2matching} we have found two families of exact solutions of the matching conditions. In figure
\ref{diskss} we show the arrangement of the two disks for these families. The light and dark shading corresponds to
constituents of plus and minus magnetic charge. One example of what we called the
'rectangular' configuration is on the left, a 'crossed' configuration is on the right, with
the curves indicating the approximate constituent locations as $\alpha$ is varied from $\pi$ to $0$, see section \ref{charge2matching}
for more details.

The expression (\ref{laplacedisk}) allows for the integral representation,
\bea
\label{integralv}
\curly V(\x) = \frac{2}{r} + \frac{D}{\pi} \int_0^{2\pi} d\varphi \int_0^{\frac{D}{2}} \frac{dp}{\sqrt{D^2-4p^2}} \frac{\d}{\d p}
\frac{1}{\sqrt{\left(x_1-\k^\vesszo p \cos\varphi\right)^2+x_2^2+\left(x_3 - p \sin\varphi\right)^2}}\,,
\eea
which can be used to check the $\k\to1$ or equivalently $\k^\vesszo\to0$ limit. For an arbitrary test function $f(\x)$ we have
\bea
\label{distr}
\int d^3x \rho(\x) f(\x) = 2 f(0) + \frac{D}{\pi} \int_0^{2\pi} d\varphi \int_0^{\frac{D}{2}} \frac{dp}{\sqrt{D^2-4p^2}} \frac{\d}{\d p}
f\left(\k^\vesszo p \cos\varphi,0,p\sin\varphi\right)\,,
\eea
which is smooth in $\k^\vesszo$ and for $f(\x)=1$ verifies the proper normalization, $\int d^3x \rho(\x) = 2$.
Evaluation at $\k^\vesszo = 0$ gives,
\bea
\label{kvesszo0}
\int d^3 x \rho(\x) f(\x) = 2f(0) + \frac{D}{\pi} \int_{-\frac{D}{2}}^{\frac{D}{2}} dx_3 \int_{-\sqrt{\frac{D^2}{4}-x_3^2}}^{\sqrt{\frac{D^2}{4}-x_3^2}}
\frac{dx_1}{x_1^2+x_3^2} \frac{x_3\d_3 f(0,0,x_3)}{\sqrt{D^2-4(x_1^2+x_3^2)}}=\nn\\
= 2f(0)+\int_{-\frac{D}{2}}^{\frac{D}{2}} dx_3\,\sign(x_3) \d_3 f(0,0,x_3) = f\left(0,0,\frac{D}{2}\right) + f\left(0,0,-\frac{D}{2}\right)\,,
\eea
which is indeed the expected result from (\ref{pointlike}). We conclude that the rather peculiar
distribution of zero-modes that for generic $\k$ values is supported on an exteded disk smoothly tends to the point-like
limit for $\k\to1$. In this limit the minor axis of the ellipse shrinks to zero while the major axis stays $D$,
eventually tending to a segment of length $D$, with support for the zero-modes at its two ends.
We recall from (\ref{kdblabla}) that for large separation $D$ between the constituents $\k$ tends to unity,
i.e.\ for large separation the constituents indeed become point-like.

\subsection{Higgs field and zero-modes}
\label{higgsfieldandzero}

We have derived a formula for the Higgs field $\Phi(\x)$ of charge 2 calorons in the abelian limit
in section \ref{su2abelianlimit}. Its form
that naturally emerged from the Nahm transform entangles the contributions from the 2 types of monopoles.
Nevertheless, since linear superposition preserves self-duality for abelian fields, we expect that it can
be factorized into two parts, each carrying Nahm data specific to one type of monopole. Now it was shown
in \cite{Hurtubise:1985af} that the algebraic tail of the Higgs field is harmonic almost everywhere and
has the same kind of extended support on disks as the one we have found for the zero-modes. This makes
us conjecture that the potential $\curly V_A(\x)$ for monopoles of type $A$ actually equals the
Higgs field of these monopoles in the abelian limit. Since we did compute the Higgs field for
the caloron, see (\ref{higgs}) and (\ref{asympthiggs}), we have verified the relation
\bea
\label{fakt}
\Phi(\x) = \curly V_1(\x) - \curly V_2(\x)
\eea
numerically and found agreement to high precision. This results in the exact identity
\bea
\label{csicsi}
\tr \psi_z^\dagger(\x) \psi_z(\x) = -\frac{1}{8\pi^2} \Delta \Phi_A(\x)\,,
\eea
for $\mu_A < z < \mu_{A+1}$. In other words, the fermion zero-mode density equals the abelian charge distribution
of the monopole gauge field. Such a relation is at the heart of using chiral fermion zero-modes as filters to isolate
the underlying topological lumps from rough lattice Monte Carlo configurations
\cite{Gattringer:2002wh, Gattringer:2002tg, Gattringer:2003uq, Gattringer:2004pb}.

\section{Exact results}
\label{exactresults}

In the previous section we have seen that in the abelian limit, where exponential contributions coming
from the non-abelian cores of the massive constituent monopoles are dropped, the zero-modes
of the Dirac operator develop a singularity structure. In a full non-abelian solution, naturally, everything
is smooth. We have computed the Green function for charge 2 exactly, and using (\ref{greendensity}) we can now
inspect how the singularity is resolved in the full solution. Exact zero-modes for unit
topological charge have been computed in \cite{Chernodub:1999wg, GarciaPerez:1999ux}.

\begin{figure}[htb]
\vspace{9.9cm}
\includegraphics{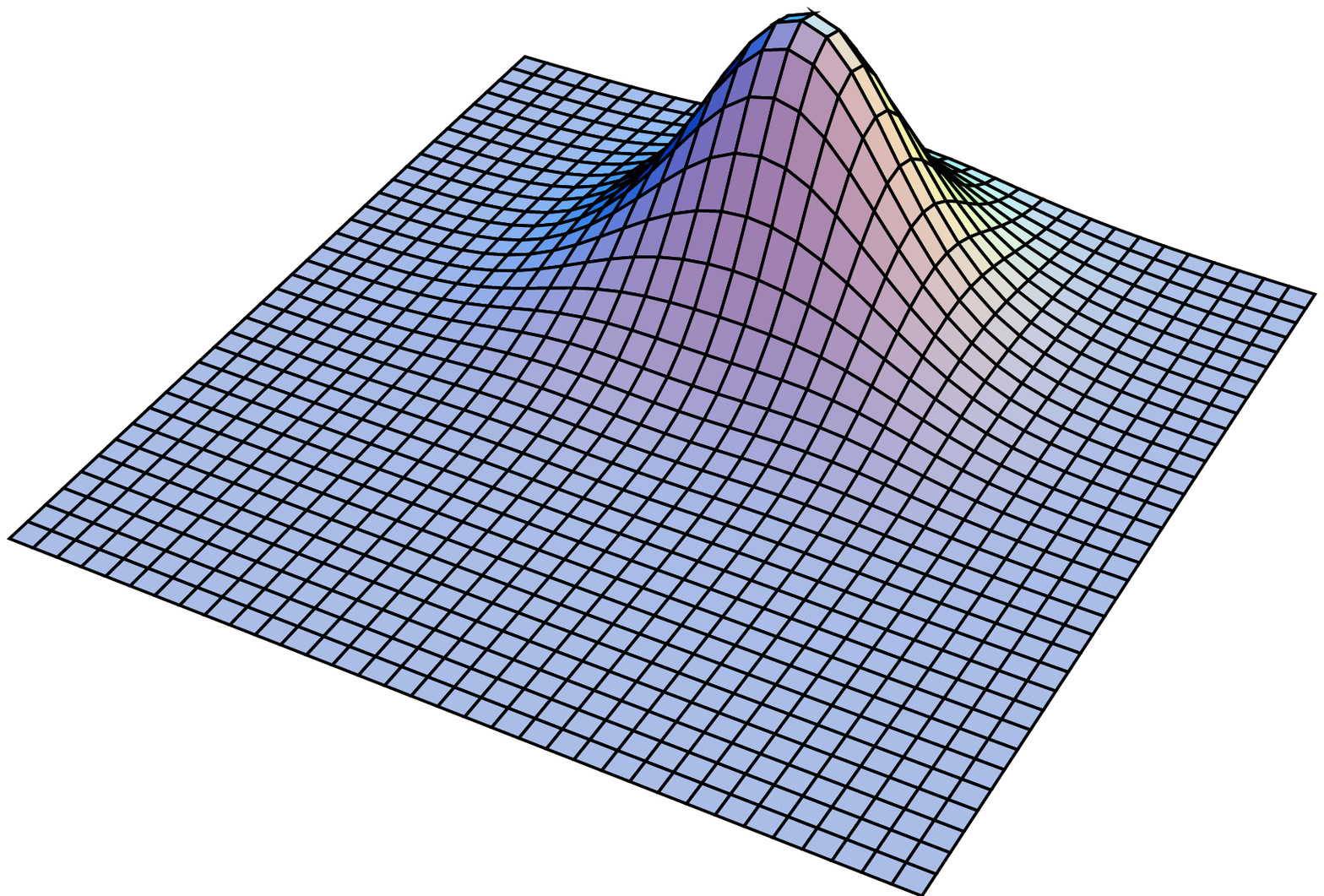}
\includegraphics{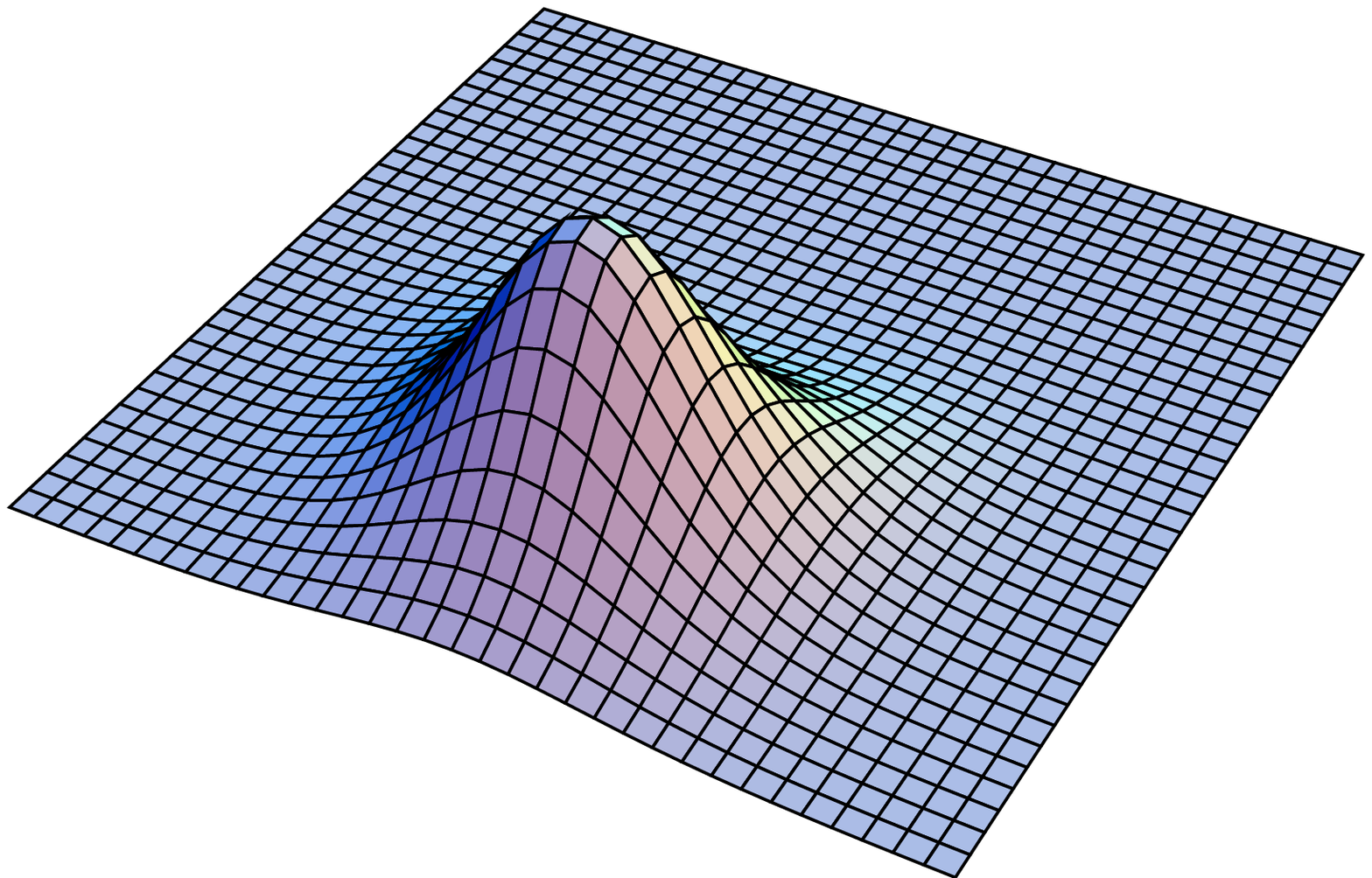}
\includegraphics{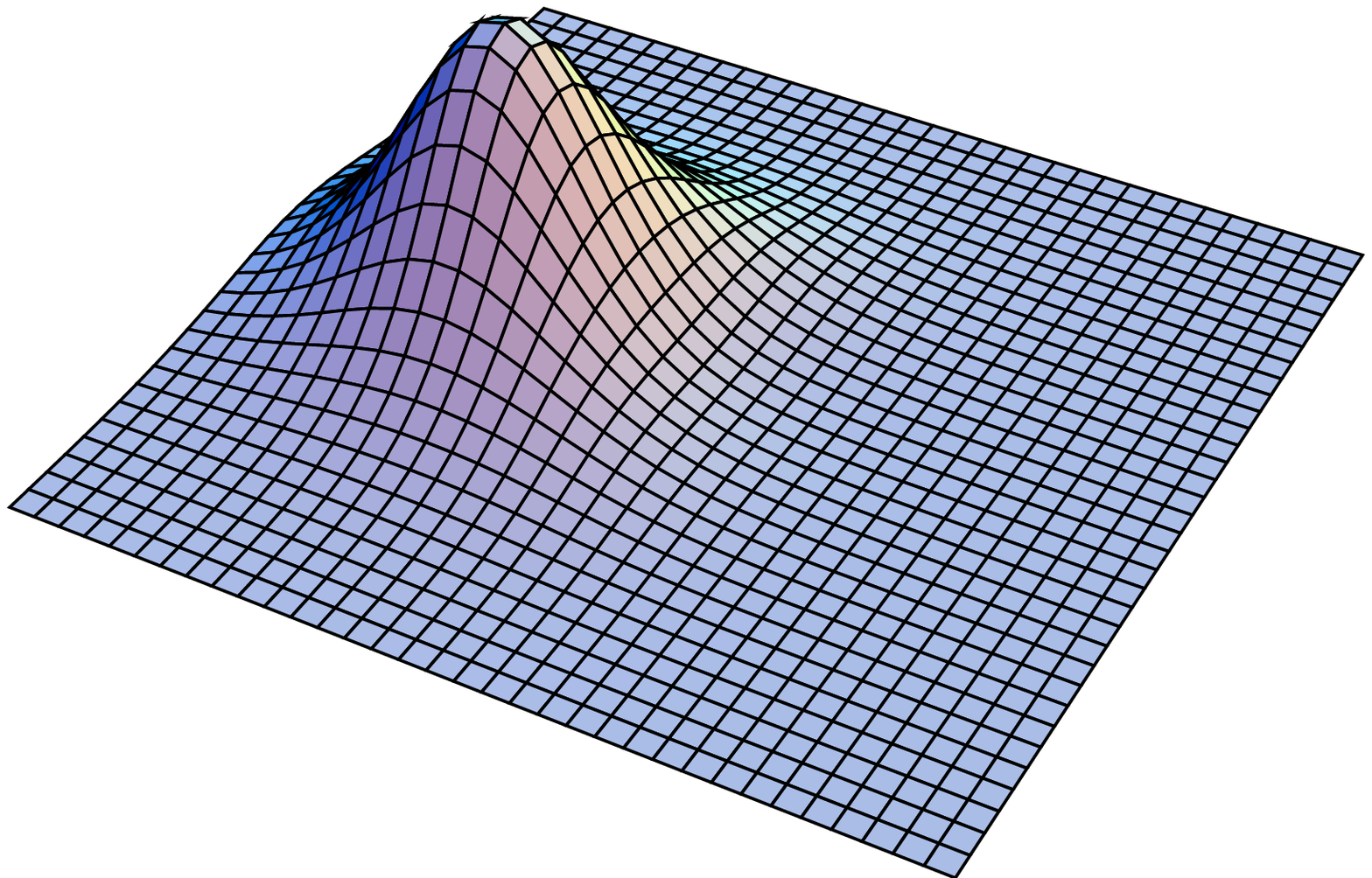}
\includegraphics{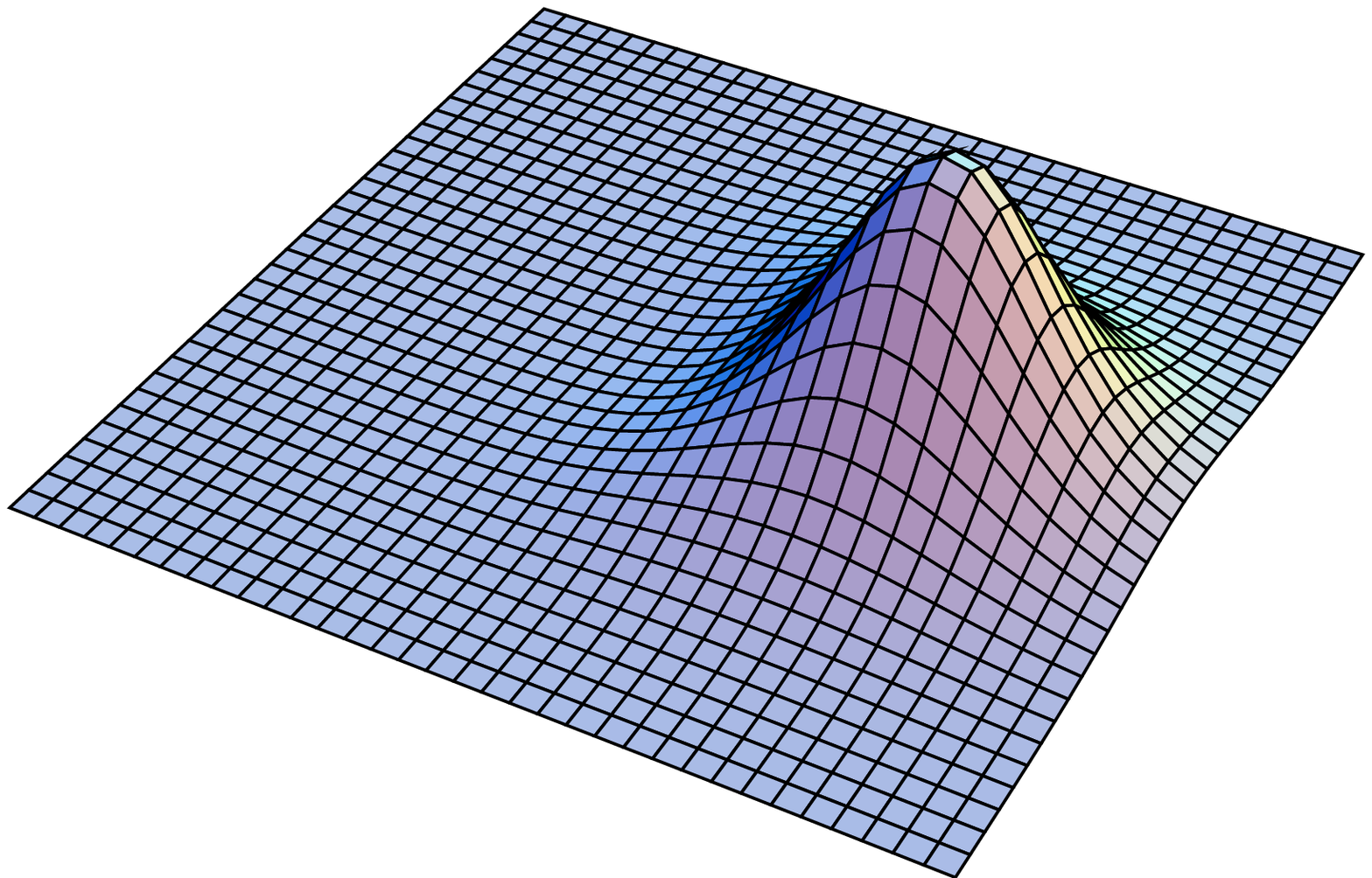}
\includegraphics{caloron_ad.eps}
\caption{\label{crossedzm}\footnotesize An $SU(2)$ caloron of topological charge 2, with the 2 periodic
zero-modes on the left, the 2 anti-periodic zero-modes on the right and the action density in the middle.}
\end{figure}

\begin{figure}[htb]
\vspace{9.3cm} 
\includegraphics{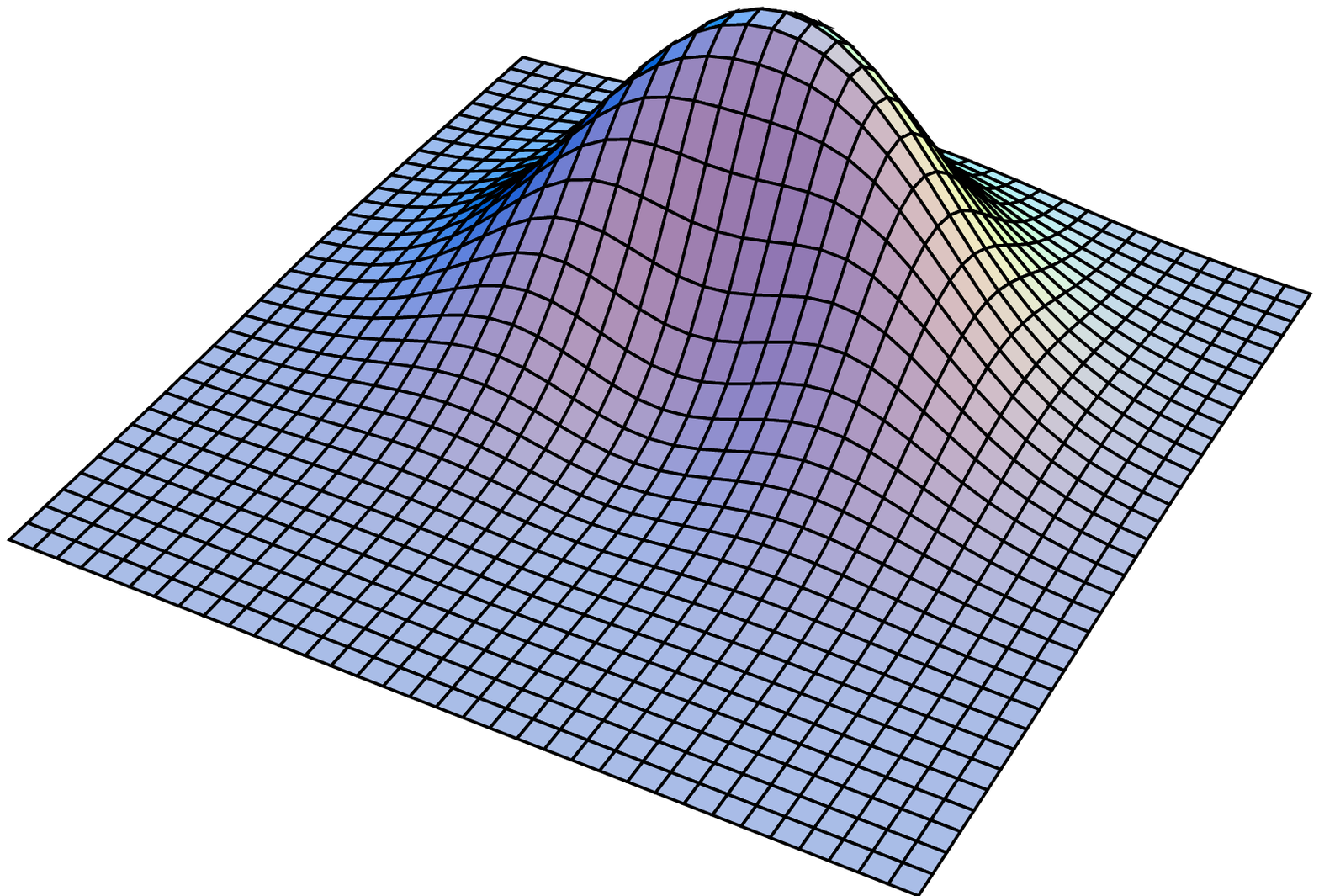}
\includegraphics{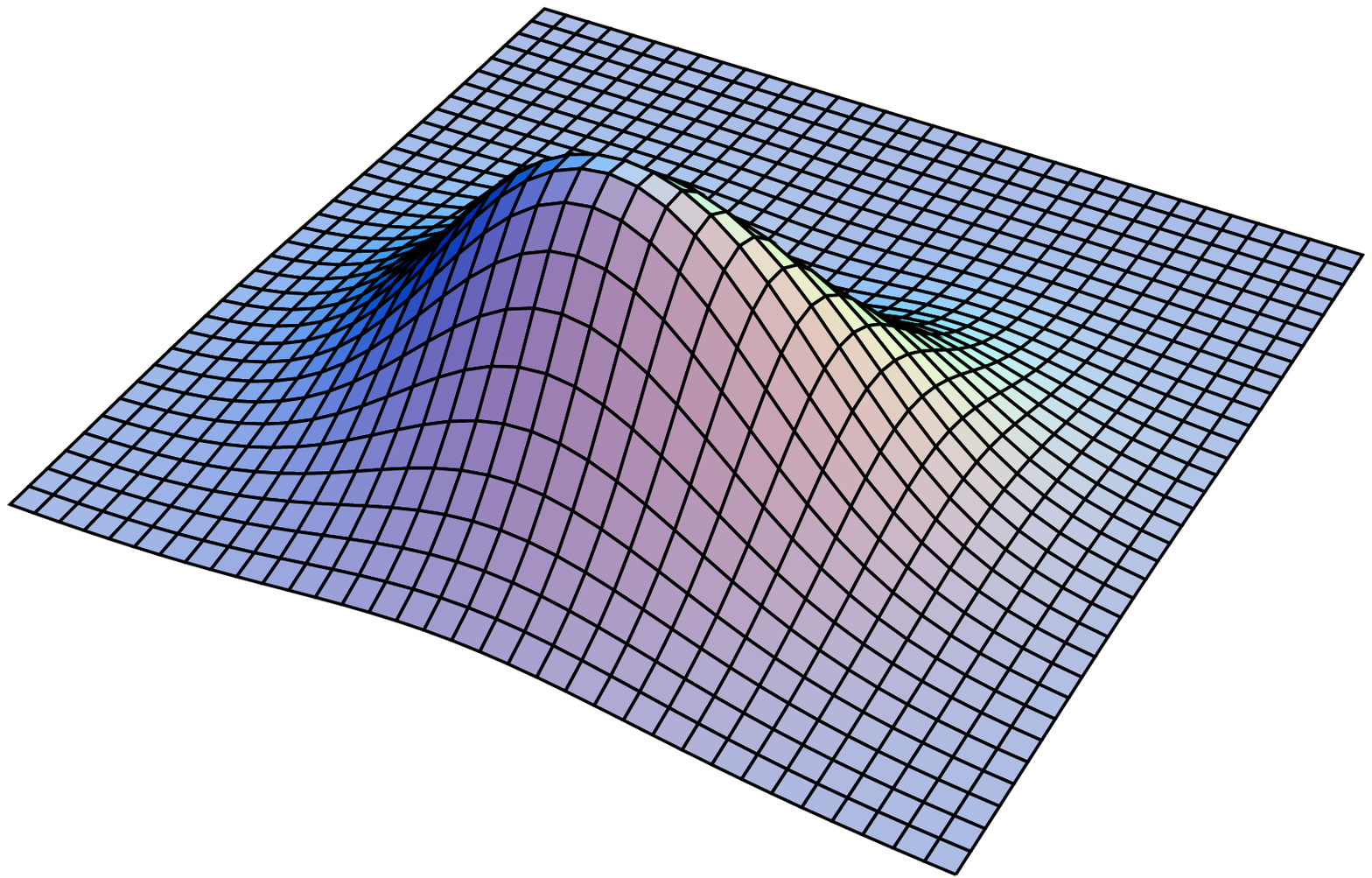}
\includegraphics{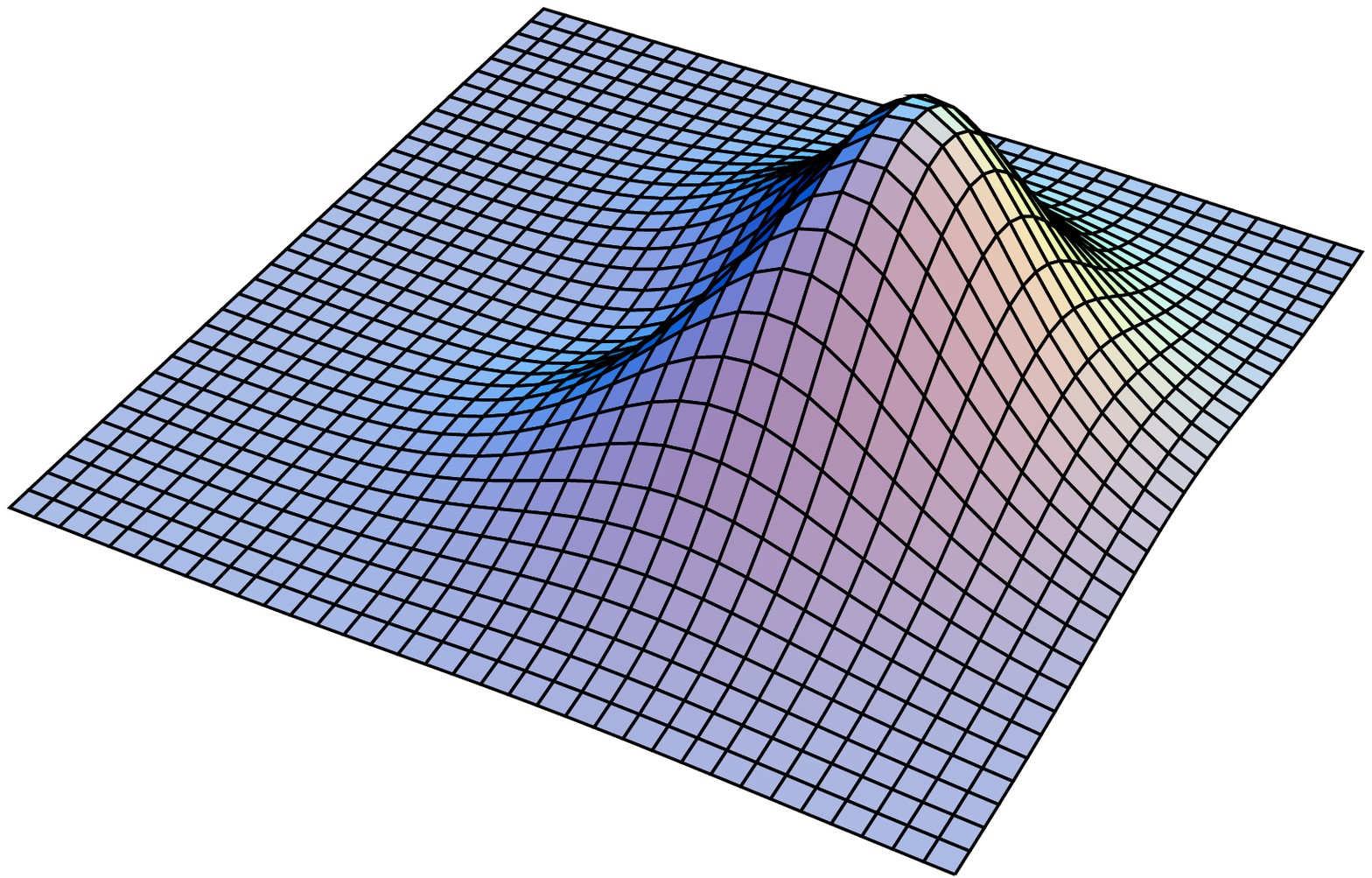}
\includegraphics{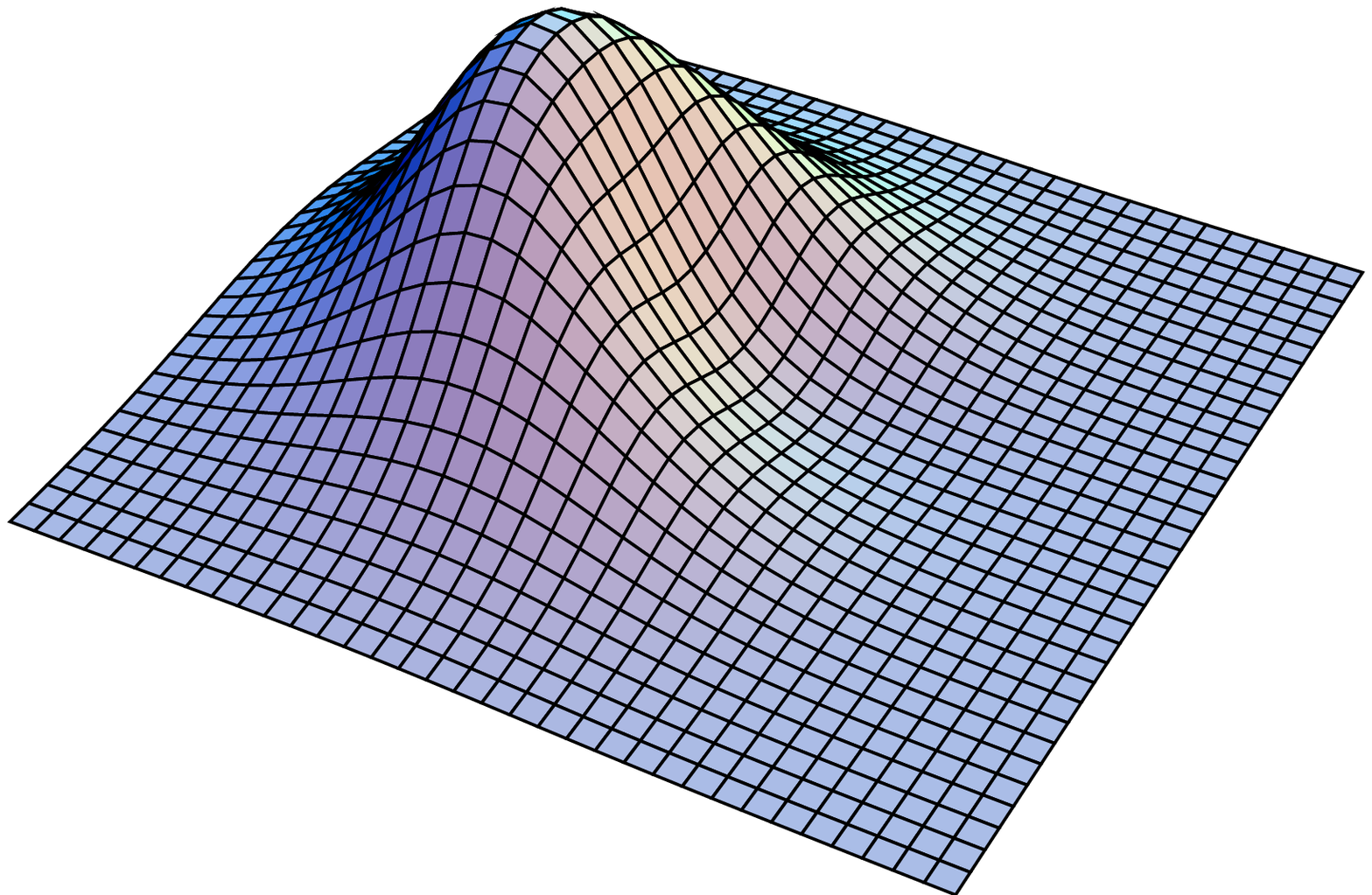}
\includegraphics{caloron2_ad.eps}
\caption{\label{crossedzm2}\footnotesize An $SU(2)$ caloron of topological charge 2, with considerably overlapping constituents. Even
though the action density in the middle does not show the constituents separately, the zero-modes do. The 2 periodic
zero-modes are on the left, the 2 anti-periodic are on the right.}
\end{figure}

We have plotted the action densities in section \ref{su2gaugefield} for two particular cases. Now we will
compare these with the behaviour of the zero-modes. The parameter values for figure \ref{crossedzm} are
the same as for \ref{crossedad}, but the scale of the zero-modes is enhanced by a factor of 5 relative to the action density.
The zero-modes are seen to follow the action density, with approximately coinciding peaks. This is due to having
a $\k$ value close to unity, however, the configuration is not static, monopoles of opposite charge are close enough
to produce time dependence.
\newpage
It is instructive to plot the configuration with strongly overlapping constituents as well, where the action density does not
show the 4 constituent peaks. Nevertheless, the zero-modes reveal the presence of constituents.
Since monopoles of opposite magnetic charge are closer than in the previous example,
this configuration shows a stronger time dependence. The parameter values for figure \ref{crossedzm2} are
the same as for \ref{crossedad2}.

\begin{figure}[htb]
\vspace{7.1cm} 
\includegraphics{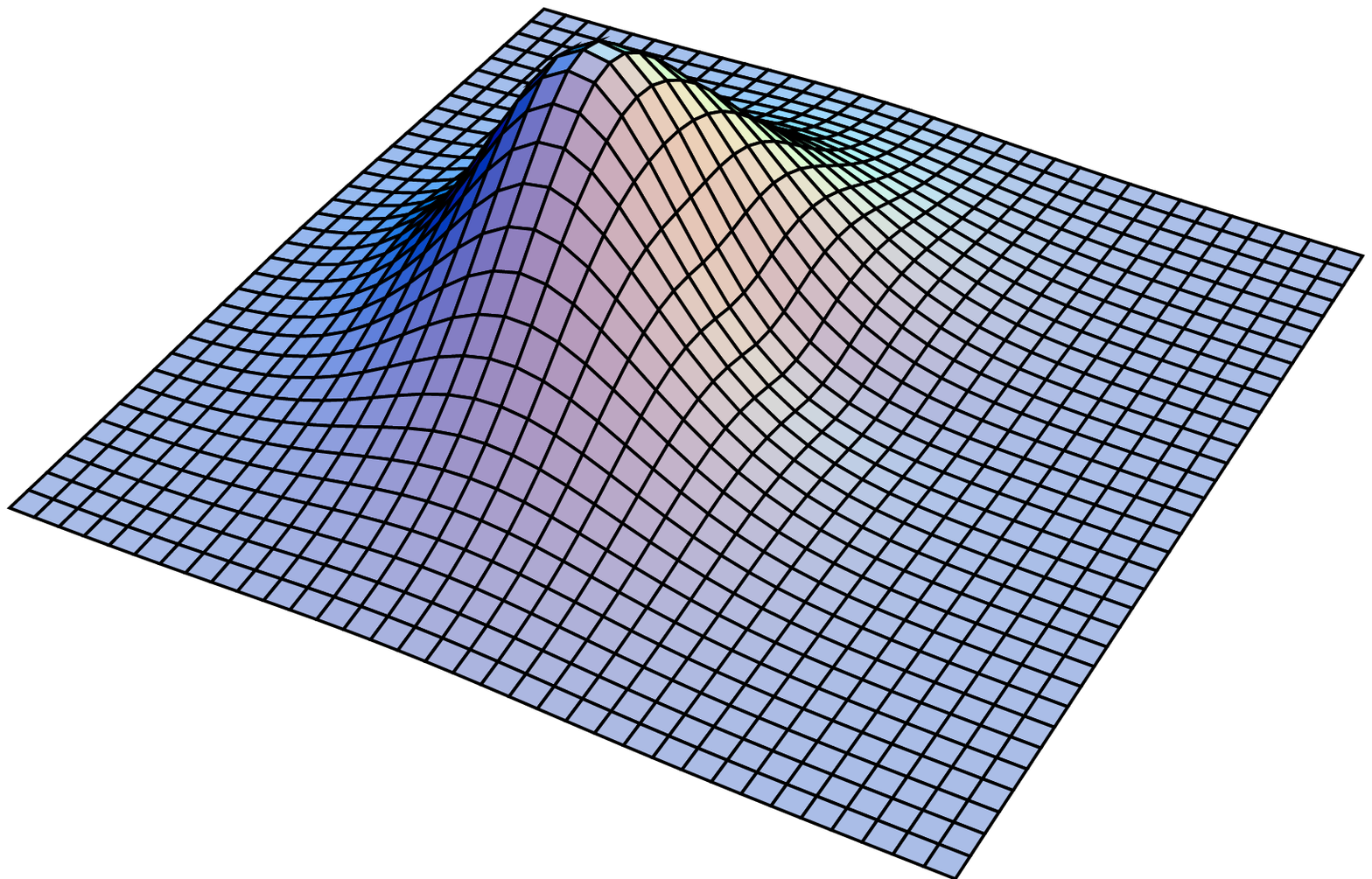}
\includegraphics{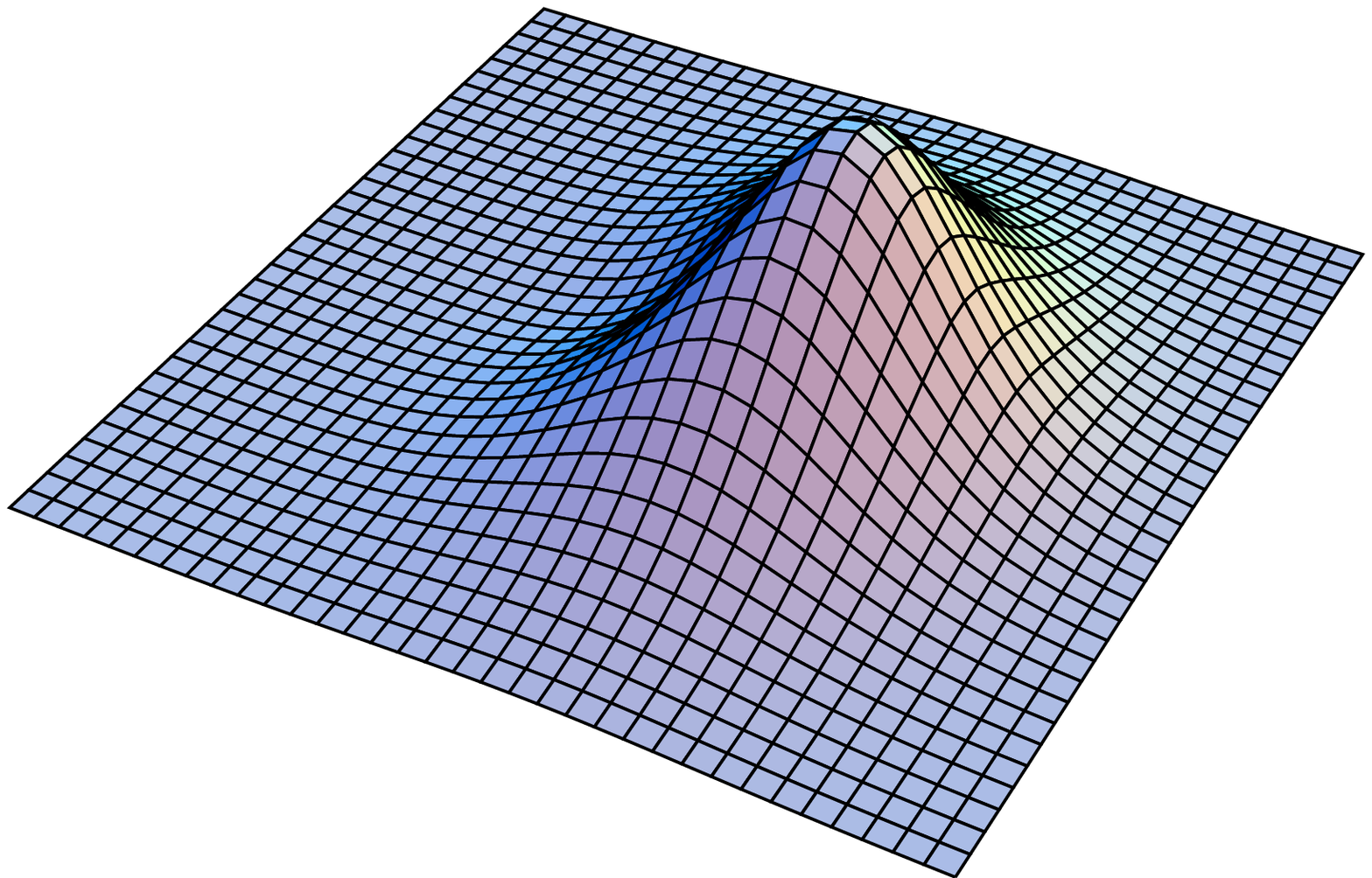}
\includegraphics{monopole_ad.eps}
\caption{\label{monopolezm}\footnotesize A charge 2 monopole for $SU(2)$ as obtained from a caloron solution in an appropriate
limit. The action density is in the middle and the two zero-modes on the sides.}
\end{figure}

We have derived in section \ref{monopolelimit} the Green function for pure magnetic monopoles as well and have plotted
the action density of a typical charge 2 monopole in figure \ref{monopolead}. In figure
\ref{monopolezm} the 2 zero-modes are shown together with the action density for the same configuration.
\newpage

\chapter{Twistors and moduli}
\label{twistors}

The moduli space of self-dual fields over various flat spaces (monopoles over $\R^3$, instantons over $\R^4$,
calorons over $S^1\times \R^3$ or Nahm-solutions over $S^1$, etc.) carry natural metrics. These are
inherited from the $L^2$-norm of the corresponding gauge field, restricted to modes transverse to the moduli space.
The geometrical properties of the spaces over which the equations are defined are
reflected by geometrical properties of the moduli space. In particular, both $\R^4$ and $S^1\times \R^3$ are hyper\kahler
and as a result the moduli spaces will carry a hyper\kahler metric. 
In the following section the main ingredients of the apparatus needed will be summarized, followed by an elementary
account of the hyper\kahler quotient method. For a review on hyper\kahler geometry see \cite{Hitchin:1986ea,Hitchin}.

We will be concerned with the moduli space $\curly M$ of $SU(n)$ calorons of arbitrary topological charge $k$ and vanishing
over-all magnetic charge with maximal symmetry breaking,
or equivalently with non-trivial holonomy at spatial infinity, and its twistor space.

As to the algebraic geometry of the
moduli space we will derive a correspondence
with stable holomorphic bundles over the projective plane $\cp$ that are trivial on two lines as opposed to one line as in the case of instantons.
The difference is due to the different way $\R^4$ and $S^1\times \R^3$ compactify to $\cp$. This correspondence is very much
along the lines of geometric invariant theory as applied to instantons \cite{Donaldson:1984tm}. We arrive at it from
an explicit parametrization of the moduli space by a number of finite dimensional complex matrices subject to a constraint, similarly to
the quadratic ADHM equations for instantons. As this correspondence both
for instantons and calorons is independent of the twistor construction, it is perhaps more fundamental than the usual correspondence
between the moduli spaces and stable holomorphic bundles over the twistor space of the base manifold (or compactifications thereof).

In general the (framed) caloron moduli space\footnote{We will be only concerned with the framed moduli space that includes the moduli
corresponding to a constant gauge rotation.} will be of real dimension $4nk$. This moduli space, just by naive counting of dimensions,
can incorporate $nk$
BPS monopoles of unit charge as every such monopole has 4 parameters, 3 for the location and 1 for a phase. We show that
this is indeed the case, as in the limit of large separations the moduli space factorizes into $nk$ copies
of $S^1\times \R^3$. For calorons of unit charge the exact metric was first conjectured in \cite{Lee:1997vp}
based on string theoretic arguments and later proved in \cite{Kraan:1998kp} for gauge group $SU(2)$ with extension to
$SU(n)$ in \cite{Kraan:1998pn}. 

\section{Hyper\kahler geometry and twistor theory}
\label{hyperkahlergeometry}

Through the prototypical example of $\R^4$ one can motivate twistorial ideas as follows \cite{AtiyahFermi}. In two dimensions
one can introduce complex coordinates $z = x + iy$ thereby identifying $\R^2 \simeq \C$. Since $SO(2) \simeq U(1)$
there is essentially a unique way to do this in a manner compatible with the metric. Studying two dimensional
problems in a single complex coordinate is a useful thing and one wants to study four dimensional problems using two
complex coordinates. However, the situation in four dimensions is different because $SO(4)$ is much bigger
than $U(2)$, in fact $SO(4)/U(2) \simeq S^2$, which we will parametrize by $\zeta$. There is a whole family of choices to identify $\R^4$ with $\C^2$,
one of which is $z_1 = x_1+ix_2$, $z_2=x_3+ix_4$. However, one does not want to single this out, for example
any other choice with the coordinates permuted is just as good. Singling out a particular choice
might spoil some symmetry of the problem or hide some features which are manifest in the formulation
with 4 real coordinates.

Thus one is lead to study problems in four dimensions by 3 complex coordinates \newline $(z_1(\zeta),z_2(\zeta),\zeta)$ where $z_1$
and $z_2$ themselves depend on $\zeta$ because they are the complex coordinates obtained by
the identification that corresponds to the choice $\zeta$. If the original problem really lives on $S^4$
then the above description gives a parametrization of its twistor space $\cpthree$ as a fibration over $\cpone$
with fibre $S^4$. This will be the general pattern as we will see.

The idea of twistors originates from Penrose \cite{Penrose:1976js} and was invented in the Lorentzian signature context but
has proved useful for the study of self-dual Yang-Mills fields in Euclidean signature too, as developed by Atiyah and Hitchin. The general idea
is that differential geometric structures on a smooth manifold are encoded in holomorphic data on its twistor space.
For example in the case of Yang-Mills theory, an instanton solution on $\R^4$ or $S^4$ is encoded in a stable holomorphic
bundle over $\cpthree$ with some certain extra properties. In general for a hyper\kahler
manifold its twistor space will encode the hyper\kahler metric itself. Here we will describe 
only this aspect of twistor theory and neglect the original motivation of Penrose \dash general relativity.

A Riemannian manifold is said to
be hyper\kahler if it is \Kahler with respect to 3 complex structures and these
satisfy the multiplicative relations of the quaternions. Thus such manifolds come with a metric $g$, 3 integrable complex
structures $I, J$ and $K$ and the corresponding 
3 K\"ahler-forms $\omega_1 = g(I\cdot,\cdot), \omega_2=g(J\cdot,\cdot)$ and $\omega_3=g(K\cdot,\cdot)$. 
The complex structures $I,J$ and $K$ obey $I^2 = J^2 = K^2 = IJK = -1$ and are covariantly constant. It is an 
elementary fact that the dimension of such manifolds is a multiple of 4.

An equivalent characterisation can be given in terms of holonomy groups. Since parallel transport preserves
$I,J$ and $K$ the holonomy group of a $4n$ dimensional hyper\kahler manifold lies in the intersection
of $O(4n)$ and $GL(n,\Q)$. The maximal such intersection is $Sp(n)$ the group of $n\times n$ quaternionic
unitary matrices, which group is one of the possible holonomy groups on Berger's list \cite{Berger}. The group $Sp(n)$
is also the intersection of $U(2n)$ and $Sp(2n,\C)$ hence a hyper\kahler manifold is naturally a complex
manifold with a holomorphic symplectic form. This observation will be heavily used in the construction
of the twistor space and in practice will
mean that many quantities will have a holomorphic dependence simplifying their study considerably.

If $M$ is hyper\kahler and $(e_1,e_2,e_3)$ is a unit vector in $\R^3$, then $I_\zeta = e_1 I + e_2 J + e_3 K$
is again an integrable complex structure, if $\zeta\in\cpone$ corresponds to $\e\in S^2$. Thus a
hyper\kahler manifold is endowed with a whole $S^2=\cpone$ family
of complex structures. It is advantageous to study all members of the family at once just as
in the introductory example of $\R^4$.

The twistor space of a $4m$ real dimensional hyper\kahler manifold is defined to be a $2m+1$ complex dimensional manifold $Z$
together with a projection $Z\to \cpone$ such that the fiber above $\zeta$ is, as a complex manifold,
$M$ endowed with the complex structure $I_\zeta$. Topologically $Z=M\times\cpone$ but its complex structure
is non-trivial and encodes all 3 complex structures of $M$. At a point $(p,\zeta)\in Z$ 
the tangent space decomposes as $T_pM \oplus T_\zeta \cpone$ and the complex structure is $I_\zeta \oplus i$,
where $i$ denotes the standard complex structure of $\cpone$. The projection $Z\to \cpone$ is in fact holomorphic.
The 3 K\"ahler-forms combine into $\omega = \omega_2 + i \omega_3 + 2\zeta\omega_1 - \zeta^2(\omega_2-i\omega_3)$,
a holomorphic symplectic form with respect to $I_\zeta$ on each fibre of $Z$. Because of the quadratic
dependence on $\zeta$ (compare with (\ref{zetaparam} - \ref{curve})) it takes values in $\otwo$. Here we use the same symbol $\otwo$ for the bundle
over $\cpone$ and its pull back to $Z$. Following standard notation, we denote by $\ok$ the bundle on $\cpone$ with 
transition function $1/\zeta^k$.

On $\cpone$ there is the antipodal map $\zeta \to -1/{\bar\zeta}$, which induces an antiholomorphic involution,
or real structure on $Z$. Let us see how points in $M$ are represented in $Z$. One can think of any point $p\in M$
as a holomorphic section $\{p\}\times \cpone$ of the projection $Z\to \cpone$. They are obviously real, that is are
left invariant by the real structure. One can show
that the normal bundle of such sections is $\C^{2n}(1)$. Here and henceforth we use the standard notation ${\curly F}(k)$
for any bundle $\curly F$ twisted by $\ok$ and for the trivial bundle with fibre $V$ twisted by $k$ we also simply write
$V(k)$.

The crucial fact is that the above construction can be reversed. Suppose that the following are given,
\begin{itemize}
\item a holomorphic projection $Z\to \cpone$ where $Z$ has complex dimension $2m+1$,
\item a holomorphic symplectic form on each fibre with values in $\otwo$,
\item a family of holomorphic sections each with normal bundle $\C^{2m}(1)$,
\item a real structure on $Z$ which induces the antipodal map on $\cpone$,
\end{itemize}
and the projection, holomorphic symplectic form and the family of sections are compatible with
the real structure. Then the parameter space of real sections of $Z$ is a hyper\kahler manifold
of real dimension $4m$ whose twistor space is $Z$. This is the precise statement how holomorphic data on $Z$ encodes
the hyper\kahler metric of $M$.

We will make use of the following result that is a direct consequence of the definition. If $\Omega$ is a
flat hyper\kahler manifold, then its twistor space is $Z=\Omega(1)$. 

\section{Self-duality and hyper\kahler quotient}
\label{selfdualityandhyper}

The self-duality equations \dash which are 3 independent equations \dash can be written as the 3 components of
a hyper\kahler moment map set to zero. In this section we first review briefly the hyper\kahler quotient construction
and then demonstrate how it applies to self-duality.

If $G$ is a compact Lie group acting freely on a hyper\kahler manifold $M$ preserving the 3 complex structures $I, J$ and $K$, then
it also preserves the 3 \Kahler forms $\omega_i$. Thus it is possible to define 3 moments maps, $m_i:M\to\gotg^*$, for each in
the standard way \cite{choquet}. We will usually identify the dual $\gotg^*$ with $\gotg$. One can show that the induced metric on the quotient
\bea
\label{nunuke}
\left. \bigcap_i m_i^{-1}(0) \right/ G
\eea
is then also hyperk\"ahler. This manifold is called the hyper\kahler quotient of $M$ with respect to $G$.

Focusing first on the complex structure $I$, let us combine $m_2$ and $m_3$ into the complex moment map $m = m_2+im_3$.
It can be shown to be a holomorphic function with respect to $I$ and furthermore to be preserved by the complexified group
$\Gc$. The corresponding complex symplectic form, $\omega = \omega_2 + i \omega_3$, can also be shown to be holomorphic
with respect to $I$. In fact, $m$ is the holomorphic moment map with respect to the action of $\Gc$ and holomorphic
symplectic form $\omega$. Then the quotient in (\ref{nunuke}) can equivalently be constructed as
\bea
\label{hnunuke}
\left. m^{-1}(0) \right/ \Gc\,,
\eea
in other words a hyper\kahler quotient can be realized as an ordinary symplectic \dash or Marsden-Weinstein \dash
quotient but in a holomorphic setting \cite{Hitchin:1986ea}. The equation $m_1=0$ is called the real equation,
whereas $m=0$ is called the complex equation.

It is easy to see explicitly how the two descriptions are equivalent. Setting $m=0$ is the same as setting
$m_2=m_3=0$. In the holomorphic setting we solve $m=0$ up to $\Gc$, whereas $\Gc/G$ can be used to
solve $m_1=0$ as well, since $m_1$ is not invariant with respect to $\Gc$ only to $G$. Thus solving $m=0$
up to $\Gc$ is the same as solving $m_1=m_2=m_3=0$ up to $G$.

Let us count the dimensions. Setting the 3 real moment maps to zero
reduces the dimension of $M$ by $3\dim G$, factoring by $G$ then reduces the dimension by $\dim G$,
leaving over-all a manifold of dimension $\dim M - 4\dim G$. In the holomorphic description setting $m=0$
reduced the dimension by $\dim \Gc = 2\dim G$, factoring by $\Gc$ further reduces the dimesion by $2\dim G$, resulting
in again a $\dim M - 4\dim G$ dimensional manifold, as it should.

What we have done above was specific to singling out $I$. 
In the previous section we have seen that it is advantageous to combine the 3 complex structures and study them all at once.
Let us now describe the hyper\kahler quotient construction
for any choice of complex structure $I_\zeta$, in other words we will consider it on the level of the
twistor space. We have mentioned that $\omega_Z = \omega_2 + i\omega_3 + 2\zeta \omega_1 -\zeta^2(\omega_2-i\omega_3)$
is holomorphic with respect to $I_\zeta$ on each fiber of $Z$ and we now analogously define
\bea
\label{twistormoment}
m_{\rm Z} = m_2 + im_3 + 2\zeta m_1 - \zeta^2(m_2-im_3) : Z\nyil\gc(2)\,,
\eea
as a $\gc(2)$-valued function on the twistor space of $M$. This complex moment map is holomorphic on $Z$ and we see that the choice $\zeta=0$ corresponds
to the discussion above when we have singled out $I$.

Just as before, $\omega_Z$ is preserved by the complexified group $\Gc$ and we are led to the
following description of the twistor space of the hyper\kahler quotient of $M$,
\bea
\label{twistorquotient}
\left.m_Z^{-1}(0)\right/\Gc\,.
\eea
Clearly, (\ref{twistorquotient}) is a generalization of (\ref{hnunuke}) which was valid in the fiber of $Z$ over $\zeta=0$ and now we
have a description for the whole of $Z$.

This concludes our brief review and now we turn to the (anti)self-duality equations on $\R^4$.
First note that $\R^4$ is flat and hyper\kahler with hyper\kahler structure
\bea
\label{r4hyper}
\omega_i = \half \eta^i_{\mu\nu} dx_\mu \wedge dx_\nu\,.
\eea
By the same token the set of gauge potentials $A_\mu$ is also hyper\kahler and flat, although infinite dimensional. Gauge transformations
with gauge group $G$ act as given in (\ref{gaugetransformation}) and preserve the hyper\kahler structure. Thus we can
invoke the hyper\kahler quotient construction. The 3 moment maps are easily seen to be
\bea
\label{momentgauge}
m_i = \half \eta^i_{\mu\nu} F_{\mu\nu} = -F_{0i} - \half \varepsilon_{ijk} F_{jk}\,.
\eea
Thus setting $m_i =0$ is equivalent to the anti-self-duality equations.
The reason for being interested in anti-self-duality now, rather then self-duality, is simply that \dash following
the literature \dash we want to arrive at holomorphic quantities, rather then anti-holomorphic. The choice is a matter
of convention and in this chapter we will stick to anti-self-duality and calorons will be anti-self-dual gauge fields
on $S^1\times \R^3$. Time reversal interchanges self-dual and anti-self-dual gauge fields and leaves the moduli space
metric invariant.

First let us introduce fixed complex coordinates $z_1=x_0-ix_1$ and $z_2=x_2-ix_3$ as well as $\alpha = A_0 + iA_1$ and $\beta = A_2 + iA_3$.
The moment maps in terms of these coordinates can be written,
\bea
\label{holgauge}
m &=& m_2 + i m_3 = -\frac{\d}{\d z_1} \beta - [\alpha,\beta]+\frac{\d}{\d z_2} \alpha\nn\\
2im_1 &=& \frac{\d}{\d \bar{z}_1} \alpha + \frac{\d}{\d z_1} \alpha^\dagger + [\alpha,\alpha^\dagger]+
\frac{\d}{\d \bar{z}_2} \beta + \frac{\d}{\d z_2} \beta^\dagger +[\beta,\beta^\dagger]\,.
\eea
Clearly, $m$ is holomorphic in $\alpha$ and $\beta$ and the set $m=0$ is preserved by $\Gc$-valued gauge transformations.
On the other hand $m_1$ is not holomorphic and the set $m_1=0$ is only preserved by the original $G$-valued gauge transformations.

This completes our review of the hyper\kahler geometry of the anti-self-duality equations on $\R^4$. These ingredients will be used for
the construction of the moduli space of calorons and its twistor space in the next section where we discuss
more details specific to our application.

\section{Moduli of calorons}
\label{moduliofcalorons}

We have seen that the gauge equivalence class of caloron solutions is in a one-to-one
correspondence with Nahm data $(\ahat_\mu,\lambda)$ satisfying Nahm's equation on the dual circle,
modulo dual gauge transformations. Let us recall
how these dual gauge transformations $g=g(z)$ act on the Nahm data,
\bea
\label{modulitr}
\ahat_0 &\nyil& g\ahat_0g^{-1} - g^\vesszo g^{-1}\,,\qquad\ahat_i \nyil g\ahat_ig^{-1}\\
\label{alone}
\lambda^A &\nyil& g(\mu_A) \lambda^A\,,
\eea
where $\lambda^A$ is the $A^{th}$ row of the $n\times k$ matrix of 2-component spinors $\lambda$. Simply
dimensionally reducing to one dimension the formulae in the previous section we obtain a flat hyper\kahler
description for the dual gauge field $\ahat_\mu$. The space of all $\lambda$ is $\C^{2nk}$ which is also hyper\kahler
and flat. It is easy to see that the jumps $i\rho_j^A$, which are quadratic in $\lambda$, are the hyper\kahler
moment maps with respect to the action (\ref{alone}) at $z=\mu_A$.

From the previous section it then follows that
the moduli space of calorons is the hyperk\"ahler quotient of the space of all Nahm data $(\ahat_\mu,\lambda)$
with respect to the action (\ref{modulitr}-\ref{alone}). This moduli space $\curly M$ will be our main object of study.

Let us set $G=U(k)$, $\; \Gc = \GL$ for its complexification and $\gotg = u(k)$ and $\gc = \gl$ for
the corresponding Lie algebras. 

The above quotient construction can be performed in two stages. First quotienting out by gauge transformations which are the identity
at the jumping points $z=\mu_A$, followed by quotienting out by the remnant gauge transformations located at the jumping points.
In the first stage the jumping data does not play a role as it is left invariant, hence one can concentrate on the
dual gauge field $\ahat$ alone. On each $(\mu_A,\mu_{A+1})$ interval it satisfies the homogeneous Nahm equation subject to
the boundary condition that at the endpoints $\ahat(\mu_A)$ and $\ahat(\mu_{A+1})$ are finite.
Henceforth we will restrict our attention to this fixed interval only.

This situation was analyzed in \cite{Kronheimer}, see also \cite{Dancer:1992yg}; we will follow the same argument.
Using the complex coordinates $\alpha$ and $\beta$ introduced in the previous section, dimensional
reduction to one dimension of (\ref{holgauge}) leads to
\bea
\label{real}
\left( \alpha + \alpha^\dagger \right)^\vesszo + [\alpha,\alpha^\dagger] + [\beta,\beta^\dagger] &=& 0 \\
\label{complex}
\beta^\vesszo + [\alpha,\beta] &=& 0 \, .
\eea
We have seen that the space of solutions of equations (\ref{real}-\ref{complex}) modulo gauge transformations by $G$
is the same as the space of solutions of the complex equation (\ref{complex}) alone, modulo $\Gc$ gauge transformations \cite{Donaldson:1985id}.
This general principle of hyperk\"ahler geometry can be shown explicitly to hold in this setting as follows.
The complex equation (\ref{complex}) can be solved by
\bea
\label{gamma}
\alpha &=& -\gamma^\vesszo \gamma^{-1} \nn \\
\beta &=& \gamma B \gamma^{-1}
\eea
with a constant $B=\beta(\mu_A) \in \gc$ and $\Gc$-valued $\gamma(z)$ on which we impose $\gamma(\mu_A)=1$ for definiteness.
$\Gc$ gauge transformations act on $\gamma$ as $\gamma(z) \nyil g(z) \gamma(z)$,
hence the only gauge invariant quantity
is $h = \gamma(\mu_{A+1})$ (remember that we only quotient at this stage with gauge transformations which are
the identity at the endpoints of the interval). Plugging this form into the real equation (\ref{real}) gives for $V=\gamma^\dagger \gamma$,
\bea
\label{P}
\left( V^\vesszo V^{-1} \right)^\vesszo + [B^\dagger,V B V^{-1}] = 0 \, ,
\eea
subject to the boundary condition $V(\mu_A) = 1$ and $V(\mu_{A+1}) = h^\dagger h$. Such a solution\footnote{
Note that eq.\ (\ref{P}) is the Euler-Lagrange equation for 
$S = \int_{\mu_A}^{\mu_{A+1}} \tr \left( \half \left( V^\vesszo V^{-1} \right)^2+B^\dagger V B V^{-1} \right)dz$, that is
a 1-dimensional WZNW-like model with a potential.} for $V$
on the interval $[\mu_A,\mu_{A+1}]$ is unique and we can reconstruct $\gamma$ from $V$
up to a $G$ gauge transformation. In fact, $V$
can be thought of as being $\Gc / G$-valued, in accordance with the general discussion in the previous section. 
Hence indeed
the moduli space for eqs. (\ref{real}-\ref{complex}) with $G$ invariance is the same as the moduli space
for the complex equation alone with $\Gc$ invariance.

In addition we have also obtained that this moduli space can be
parametrized by $(h,B) \in \Gc \times \gc \simeq T^* \Gc$, the cotangent bundle of $\Gc$. In terms of $\alpha$ and $\beta$ they are
given by
\bea
\label{tttt}
h&=&{\rm P} \exp \left(-\int_{\mu_A}^{\mu_{A+1}} \alpha(z) dz \right)\nn\\
B&=&\beta(\mu_A)\,.
\eea

The above discussion applies to each of the $n$ intervals separately, giving reduced Nahm data $(h_A,B_A)$ for each interval labelled by $A$.
Now we have to incorporate the jumping data. At each $z=\mu_A$ these are 2-component spinors in the fundamental
representation of $G$, which can be cast into a $k$ dimensional column vector $u_A$, and $k$ dimensional
row vector $w_A$, in such a way that $\rho_2^A + i \rho_3^A = u_A \otimes w_A$. This is because we have
seen in section (\ref{structureofthejumps}) that whenever $\y$ is null, $y_i \rho_i^A$ is of rank 1 and
now we take $\y = (0,1,i)$.

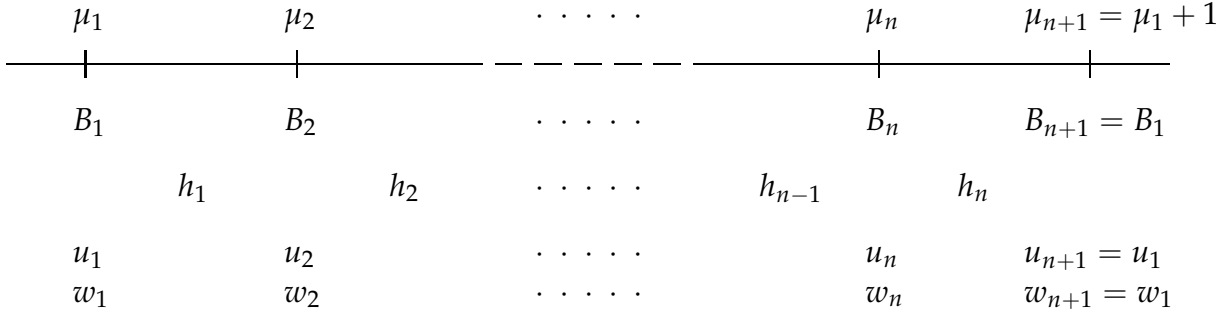
\begin{figure}
\setlength{\unitlength}{1pt}
\begin{picture}(260,140)
\put (0,90){\line(1,0){180}}
\put (185,90){\line(1,0){10}}
\put (200,90){\line(1,0){10}}
\put (215,90){\line(1,0){10}}
\put (230,90){\line(1,0){10}}
\put (245,90){\line(1,0){10}}
\put (260,90){\line(1,0){180}}

\put (30,85){\line(0,1){10}}
\put (110,85){\line(0,1){10}}
\put (330,85){\line(0,1){10}}
\put (410,85){\line(0,1){10}}

\put (25,105){$\mu_1$}
\put (105,105){$\mu_2$}
\put (200,105){$\cdot\;\cdot\;\cdot\;\cdot\;\cdot$}
\put (325,105){$\mu_n$}
\put (385,105){$\mu_{n+1}=\mu_1+1$}

\put (25,65){$B_1$}
\put (105,65){$B_2$}
\put (200,65){$\cdot\;\cdot\;\cdot\;\cdot\;\cdot$}
\put (325,65){$B_n$}
\put (385,65){$B_{n+1}=B_1$}

\put (65,40){$h_1$}
\put (145,40){$h_2$}
\put (200,40){$\cdot\;\cdot\;\cdot\;\cdot\;\cdot$}
\put (285,40){$h_{n-1}$}
\put (360,40){$h_n$}

\put (25,15){$u_1$}
\put (105,15){$u_2$}
\put (200,15){$\cdot\;\cdot\;\cdot\;\cdot\;\cdot$}
\put (325,15){$u_n$}
\put (385,15){$u_{n+1}=u_1$}

\put (25,0){$w_1$}
\put (105,0){$w_2$}
\put (200,0){$\cdot\;\cdot\;\cdot\;\cdot\;\cdot$}
\put (325,0){$w_n$}
\put (385,0){$w_{n+1}=w_1$}
\end{picture}
\vspace{5mm}
\caption{\label{figura}Reduced Nahm data on the dual circle.}
\end{figure}

These vectors make up the flat quaternionic space $\Q^{nk}$. In
figure \ref{figura} the reduced Nahm data is summarized showing that the matrices $B_A$ and vectors $u_A, w_A$
are located at the jumping points with $h_A$ giving the propagation in between.

Now we proceed to the second stage of the reduction, namely quotienting out with the constant gauge
transformations at the jumping points. These remnant complexified gauge transformations
$g_A \in \Gc$ act as
\bea
\label{remnant}
B_A \nyil g_A B_A {g_A}^{-1}\, , \quad h_A \nyil g_{A+1} h_A g_A^{-1}\, , \quad u_A \nyil g_A u_A\, , \quad w_A \nyil w_A g_A^{-1}\, ,
\eea
with associated complex moment map equations
\bea
\label{chain}
B_{A+1} - h_A B_A h_A^{-1} = u_{A+1} \otimes w_{A+1}\, .
\eea
At this point the moduli space $\curly M$ consists of elements in $(T^* \Gc)^n \times {\mathbb H}^{nk}$ subject to eq.\ (\ref{chain})
modulo the action of $\Gc^n$ given by (\ref{remnant}), with
\bea
\label{notacio}
(T^* \Gc)^n&=& \underbrace{T^*\Gc \times \cdot\cdot\cdot \times T^*\Gc}_{n}\nn\\
\Gc^n&=&\underbrace{\Gc\times \cdot\cdot\cdot \times \Gc}_{n}\,.
\eea

However, we can explicitly solve for $n-1$ of the $B_A$
variables using (\ref{chain}) and can gauge away the corresponding $h_A$ variables using the gauge transformations
at $\mu_2,\, \mu_3,$ $\ldots, \mu_n$. The remaining variables are
$B=B_1,\, h=h_n h_{n-1}\cdots h_1$ and some gauge transforms of the original $u_A$ and $w_A$ vectors which
we will continue to label by $u_A$ and $w_A$ and assemble into a $k\times n$ and an $n\times k$ matrix
$u=(u_{aA})$ and $w=(w_{Aa})$. The remaining symmetry by $\Gc$ is
\bea
\label{g}
B\nyil g B g^{-1}\, , \qquad h \nyil g h g^{-1}\, , \qquad u \nyil g u\, , \qquad w \nyil w g^{-1}\, ,
\eea
where $g=g_1$. This action gives the complex moment map equation
\bea
\label{B}
B - h B h^{-1} = u w\, .
\eea

The result of the present section is that the moduli space of $SU(n)$ calorons of topological charge $k$ and zero overall magnetic charge with maximal
symmetry breaking can be parametrized by matrices $(h,B,u,w)$ subject to eq.\ (\ref{B}) modulo the $\Gc$-action (\ref{g}).

Now recall that the ADHM construction gives the moduli space of $SU(n)$ instantons on $\R^4$ (or $S^4$) of topological charge $k$ as the space of
four $k\times k$ hermitian matrices $B_\mu$ and a 2-component spinor of $n\times k$ matrices
$\lambda_\alpha$, subject to the quadratic equations
\bea
\label{moduliadhm}
~[B_0,B_1]+[B_2,B_3]&=&\frac{1}{2i}\left( \lambda_1^\dagger \lambda_2 + \lambda_2^\dagger\lambda_1\right)\nn\\
~[B_0,B_2]+[B_3,B_1]&=&\half\left(\lambda_2^\dagger\lambda_1-\lambda_1^\dagger\lambda_2\right) \\
~[B_0,B_3]+[B_1,B_2]&=&\frac{1}{2i}\left(\lambda_1^\dagger\lambda_1-\lambda_2^\dagger\lambda_2\right)\,,\nn
\eea
modulo a natural $U(k)$ action. Note a sign change relative to (\ref{adhm}) as we are now describing anti-self-dual,
rather then self-dual fields.

It follows from (\ref{holgauge}) by dimensional reduction to zero dimension
that the ADHM equations (\ref{adhm}) can also be seen as setting a hyperk\"ahler moment map to zero. In terms of
the complex coordinates
\bea
\label{adhmcomplexcoord}
a = \frac{\lambda_1^\dagger + \lambda_2^\dagger}{\sqrt{2}}\,,&&\qquad \nn\alpha = B_0 + i B_1\nn\\
b = \frac{\lambda_1 - \lambda_2}{\sqrt{2}}\,,&&\qquad \beta = B_2 + i B_3\,,
\eea
where $a$ is $k\times n$ and $b$ is $n\times k$, they turn into,
\bea
\label{adhmc}
~[\alpha,\beta] + ab &=&0\\
\label{adhmr}
~[\alpha,\alpha^\dagger]+[\beta,\beta^\dagger] +aa^\dagger - bb^\dagger&=& 0\,,
\eea
where again (\ref{adhmc}) is called the complex, (\ref{adhmr}) the real equation. Applying the general principle of hyperk\"ahler
geometry to this case implies that the moduli space of instantons can be identified with the space
of matrices $(\alpha,\beta,a,b)$ subject to the complex equation $[\alpha,\beta] + ab =0$ alone, modulo the
complexified group $\GL$ \cite{Donaldson:1984tm}.

What we have found is that the moduli space of calorons on $S^1 \times \R^3$ is almost the same space upon the
identification
\bea
\label{id}
\alpha=h,\qquad \beta=B,\qquad a=u,\qquad b=wh
\eea
except for the condition that one of the two $k\times k$ matrices should be non-degenerate as $\det h \neq0$. Hence we have found an embedding
of the caloron moduli space into the instanton moduli space as an open subset. The identification (\ref{id}) gives a 
dictionary how to translate the $4nk$ instanton moduli consisting of $k$ scales and 4-dimensional locations plus some
gauge orientations into the same number of caloron moduli consisting of $nk$ 3-dimensional locations and phases characterizing
the $nk$ constituent monopoles. 

\subsection{Stable bundles on the projective plane}
\label{stable}

Based on the above result a description in terms of stable holomorphic bundles on $\cp$ is possible. The relationship
between stable bundles on projective space and instantons goes back to \cite{Atiyah:1978ri,
Drinfeld:1978xr, Atiyah:1977bu, Hartshorne:1978vv}.
The original construction gave a correspondence between instantons and bundles on the twistor space of $S^4$, which is
$\mathbb C \rm P^3$. The analogous correspondence for calorons was described in \cite{Garland:1988bv} giving
bundles over the twistor space of $S^1\times \R^3$.

Later it was found in \cite{Donaldson:1984tm} that compactifying the Euclidean 4-space to the projective plane $\cp$ instead
of $S^4$ gives the instanton moduli spaces an interpretation in terms of bundles on $\cp$
without reference to twistor methods. In some sense this relationship is more fundamental than the twistor theoretic one.
Our construction is a simple extension of \cite{Donaldson:1984tm} and we summarize its essential ingredients below,
for more details see \cite{Barth}.

As said above, $\R^4$ is compactified to $\cp$ by adding a "line at infinity". For any $X \in \cp$ let
$[z_1:z_2:z_3]$ denote its homogeneous coordinates. Then a monad is a sequence,
\bea
\label{monad}
\C^k \overset{C_X}{\nyil} \C^{2k+n} \overset{D_X}{\nyil} \C^k
\eea
where the linear maps $C_X$ and $D_X$ depend linearly on $z_i$, $C_X$ is injective, $D_X$ is surjective and in
addition $D_X C_X=0$. Because of the linear dependence on the coordinates we can write $D_X = z_i D_i,\; C_X = z_i C_i$ and
the $D_X C_X =0$ condition gives six quadratic equations $D_i C_j + D_j C_i =0$. Every such monad defines
a rank-$n$ holomorphic bundle on $\cp$ by assigning to each point $X \in \cp$ the fibre ${\rm ker} D_X / {\rm im} C_X$
and this bundle will have second Chern class $k$. Conversely,
it can be shown that any bundle on $\cp$, that is trivial on a line comes from a monad.

The condition of triviality on a fixed line, say $[z_1:z_2:0]$, means for the matrices $C_i$ and $D_i$ that $D_1 C_2 = -D_2 C_1$
is non-degenerate. In this case by an appropriate choice of bases for the three vector spaces in (\ref{monad}) one
can achieve that $D_1 C_2 = 1$ and also that
\bea
\label{blocks}
C_1 = \left( \begin{array}{c} 1 \\ 0 \\ 0 \end{array} \right) \qquad C_2 &=& \left( \begin{array}{c} 0 \\ 1 \\ 0 \end{array} \right) \qquad
C_3 = \left( \begin{array}{c} \alpha \\ \beta \\ b \end{array} \right) \nn \\ \\
D_1 = \left( \begin{array}{ccc} 0 & 1 & 0 \end{array} \right) \quad D_2 &=& \left( \begin{array}{ccc} -1 & 0 & 0 \end{array} \right) \quad
D_3 = \left( \begin{array}{ccc} -\beta & \alpha & a \end{array} \right) \nn
\eea
for some $\alpha,\, \beta,\, a$ and $b$ matrices where the first two components of the $C_i$ and $D_i$
are $k\times k$, the last component of $C_i$ is $n\times k$ and
the last component of $D_i$ is $k\times n$. From the six quadratic constraints only one remains, namely $[\alpha,\beta]+ab=0$.
This is the same as the complex ADHM equation (\ref{adhmc}) hence showing the aforementioned correspondence
between instantons on $\R^4$ and holomorphic bundles on the projective plane.

Now we are in a position to determine what the extra condition $\det h \neq 0$
means in terms of holomorphic bundles. Upon the identification (\ref{id}) we obtain the following monad data
\bea
\label{monaddata}
C_3 &=& \left( \begin{array}{c} h \\ B \\ wh \end{array} \right) \nn \\ \\
D_3 &=& \left( \begin{array}{ccc} -B & h & u \end{array} \right) \nn
\eea
with $C_{1,2}$ and $D_{1,2}$ as before. Using that $D_3 C_2 = - D_2 C_3 = h$ and the same argument \cite{Donaldson:1984tm} that
led to the conclusion that triviality on the line $[z_1:z_2:0]$ means that $D_1 C_2 = -D_2 C_1$ is non-degenerate,
we conclude that the bundle has to be trivial on the line $[0:z_2:z_3]$ as well.

This is the result of the present section; there is a one-to-one correspondence between the moduli space of
$SU(n)$ calorons of charge $k$ with maximal symmetry breaking and zero magnetic charge and the moduli space of stable rank-$n$ holomorphic
bundles on the projective plane having second Chern class $k$ which are trivial on {\em two} distinct lines. 

A simple geometric picture clarifies how triviality on two distinct lines in the holomorphic language
and the base manifold $S^1\times \R^3$ in the gauge theory language are related. Two lines in the projective
plane intersect in a single point. Our holomorphic bundle is trivial on both hence we can decompactify
by removing them. The first removal gives $\cp - \cpone \simeq \C^2$ just as in the original
instanton construction but now we have to remove the other $\cpone$ as well. They intersect in
a point that has been removed already by the first $\cpone$ leaving a $\cpone - \{*\} \simeq \C$
behind from the second line. Now if we parametrize $\C^2$ by $(z_1,z_2)$ and remove the remaining plane $(0,z_2)$ as well, we obtain
$\C^2 - \C \simeq \C^* \times \C \simeq S^1 \times \R^3$.

Thus, what we have shown is essentially that the gauge potential $A_\mu(x)$, defined on $S^1\times\R^3 \simeq \C^*\times\C$,
extends holomorphically to $\cp$. An analogous extension was proved for calorons in the twistor theoretic correspondence
in \cite{Garland:1988bv}. We would like to emphasize again, that our construction \dash just as Donaldson's
construction for instantons \dash does not rely on the twistor correspondence.

\subsection{Twistor space and spectral data}
\label{twistor}

So far we have identified the caloron moduli space $\curly M$ as a complex manifold but we have not said anything about
the induced hyperk\"ahler metric on it. This is encoded in its twistor space and has a convenient description
in terms of spectral data, similarly to $SU(2)$ BPS monopoles as mentioned in secion \ref{magnetic}. The spectral
data will be derived from the twistor construction \cite{Hitchin:1986ea} which will be the subject of the
present section.

In order to find the twistor space $\curly Z$ one has to redo most of the first part of section \ref{moduliofcalorons}
with an arbitrary choice of complex
structure labelled by $\zeta \in \cpone$ and trace the dependence on $\zeta$. This dependence can be expressed by
transition functions from the patch $U=\{\zeta \in \cpone | \zeta \neq \infty\}$ to the patch $V=\{\zeta \in \cpone | \zeta \neq 0\}$
under $\zeta \nyil \tilde{\zeta}=\frac{1}{\zeta}$ since there exist holomorphic trivializations over both $U$ and $V$.
Quantities defined over $V$ will be denoted by a tilde.

It turns out that for this purpose the parametrization (\ref{g}) is not very useful, it is better to go back
to eq.\ (\ref{chain}) with variables in $(T^* \Gc)^n \times {\mathbb H}^{nk}$ and symmetry $\Gc^n$. First we will
describe the twistor space of $T^* \Gc$ that corresponds to a given interval, then take $n$ copies
together with the jumping data ${\mathbb H}^{nk}$ and carry out the reduction by $\Gc^n$. The hyperk\"ahler structure on
$T^* \Gc$ and its twistor space have been described in \cite{Kronheimer}, below we will reproduce the ingredients we need.

To identify the twistor space of $T^* \Gc$ recall that it came from an infinite dimensional hyperk\"ahler
reduction of Nahm data on an interval with regular endpoints. The vector space $\Omega$ of $(\alpha,\beta)$ pairs
is flat, hence its twistor space is $\Omega(1) \to \cpone$. 

The fact that the twistor space is $\Omega(1)$ means that one can pick the complex structure corresponding to $\zeta$
in such a way that $(\alpha,\beta)$ change on the overlap of $U$ and $V$ according to
\bea
\label{abtransform}
\alpha &\nyil& \tilde{\alpha} = \frac{\alpha}{\zeta}\nn\\
\beta &\nyil& \tilde{\beta} = \frac{\beta}{\zeta}\,.
\eea
Complexified gauge transformations (which are the identity at the endpoints of the interval) act over $U$ as
\bea
\label{zetagauge}
\alpha &\nyil& g \alpha g^{-1} - g^\vesszo g^{-1}\nn\\
\beta &\nyil& g \beta g^{-1} - \zeta g^\vesszo g^{-1}
\eea
and over $V$ as
\bea
\label{zetagaugetilde}
\tilde{\alpha} &\nyil& g \tilde{\alpha} g^{-1} - \tilde{\zeta} g^\vesszo g^{-1}\nn\\
\tilde{\beta} &\nyil& g \tilde{\beta} g^{-1} - g^\vesszo g^{-1}\,.
\eea
The corresponding complex moment map equation (generalizations of eq.\ (\ref{complex})) are easily computed
over $U$ and $V$ to be
\bea
\label{zetacomplex}
\beta^\vesszo + [\alpha,\beta]+\zeta \alpha^\vesszo &=& 0\nn\\
{\tilde{\alpha}}^\vesszo + [\tilde{\alpha},\tilde{\beta}]+\tilde{\zeta}{\tilde{\beta}}^\vesszo &=&0\,.
\eea
Their solution (generalization of eq.\ (\ref{gamma})) over $U$ is
\bea
\label{zetagammaU}
\alpha &=& -\gamma^\vesszo \gamma^{-1}\nn\\
\beta &=& \gamma B \gamma^{-1} + \zeta \gamma^\vesszo \gamma^{-1}\, ,
\eea
and over $V$
\bea
\label{zetagammaV}
\tilde{\alpha} &=& \tilde{\gamma} \tilde{B} {\tilde{\gamma}}^{-1} - \tilde{\zeta} {\tilde{\gamma}}^\vesszo {\tilde{\gamma}}^{-1}\nn\\
\tilde{\beta} &=& {\tilde{\gamma}}^\vesszo {\tilde{\gamma}}^{-1}\, ,
\eea
with constant matrices $B, \tilde{B}$ and $\Gc$-valued functions $\gamma(z), \tilde{\gamma}(z)$ subject to
the initial conditions $\gamma(\mu_A)={\tilde{\gamma}}(\mu_A)=1$. From the latter we define
the gauge invariant variables $h=\gamma(\mu_{A+1}),\,$ $\tilde{h}=\tilde{\gamma}(\mu_{A+1})$, both in $\Gc$,
analogously to section \ref{moduliofcalorons}.

To specify the twistor
space of $T^* \Gc$ the transition function $(h,B) \to (\tilde{h},\tilde{B})$ is needed. Substituting
$(\tilde{\alpha},\tilde{\beta})$ from (\ref{abtransform}) and $(\alpha,\beta)$ from (\ref{zetagammaU}) into (\ref{zetagammaV})
gives
\bea
\label{gammabtilde}
\tilde{\gamma}(z) &=& \gamma(z) \exp\left({(z-\mu_A)\frac{B}{\zeta}}\right)\\
\tilde{B} &=& \frac{B}{\zeta^2}\nn
\eea
where the factor of $\mu_A$ appeared in order to maintain $\tilde{\gamma}(\mu_A)=1$, once $\gamma(\mu_A)=1$. From
(\ref{gammabtilde}) we obtain the desired transition function for $(h,B)$,
\bea
\label{hbtilde}
\left(\tilde{h},\tilde{B}\right)&=&\left(h \exp\left({\nu_A\frac{B}{\zeta}}\right),\frac{B}{\zeta^2}\right)
\eea
where $\nu_A = \mu_{A+1} - \mu_A$ is the length of the interval and is related to the masses of the constitutent monopoles
of type $A$. Now we are able to identify the twistor space of
$T^*\Gc$ as
\bea
\label{z}
Z\nyil\gc(2)\nyil\cpone\, ,
\eea
where the first arrow is a principal $\Gc$ bundle over the total space of $\gc(2)$ with transition
function $\exp\left(\nu_A B/\zeta\right)$ and $h$ is a section of this principal bundle, while $B$ is
a section of $\gc(2)$.

The holomorphic symplectic form on each fibre of $Z$ is the natural invariant 2-form 
$\omega = d \, \tr \left( B h^{-1} dh\right)$. Its transformation rule is
\bea
\label{omegazeta}
\tilde{\omega} &=& d \, \tr \left( \tilde{B} \tilde{h}^{-1} d\tilde{h}\right) =
d \, \tr \left( \zeta^{-2} B \exp\left(-\nu_A\zeta^{-1}B\right) h^{-1} d\left(h\exp\left({\nu_A\zeta^{-1}B}\right)\right)\right)=\nn\\
&=&\zeta^{-2} d \, \tr \left(Bh^{-1}dh\right) + \zeta^{-2} d\, \tr \left( \exp\left(-\nu_A\zeta^{-1}B\right) d \exp\left(\nu_A\zeta^{-1}B\right) B\right)=\\
&=&\zeta^{-2} d\,\tr\left(Bh^{-1}dh\right) + \zeta^{-3} \nu_Ad\,\tr \left(B d\,B\right)=\zeta^{-2} d\,\tr\left(Bh^{-1}dh\right)=\zeta^{-2}\omega\, ,\nn
\eea
hence it is a globally defined $\otwo$ valued 2-form along the fibres, as it should be.

The above discussion applies to every interval labelled by $A$ separately and we obtain twistor spaces $Z_A$
for each. Note that they are not identical, they have different transition functions as the masses
of the constituent monopoles may vary.

Now, parallel to section \ref{moduliofcalorons}, we can incorporate the $u_A$ and $w_A$ jumping data that makes up the flat $\mathbb H^{nk}$
space. Again, because of flatness its twistor space is $\mathbb H^{nk}(1)\to\cpone$.

The action of $\Gc^n$ on the reduced Nahm data is the same as in (\ref{remnant}) giving $\gc(2)$ valued
moment maps which when set to zero give eq.\ (\ref{chain}) just as before.

Upon quotienting with $\Gc^n$ we recover the well-known spectral data of calorons \cite{Garland:1988bv,Nahm:1982jt} in the following way. Part of the gauge
invariant quantities are the spectral curves $S_A$ in $\otwo$ associated to each interval defined by
$S_A = \{ (\zeta,\eta_A)\in \otwo\;|\; \det (\eta_A - B_A(\zeta)) = 0 \}$, where $B_A$ is a section of
$\gc(2)$ and $\eta_A$ is a section of $\otwo$. These curves have
genus $(k-1)^2$. Also gauge invariant are the sections over $\otwo$
\bea
\label{psi}
\psi_A = w_{A+1} h_A \adj \left(\eta_A - B_A\right) u_A\,.
\eea
Since $((h_A,B_A),(u_A,w_A))$ is a
section of $Z_A \times \mathbb H^k(1)$ with known transition functions, the transformation of $\psi_A$ under $\zeta\to\frac{1}{\zeta}$ is
\bea
\label{psitrans}
\tilde{\psi}_A = \zeta^{-2k} \exp\left(\nu_A\frac{\eta_A}{\zeta}\right)\psi_A\, ,
\eea
once the $\psi_A$ is restricted to the curve $S_A$. In other words the $\psi_A$ are sections of the line bundle $L^{\nu_A}(2k)|_{S_A}$
first introduced in \cite{Hitchin:1982gh} in the context of magnetic monopoles, see also \cite{Nahm:1982jt}. The transition
function in (\ref{psitrans}) can actually be taken as the definition of the line bundle $L^{\nu_A}(2k)$ over $\otwo$.

The invariants $\psi_A$ do not exhaust the list of gauge invariant quantities, the combinations
\bea
\label{psichain}
\psi_{AC} = w_{A+1} h_A \adj \left(\eta_A - B_A\right) \cdot\cdot\cdot h_C \adj \left(\eta_C - B_C\right) u_C, \quad A\geq C
\eea
are all gauge invariant and obviously $\psi_A = \psi_{AA}$. One can think of the $\psi_{AC}$ as being associated to the
interval $(\mu_C,\mu_{A+1})$ which for $A-C\geq1$ contains several jumping points, while the $\psi_A$ invariants are associated to a basic interval
$(\mu_A, \mu_{A+1})$. Again, the transition functions for the invariants $\psi_{AC}$
follow from eq.\ (\ref{hbtilde}),
\bea
\label{psichaintrans}
\tilde{\psi}_{AC} = \zeta^{-2(A-C+1)(k-1)-2} \exp\left(\frac{\nu_C\eta_C + \cdot\cdot\cdot + \nu_A \eta_A}{\zeta}\right)\psi_{AC}\, ,
\eea
once they are restricted to $S_D$ in each variable $\eta_D$ for $D=C,\ldots,A$. Here and throughout $A-C$ stands for the
difference modulo $n$.

The jumping conditions (\ref{chain}) impose constraints on the sections $\psi_A$. Specifically,
\bea
\label{vanish}
\psi_{A+1} \psi_A = \left(\eta_{A+1}-\eta_A\right) \psi_{A+1,A}\, ,
\eea
where we have used the fact that
$\psi_A$ is a section over $S_A$, thus $\det (\eta_A - B_A) = 0$ and similarly for $A+1$.
This means that over the intersection of neighbouring spectral curves $S_A$ and $S_{A+1}$, where $\eta_A = \eta_{A+1}$, either $\psi_A$ or
$\psi_{A+1}$ vanishes. Analogously to eq.\ (\ref{vanish}) the jumping conditions (\ref{chain}) give
\bea
\label{twistorrel}
\psi_{A,D+1} \psi_{DC} = \left(\eta_{D+1}-\eta_D\right)\psi_{AC} \qquad (\rm{no\; sum})\, .
\eea
for $A-C\geq 1$. From the relations (\ref{vanish}-\ref{twistorrel}) it follows directly that
\bea
\label{simpler}
\prod_{D=C}^A \psi_D = \psi_{AC}\prod_{D=C}^{A-1} \left(\eta_{D+1}-\eta_D\right) \, ,
\eea
in other words, if the interval $(\mu_C,\mu_{A+1})$ contains several jumping points and hence can be broken into $A-C+1$
basic intervals $(\mu_C,\mu_{C+1}), (\mu_{C+1},\mu_{C+2}), \ldots, (\mu_A,\mu_{A+1})$ then there is a relation
between the invariant $\psi_{AC}$ and its "constituents" $\psi_D$ associated to $(\mu_D,\mu_{D+1})$ for $D=C,\ldots,A$.
In fact once (\ref{simpler}) holds the relations (\ref{vanish}-\ref{twistorrel}) follow.
It is clear from eq.\ (\ref{simpler}) that locally the invariants $\psi_{AC}$ can be expressed by the $\psi_A$, which
implies that
the spectral curves $S_A$ and the sections $\psi_A$ over them can be used as local coordinates for the
twistor space $\curly Z$. 

More concretely, each spectral curve $S_A$ is given by a polynomial of order $k$ with
leading coefficient $1$, which may be parametrized by its roots $\eta_{Aa}$.
The sections $\psi_A$ are restricted to these curves, thus can always be written as a polynomial of order $k-1$,
\bea
\label{psiseries}
\psi_A = \sum_{a=1}^k \xi_{Aa} \eta_A^{a-1} \, ,
\eea
with $nk$ coefficents $\xi_{Aa}$. Then the parameters $(\zeta,\eta_{Aa},\xi_{Aa})$ are local coordinates
for the $2nk+1$ complex dimensional twistor space $\curly Z$.

These serve only as {\em local} coordinates because there is an additional set of invariants, specific
to the fact that for calorons the Nahm equation has periodic boundary conditions, namely the $k$
invariants of the dual holonomy $h=h_n h_{n-1} \cdot\cdot\cdot h_1$. For these invariants there are
additional constraints, which can be solved {\em locally} in terms of $(\eta_{Aa},\xi_{Aa})$ but not {\em globally}.

The spectral data we have obtained from the Nahm transform recovers the spectral data of magnetic monopoles,
as described in section \ref{magnetic}. The $n$ rational functions are simply
\bea
\label{rationalfunccal}
r_A(\eta_A) = \frac{\psi_A(\eta_A)}{\det(\eta_A-B_A)} = w_{A+1}h_A(\eta_A-B_A)^{-1}u_A\,.
\eea
These rational functions involve all independent local coordinates, thus invoking the same argument as in
section \ref{magnetic} for large separations between the constituent monopoles, we obtain a parametrization
of $\curly M$ in terms of $nk$ 3-dimensional locations and $nk$ phases, describing all together $nk$ monopoles.

Although above we have obtained only a local parametrization of the twistor space, the spectral data that
emerged is sufficient to calculate the exact hyper\kahler metric on $\curly M$. This is done using the generalized
Legendre transform \cite{Hitchin:1986ea} and is work in progress, close to completion.

\chapter{Lattice aspects}
\label{lattice}

Lattice gauge theory provides a non-perturbative framework to investigate the relevance of calorons in a dynamical context.
This is necessary, because the mere existence of our new solutions does not
guarantee that they play any role dynamically. Nevertheless, the non-trivial constituent nature of the solutions is intimately
tied to the average Polyakov loop, which {\em is} set dynamically as discussed in section \ref{finitetemperature}.
Also a semi-classical calculation, based on the 1-loop determinant in a caloron background, provides evidence in
favour of their dynamical relevance \cite{Diakonov:2004jn}.

There are several lattice methods to investigate the low-energy behaviour of a gauge ensemble. Two main methods
are cooling and the study of the zero-modes of the improved chiral Dirac operator.
The first, cooling, only deals with the bosonic sector and
probes the long range correlation in the gauge fields, their topological content in particular
\cite{GarciaPerez:1999hs,Ilgenfritz:2002qs,Ilgenfritz:2003cr}. Our exact results on the
caloron gauge field may be compared with the lattice findings.

The second method probes the fermionic sector and
uses the same philosophy as we have for motivating our study of the fermion zero-modes. That is, the low lying
spectrum of the Dirac operator carries information on the topological content of the underlying gauge field and is
responsible for several non-perturbative aspects of QCD. We have shown how continously changing the boundary condition
in the compact time direction for a zero-mode makes it hop between constituent monopoles. The surprising fact
is that on the background of Monte Carlo generated configurations \dash which are usually very rough \dash
the zero-modes of the improved chiral lattice Dirac operator show very similar behaviour. This property can be used
to probe the monopole content of the gauge fields \cite{Gattringer:2001yu,Gockeler:2001hr}.

The results of our exploratory lattice investigations \dash which nicely show how numerical simulations complement
the study of the formal structures \dash is summarized in this chapter. The computations were done with
modest computer resources and are sufficiently accurate for our purposes. For dedicated large scale Monte
Carlo simulations we refer to the works mentioned above.

\section{Lattice gauge theory}
\label{latticegaugetheory}

In order to simulate $SU(n)$ Yang-Mills theory on a computer first and foremost spacetime and the gauge fields must be discretized.
This is usually done by replacing spacetime by a 4-dimensional square lattice with 4 group valued fields defined
on each oriented link. These $U_\mu(x)$ link variables represent parallel transport \dash or path ordered exponential
\dash from $x$ to $x+\hat\mu$ where $x$ is a site and $x+\hat\mu$ is the next site in the $\mu$ direction. The distance
between $x$ and $x+\hat\mu$ is the lattice spaceing $a$. The gauge potential of the continuum theory
is encoded in the link variables. In order to make the continuum limit more explicit
one can define a lattice gauge potential by
\bea
\label{latticepot}
U_\mu(x) = \exp (a g_{\rm YM} A_\mu(x))\,.
\eea
Gauge transformations $g(x)$ are defined at each site and act on the link variables by
\bea
\label{latticegaugetr}
U_\mu(x) \nyil g(x+\hat\mu) U_\mu(x) g(x)\,,
\eea
the form one would expect from a path ordered exponential from $x$ to $x+\hat\mu$ of a continuum gauge field $A_\mu(x)$. The assignment that
the inverse is associated to a link with opposite orientation is also quite natural. 

From the transformation property above it is clear that the trace of any ordered product of link variables along a closed loop
is gauge invariant in exactly the same way as a Wilson or Polyakov loop is gauge invariant in the continuum. In particular
the trace of a basic plaquette of sides $a$,
\bea
\label{plaquette}
U_{\mu\nu}(x) = U_\nu^{-1}(x) U_\mu^{-1}(x+\hat\nu) U_\nu(x+\hat\mu) U_\mu(x)\,,
\eea
is also gauge invariant. The action proposed by Wilson \cite{Wilson:1974sk} is then defined to be
\bea
\label{latticeaction}
S = \beta \sum \left(1-\frac{1}{n} \re \tr U_{\mu\nu}(x)\right)\,,
\eea
where the sum is taken over every possible basic plaquette in the lattice and $\beta$ is the bare coupling, not to be confused
with $1/k_BT$ as it was used in previous chapters. One can show that in the classical continuum limit, $a\to0$,
the Wilson action agrees with the continuum action corresponding to (\ref{lagrangian}) if one sets $\beta = 2n/g_{\rm YM}^2$. Here
$g_{\rm YM}$ is the bare coupling, which is subject to renormalization in the quantum theory. The above matching is purely a classical
one.

Since we are dealing with finitely many degrees of freedom the Feynman integral is an ordinary \dash yet multiple \dash integral over $SU(n)$.
The number of integration variables grows with the volume and quickly becomes too large for explicit evaluation. Instead, a Monte
Carlo method is employed to compute observables. 

Simulations at zero temperature are done in a 4-dimensional box with the same number of lattice points on each side,
provided the lattice spacing is the same in every direction. Non-zero
temperature is implemented by choosing less number of points in the time direction
than the spatial directions (assuming a homogeneous coupling in all 4 dimensions).
Once the number of lattice sites is fixed, the lattice spacing
and hence the physical temperature is controlled by $\beta=2n/g_{\rm YM}^2$. Increasing $\beta$, i.e.\ lowering $g_{\rm YM}$, reduces
the lattice spacing as dictated by the running of the coupling,
\bea
\label{latticerunning}
a\frac{dg_{\rm YM}}{\!\!da} = \frac{11n}{3} \frac{g_{\rm YM}^3}{16\pi^2}\,,
\eea
at 1-loop order.

We have already met interesting lattice observables, the Wilson or Polyakov loops, but one can easily construct many more. For
our purposes an important one, apart from the Lagrangian, is the topological charge density. A widely used lattice version is
\bea
\label{laptop}
q(x) = -\frac{1}{2^9\pi^2} \varepsilon_{\mu\nu\rho\sigma} \tr U_{\mu\nu}(x) U_{\rho\sigma}(x)\,.
\eea

\section{Phase transition}
\label{latticephasetransition}

The order parameter of the confinement -- deconfinement phase transition is $\langle p(\x) \rangle$, the vacuum expectation value of
the trace of the Polyakov loop (\ref{polyakovloop}). For $SU(2)$ it behaves as
\bea
\label{su2phase}
\langle p(\x) \rangle = \left\{ \begin{array}{l} 0\quad {\rm for}\quad \beta<\beta_c \\ 1\quad {\rm for}\quad \beta\gg\beta_c \end{array} \right.\,,
\eea
and the phase transition is second order.
For high temperatures $\beta>\beta_c$, in the deconfined phase,
the center $\Z_2$ symmetry which interchanges the vacua $p=1$ and $p=-1$ is spontaneously broken.
In the confined phase, $\beta<\beta_c$, this symmetry is restored; see section \ref{finitetemperature}. 

\begin{figure}[ht]
\begin{center}
\includegraphics{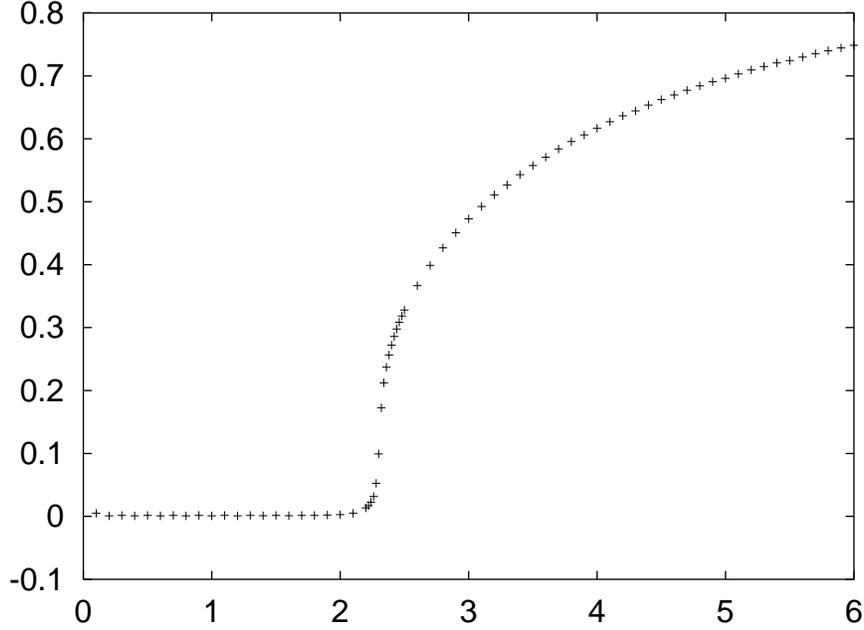}
\end{center}
\caption{\label{phasetransition}\footnotesize Phase diagram of $SU(2)$ gauge theory. The vacuum expectation value $\langle p \rangle$,
half the trace of the average Polyakov loop, is plotted as a function of $\beta$.}
\end{figure}

We will demonstrate the phase transition on a $4\times16\times16\times16$ lattice with periodic boundary conditions
using a simple heatbath Monte Carlo algorithm \cite{creutz}.
The appropriate quantity to be measured is the average $\langle p \rangle$ of $\langle p(\x) \rangle$ over the spatial lattice.
This is plotted in figure \ref{phasetransition} against $\beta$. We see that the phase transition occurs around the
critical point $\beta_c \approx 2.3$.

\section{Cooling}
\label{cooling}

In the process of
cooling, the link variables are altered such that the total action is lowered. Updating every link variable over the lattice
is called a sweep. Sweeping through the lattice many times, one is necessarily
ending up with self-dual solutions as these are minima of the action. However, due to the finite discretization,
instantons can ``fall through'' the lattice if their size becomes smaller then the lattice spacing. Following our
remark on the classical continuum limit, $a\to0$, this effect can only be $O(a^2)$. Concretely, the discretized
1-instanton solution of size $\rho$ has lattice action \cite{GarciaPerez:1993ki}
\bea
\label{latinst}
S({\rm instanton})=2\pi^2\beta\left(1 - \frac{1}{5} \left(\frac{a}{\rho}\right)^2 - \frac{1}{70} \left(\frac{a}{\rho}\right)^4 + \ldots\right)
\eea
for gauge group $SU(2)$, which will be assumed throughout. Indeed, the leading term is the continuum 1-instanton action
$8\pi^2/g_{\rm YM}^2$ and the correction is such that a decreasing size decreases the action. Thus in the process of cooling, where the action
is always lowered, instantons will tend to shrink. Their size will eventually reach the lattice spacing and assuming
a location in between lattice sites, they disappear.

\begin{figure}
\begin{center}
\includegraphics{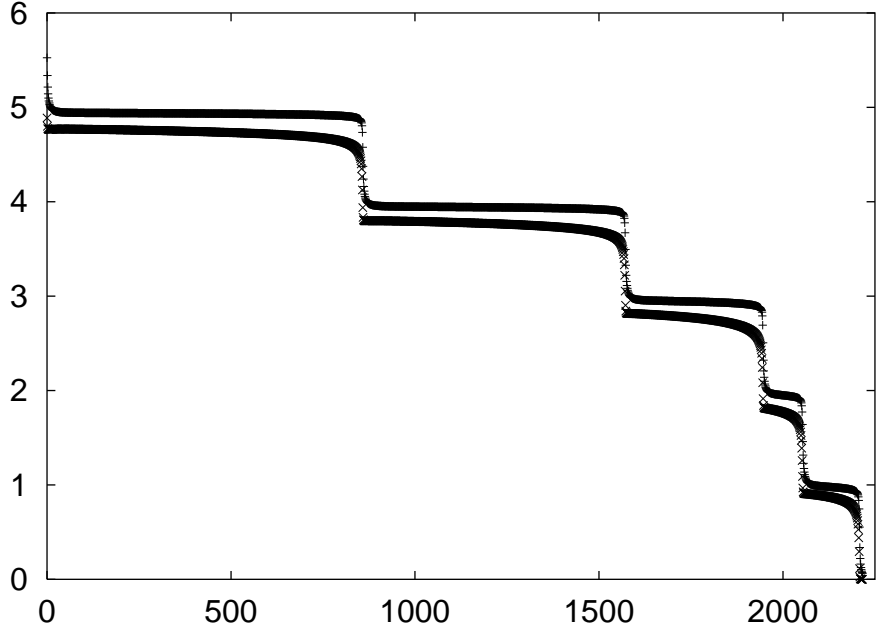}
\end{center}
\caption{\label{instantonfallthrough}\footnotesize Instantons falling through the lattice one by one. The action in units of $8\pi^2/g_{\rm YM}^2$ and
the toplogical charge is plotted against the number of cooling sweeps. The action is always slightly higher then the topological charge.}
\end{figure}

This is illustrated in figure \ref{instantonfallthrough} for a
$16\times16\times16\times16$ lattice with periodic boundary conditions. An initially random \dash or hot \dash configuration was cooled to reach
a configuration with roughly 6 instantons, after which we have plotted both the action (measured in units of $8\pi^2/g_{\rm YM}^2$) and the topological
charge against the number of cooling sweeps. Each sharp drop by one unit corresponds to one instanton falling through the lattice.
Between the falls the topological stability is reflected by long
plateaus, where the action and the topological charge roughly agree, indicating a self-dual
configuration. At the final stage the last remaining instanton disappears as well, leaving the trivial vacuum behind with each
plaquette being the identity of $SU(2)$.

If one is to investigate stable self-dual solutions on the lattice, the above shrinking phenomenon must be eliminated.
The following action serves this purpose,
\bea
\label{impaction}
S_\varepsilon = \beta\,\frac{4-\varepsilon}{3}\sum\left(1-\half\tr\plaqa\right)+\beta\,\frac{\varepsilon-1}{48}\sum\left(1-\half\tr\nagyplaqa\right)\,,
\eea
where the first term represents the Wilson action (\ref{latticeaction}), while the second means a
sum over every $2\times 2$ plaquette. Note that for $SU(2)$ all traces are real. Clearly, $S_1$ is the Wilson action.
The usefulness of $S_\varepsilon$ is that the discretized 1-instanton will have an action
\bea
S_\varepsilon({\rm instanton}) = 2\pi^2\beta\left(1-\frac{\varepsilon}{5}\left(\frac{a}{\rho}\right)^2+
\frac{12-15\varepsilon}{210}\left(\frac{a}{\rho}\right)^4+\ldots\right)\,,
\eea
which shows that once a negative value is picked for $\varepsilon$, an instanton will be stable against shrinking \cite{GarciaPerez:1993ki}.
The value $\varepsilon=0$ minimizes the lattice artefacts to this order \dash thus $S_0$ is called the improved action \dash 
and also leads to stable instantons, but the stabilizing force is
greater with a negative value of $\varepsilon$. The choice $\varepsilon=-1$ is called over-improvement.

Over-improved cooling stabilizes instantons so well, that starting from a random initial configuration with very high action, one can easily
get stuck at multi-instanton configurations with high topological charge. Thus in order to investigate self-dual configurations
with charge 1, 2 or 3, it is most useful to first apply Wilson cooling and only switch to $\varepsilon=-1$ when the action is
already below 2, 3 or 4, in units of $8\pi^2/g_{\rm YM}^2$.

\begin{figure}[ht]
\begin{center}
\includegraphics[width=8cm, height=8cm]{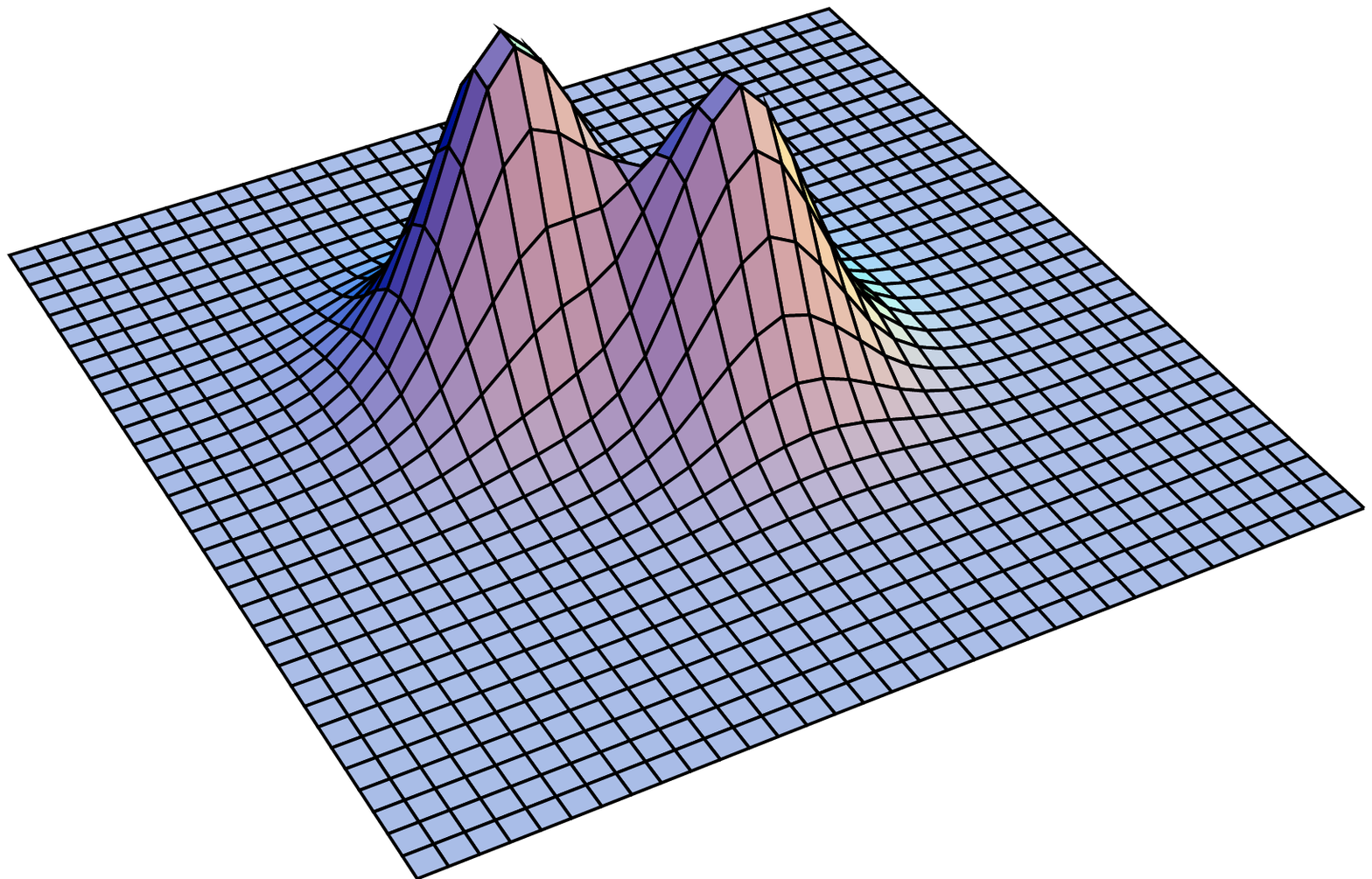}
\includegraphics[width=8cm, height=8cm]{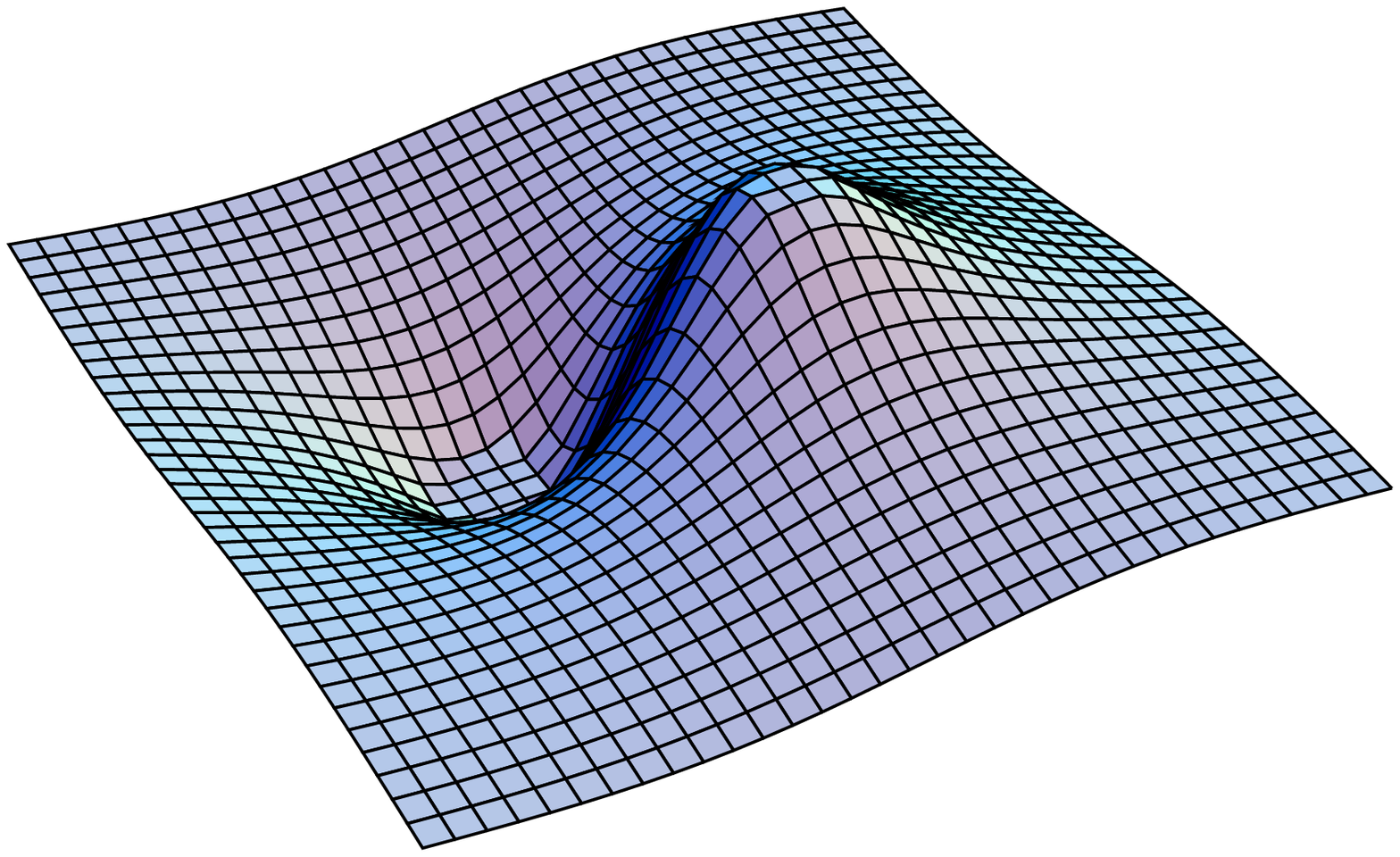}
\end{center}
\caption{\label{charge1lattice}\footnotesize An $SU(2)$ caloron of unit charge on the lattice,
obtained by over-improved cooling. The topological charge density is on the left,
half the trace of the Polyakov loop on the right, showing that the lump to the left is a monopole of charge -1 and the
lump to the right is a monopole of charge +1.}
\end{figure}

At finite temperature the situation is the same as at zero temperature. Over-improved cooling, however, will have an additional
advantageous feature, it will tend to separate the constituent monopoles of opposite charge. Thus not only the calorons will be stable against
falling through the lattice, but also their constituent nature will be revealed. We have performed simulations
on a $4\times16\times16\times16$ lattice, with periodic boundary conditions in all 4 directions.

In figure \ref{charge1lattice} a charge 1 caloron is shown. The action density is summed in one direction and plotted in the
remaining plain at a fixed time slice. It was obtained by first Wilson cooling a random initial
configuration for 665 sweeps until the action decreased below $1.8\times8\pi^2/g_{\rm YM}^2$. Then it was followed by
1200 over-improved cooling sweeps and the improved action $S_0$ became 1.014 in units of $8\pi^2/g_{\rm YM}^2$, whereas the topological charge 0.877. The
reason for the topological charge only being an integer up to 12\% is that we used the simple operator (\ref{laptop})
which has $O(a^2)$ lattice artefacts, whereas the improved action $S_0$ was deliberately chosen such, that only $O(a^4)$
artefacts are present.

\begin{figure}
\begin{center}
\includegraphics[width=8cm, height=8cm]{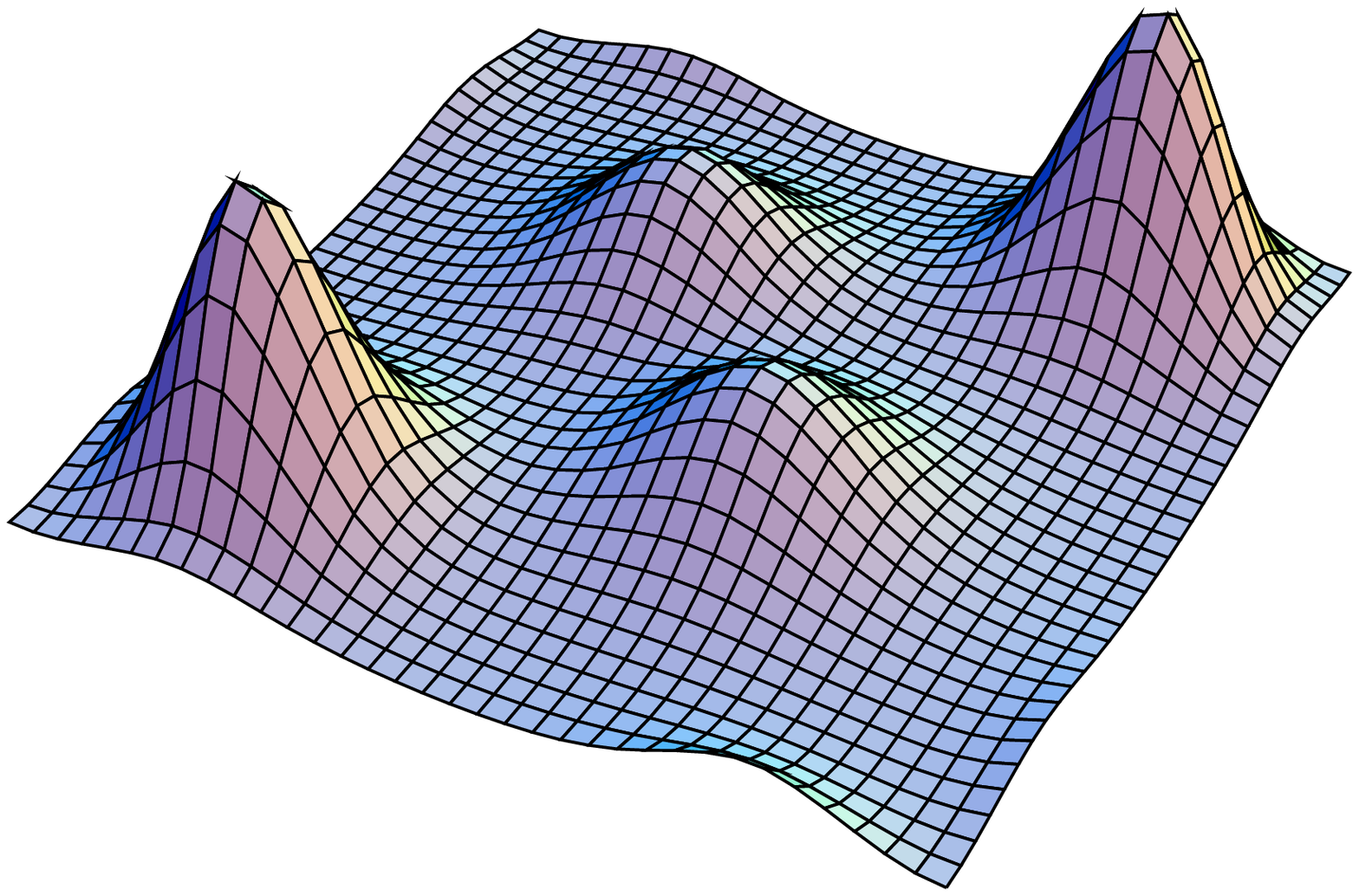} 
\includegraphics[width=8cm, height=8cm]{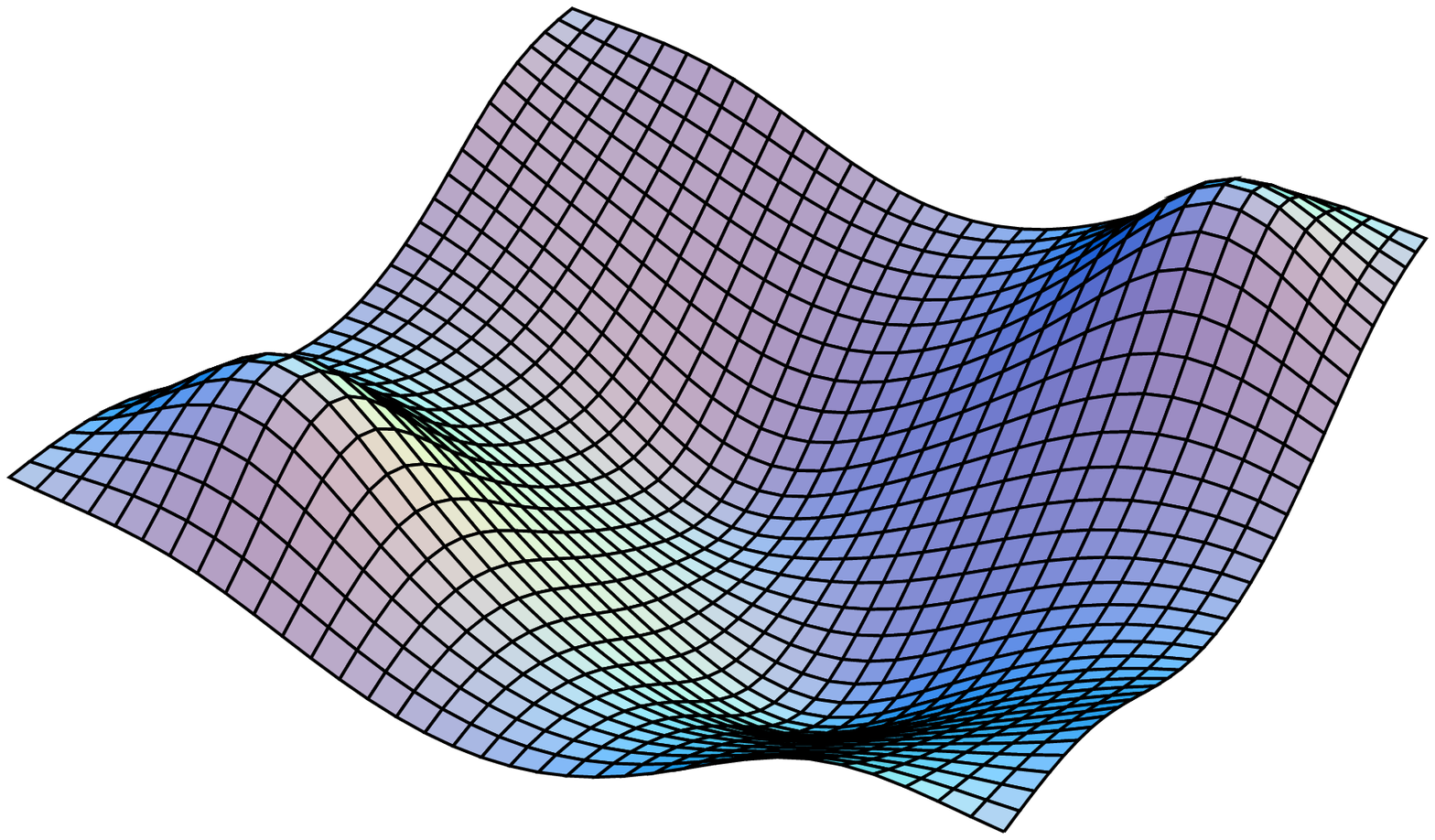}
\end{center}
\caption{\label{charge2lattice}\footnotesize An $SU(2)$ caloron of charge 2 with well-separated constituents on the lattice,
obtained by over-improved cooling. The topological charge density is on the left,
half the trace of the Polyakov loop on the right, which shows that the two smaller lumps are monopoles of charge -1 and the
two higher lumps are monopoles of charge +1.}
\end{figure}

Higher charge calorons \dash discussed in great detail in the previous chapters \dash can be illustrated as well.
Figure \ref{charge2lattice} shows a charge 2 caloron. The topological charge of this configuration is 1.82, and has
improved action 2.001 in units of $8\pi^2/g_{\rm YM}^2$. It was obtained by first Wilson cooling for 580 sweeps until the action dropped below
$2.8\times8\pi^2/g_{\rm YM}^2$ and then switching to over-improved cooling for 2150 sweeps.

This concludes the presentation of our numerical studies, complementing the exact results of the preceding chapters.
For more extensive investigations, see \cite{Ilgenfritz:2004ws,Ilgenfritz:2004zz,Bruckmann:2004ib}.

\chapter{Concluding remarks}
\label{concluding}

We have discussed the explicit construction
of multiply charged caloron solutions with non-trivial holonomy. For gauge group $SU(n)$ and topological charge $k$,
the non-trivial Polyakov loop causes the caloron to dissociate into $nk$ massive magnetic
monopoles. Thus a caloron of charge $k$ should not be seen as an approximate superposition of $k$ charge 1 calorons, but
rather as the approximate superposition of $nk$ constituent monopoles, each carrying fractional topological charge.
The constituents have their own identity, they do not know to which charge 1 caloron they belong, the non-linear
superposition of all of them gives rise to the full 4 dimensional gauge configuration. The action density clearly shows $nk$ lumps,
once the constituents are separated far enough from each other. The $nk$ monopoles come in $n$ distinct types,
each being charged under a different $U(1)$ subgroup. 

Along with the bosonic sector we have investigated the fermionic sector as well, and have found the
zero-modes of the Dirac operator in the caloron background. The compact time direction corresponding to
finite temperature necessitates a choice of boundary condition for the fermions and the index theorem
dictates the existence of $k$ zero-modes for every such choice. In order to describe
the correct partition function, physical fermions are required to be anti-periodic, however, for diagnostic purposes one may impose
for the zero-modes periodicity up to an arbitrary phase. The usefulness of employing zero-modes to identify
the constituent monopoles lies in the fact that by fixing this phase, the zero-modes localize to only one type of monopole and are blind
to the remaining $n-1$ types. By continously changing this phase one can detect all the $nk$ constituent
monopoles, where the $k$ zero-modes jump from one type to the other whenever the phase
passes through one of the eigenvalues of the asymptotic Polyakov loop.
In this way one can analyse each monopole separately, effectively deconstructing the caloron into its bare constituents.

An important tool we have employed to clearly see to what extent the constituents can be described as point-like objects,
is the abelian limit. In this limit all exponential contributions
originating from the non-abelian cores are neglected, only algebraic tails survive, giving rise to abelian fields.
The masses of the monopoles in this limit are effectively infinite, and the zero-modes localize to the non-abelian
cores.  These collapse to zero size, thus the zero-mode density exhibits a singular behaviour with singularity structure tracing the
cores. The remnants of the cores in the abelian limit are in general extended as seen by the zero-modes,
however, we have established that once the separation between constituents is large, they become point-like. 

The moduli space of calorons factorizes into $nk$ copies of $S^1\times\R^3$ for large separation,
confirming the right description in terms of $nk$ phases and $nk$ 3-dimensional locations. This is
in contrast with instantons on $\R^4$ where the correct description is in terms of $k$ scale parameters, $k$ 4-dimensional
locations and relative gauge orientations. As to the algebraic geometry of the caloron moduli space we have established
a correspondence with stable holomorphic bundles on the projective plane that are trivial on two complex lines.
This was derived from a similar correspondence between instantons on $\R^4$ and bundles that are trivial on one complex line,
which ultimately is a result of the compactification of $\R^4 \simeq \C^2$ to $\cp$.
This implies that the appropriate way of approaching the problem is not through the factorization $\R^4/\Z \simeq S^1\times \R^3$,
but rather through the embeddings $S^1\times\R^3 \simeq \C^*\times\C \subset \C^2 \subset \cp$, at least from
a holomorphic point of view.

This observation may turn out to be relevant for the study of doubly periodic instantons as well \cite{Jardim:1999mi,Ford:2000zt}. They are defined on
$T^2\times \R^2$ and traditionally \dash especially in the Nahm context \dash are studied through the factorization
$\R^4/\Z^2 \simeq T^2\times \R^2$. Now since $T^2\times \R^2 \simeq \C^*\times\C^*$ which again
can be compactified to $\cp$ by the sequence of embeddings $\C^*\times\C^* \subset \C^*\times\C\subset\C^2\subset\cp$,
we suspect the existence
of a correspondence between the moduli space of doubly periodic instantons and stable holomorphic bundles over
$\cp$ with some additional triviality constraints. 

We believe that our results on the moduli space of calorons may be useful for compactified supersymmetric gauge theories as well \cite{Davies:1999uw}.
Similarly to the formulation on $\R^4$, a number of correlation functions are saturated by instantons \dash
or calorons in the compactified case. The computation of these correlation functions
involve integration over the moduli space of instantons, which in fact can be performed by localization techniques \cite{Bellisai:2000bc}.
Understanding the metric on the caloron moduli
space is a necessary ingredient to compute these correlation functions for the compactified theories. If one
wishes to compute the contribution from every topological sector, a summation over charge is necessary, thus
the metric is needed for the moduli space of calorons with arbitrary charge. This is where we believe our results
on the twistor description of the moduli space may turn out to be useful.

We have saved our primary motivation to the end. Understanding permanent quark
confinement in QCD remains to be a challenging task. What we have set out to investigate was the
nature of the fundamental topological excitations in the confined phase. We have found that instantons
are composites and are made up of monopoles due to the non-trivial Polyakov loop background that is present
in the confined phase. The instanton liquid model \cite{Schafer:1996wv} has been very successful in
demonstrating chiral symmetry breaking and we believe that a similar model, but now with the constituent
monopoles taking the place of instantons, holds considerable promise for the future.
Especially, because the presence of magnetic monopoles in the QCD vacuum is in fact required for the dual superconductor
picture to emerge \cite{Mandelstam:1974pi,'tHooft:1977hy}. In the 3 dimensional Yang-Mills-Higgs model
monopoles have been shown to be responsible for confinement long ago \cite{Polyakov:1976fu}.

It is well-known that the dilute gas assumption of the instanton liquid model breaks down for large instantons
and also that small instantons are not sufficient for explaining confinement. In our constituent monopole
picture the scale parameter of large instantons is converted to the separation between monopoles. It
is not unlikely that the density of constituents at low temperature is so high that they form a coherent
background and as such will no longer be recognized as separate lumps. With high quark density leading
to deconfinement, it may perhaps be that a high constituent monopole density will result in confinement.

Whether or not constituent monopoles play a role in a dynamical context, remains a difficult non-perturbative question.
The natural testing ground is the lattice, which shows growing evidence in favour. Simulations were performed
very much along the lines of our investigations, first identifying the semi-classical
gauge structure \cite{Ilgenfritz:2002qs}, followed by studying the localization properties
of chiral fermion zero-modes \cite{Gattringer:2002wh,Gattringer:2002tg}. These studies \dash
some of which we have also touched in an exploratory fashion \dash seem to imply that constituent monopoles are the relevant degrees
of freedom in the confined phase.

Could it be that the non-perturbative dynamics of constituent monopoles is such, that it causes the quarks
to confine? At present, we do not have an answer to this question.

It should be noted that the idea of instantons having more fundamental building blocks is by no means new. The moduli
space of $k$-instantons in the 2-dimensional $\C{\rm P}^{n-1}$ model can be parametrized by $nk$ complex numbers (the remaining
moduli are not relevant for what follows). All of these have an interpretation as 2-dimensional locations for so-called
instanton quarks \cite{Belavin:1979fb}, or in our terminology, constituents.
The 1-loop determinant in the instanton background defines a statistical
ensemble for these $nk$ instanton quarks which may be mapped to a 2-dimensional Coulomb gas \cite{Fateev:1978xn,Berg:1979uq}.
Since the topological charge is $k$ and there are $nk$ constituents, the topological charge carried by each Coulomb particle is
fractional. 

Despite the fact that the 2-dimensional $\C{\rm P}^{n-1}$ model shares some features with QCD, such as
confinement, a similar picture is difficult to arrive at in gauge theories. Nevertheless, the results presented in this thesis
further suggest that the description in terms of instanton quarks is indeed the relevant one. 
It has been argued recently that in the confined phase of QCD the effective low energy
theory is of a Sine-Gordon type \cite{Halperin:1997bs, Halperin:1998rc}, which also has a Coulomb gas representation\cite{Jaimungal:1999wn, Son:2001jm}.
Just as for the $\C{\rm P}^{n-1}$ model, but now in four dimensions, the Coulomb
particles carry fractional topological charge \cite{Toublan:2005tn}.

We are hopeful that our results and related work mentioned above will ultimately tie together into a consistent picture of
a confining QCD vacuum.

\chapter*{Appendix}
\addcontentsline{toc}{chapter}{Appendix}
\markboth{Appendix}{Appendix}
\label{appendix}

We present below the polynomial whose root is $\curly V$ for charge 2. It follows from our discussion that this funcion $\curly V(\x)$
is harmonic almost everywhere, see section \ref{su2abelianzeromodes} for details. We have
\bea
\label{vvvv}
a_0 + a_1 \curly V^2 + a_2 \curly V^4 + a_3 \curly V^6 = 0\nn\,,
\eea
with

\begin{equation}\begin{array}{l}
a_3=\left(2x_2^6x_1^2\k^4-6x_3^6x_1^2\k^2+4x_2^2x_3^6\k^4+x_2^4x_1^4\k^4+4x_2^4x_1^4\k^2+8x_2^2x_3^4\k^4+6x_2^4x_3^2\k^6-\right.\nn\vspace{0.15cm}\\\vspace{0.15cm}
4x_2^2x_3^2\k^6-4\k^2x_3^4x_2^2+6x_1^4\k^4-2x_2^4x_1^2\k^4-2\k^6+\k^4+\k^8-6x_1^4\k^2+2x_1^2\k^2-6\k^4x_1^2+\nn\\\vspace{0.15cm}
4\k^6x_1^2+4x_1^6\k^2+4x_2^2\k^4-6x_2^4\k^4+x_2^4\k^8+x_2^8\k^4-2\k^6x_2^6+4x_2^6\k^4+2x_2^2\k^8-6x_2^4\k^6-\nn\\\vspace{0.15cm}
6\k^6x_2^2+2x_1^4x_2^2+2x_2^2x_1^6\k^2+2x_1^2\k^6x_2^4+6x_2^2x_1^4\k^4+6x_2^2x_1^2\k^6+x_3^4x_1^4\k^4+2x_3^6\k^4x_1^2-\nn\\\vspace{0.15cm}
6x_3^2x_1^4\k^4-4\k^4x_1^2x_3^4+2x_3^4\k^6x_1^2+8\k^4x_3^2x_1^2+6\k^4x_3^4x_2^4-2x_2^2x_3^2\k^8-6x_2^2\k^6x_3^4-6x_2^2x_3^6\k^2-\nn\\\vspace{0.15cm}
10x_2^2x_3^4x_1^2\k^2+6x_2^2x_3^4x_1^2\k^4+6x_2^4x_3^2x_1^2\k^4-2x_2^2x_3^2x_1^4\k^2+2x_2^2x_3^2x_1^4\k^4-x_1^4-2x_1^6+x_1^8-\nn\\\vspace{0.15cm}
6x_2^2x_1^2\k^2+2x_2^6x_1^2\k^2-2x_1^4\k^2x_2^2+10\k^4x_2^2x_1^2+6x_2^4x_1^2\k^2+2x_1^6x_2^2+x_1^4x_2^4-6x_1^2x_3^4-2x_3^2\k^2+\nn\\\vspace{0.15cm}
2x_3^2\k^4+2x_3^4\k^2+2\k^6x_3^2+2\k^2x_3^6+2\k^4x_3^6-2x_3^2\k^8-2x_3^6\k^6-6x_3^4\k^4+x_3^4\k^8+\k^4x_3^8+\nn\\\vspace{0.15cm}
2x_3^4\k^6-2\k^2x_3^8+2x_1^2x_3^2-6x_1^4x_3^2+4x_3^6x_1^2+4x_1^6x_3^2+6x_1^4x_3^4+2x_2^2x_3^4+2x_2^2x_3^6+x_2^4x_3^4+\nn\\\vspace{0.15cm}
4x_2^2x_1^2x_3^2\k^6-2x_1^2x_2^4x_3^2\k^2-6\k^2x_3^2x_2^2x_1^2-6x_2^2x_3^2x_1^2\k^4-2x_3^6+x_3^8+x_3^4-8x_3^4x_1^2\k^2-\nn\\\vspace{0.15cm}
2x_3^2x_1^6\k^2-6x_1^4x_3^4\k^2+10x_1^4x_3^2\k^2+4x_3^2x_1^2\k^2+2x_2^6x_3^2\k^2-6\k^2x_3^2x_2^4+10x_2^4x_3^2\k^4-\nn\\\vspace{0.15cm}
\left.6x_2^2x_3^2\k^2+8\k^4x_2^2x_3^2-6x_2^4x_3^4\k^2+6x_2^2x_3^4x_1^2+2x_2^4x_3^2x_1^2+4x_1^2x_2^2x_3^2+6x_1^4x_2^2x_3^2\right)^2\nn\vspace{0.15cm}
\end{array}\end{equation}

\newpage
\vspace{-3cm}
\begin{equation}\begin{array}{l}
a_2=-4\left(2x_2^6x_1^2\k^4-6x_3^6x_1^2\k^2+4x_2^2x_3^6\k^4+x_2^4x_1^4\k^4+4x_2^4x_1^4\k^2+8x_2^2x_3^4\k^4+x_2^8\k^4-\right.\nn\vspace{0.15cm}\\\vspace{0.15cm}
6x_3^2x_1^2\k^6+4x_2^6x_3^2\k^4-6x_2^4x_3^2\k^6-4x_2^2x_3^2\k^6-4\k^2x_3^4x_2^2+6x_1^4\k^4-2x_2^4x_1^2\k^4-2\k^6+\nn\\\vspace{0.15cm}
\k^4+\k^8-6x_1^4\k^2+2x_1^2\k^2-6\k^4x_1^2+4\k^6x_1^2+4x_1^6\k^2+4x_2^2\k^4+6x_2^4\k^4+x_2^4\k^8-\nn\\\vspace{0.15cm}
2\k^6x_2^6+4x_2^6\k^4+2x_2^2\k^8-6x_2^4\k^6-6\k^6x_2^2+2x_1^4x_2^2+2x_2^2x_1^6\k^2+2x_1^2x_2^4\k^6+6x_2^2x_1^4\k^4+\nn\\\vspace{0.15cm}
6x_2^2x_1^2\k^6+x_3^4x_1^4\k^4+2x_3^6x_1^2\k^4-6x_3^2x_1^4\k^4-4\k^4x_1^2x_3^4+2x_3^4x_1^2\k^6+8\k^4x_3^2x_1^2+6\k^4x_3^4x_2^4+\nn\\\vspace{0.15cm}
2x_2^2x_3^2\k^8-6x_2^2x_3^4\k^6-6x_2^2x_3^6\k^2-10x_2^2x_3^4x_1^2\k^2+2x_2^2x_3^4+2x_2^2x_3^6+6x_2^2x_3^4x_1^2\k^4+\nn\\\vspace{0.15cm}
6x_2^4x_3^2x_1^2\k^4-2x_2^2x_3^2x_1^4\k^2+2x_2^2x_3^2x_1^4\k^4+x_1^4-2x_1^6+x_1^8+8x_3^4x_1^2\k^2+6x_2^2x_1^2\k^2+2x_2^6x_1^2\k^2-\nn\\\vspace{0.15cm}
2x_1^4x_2^2\k^2-10\k^4x_2^2x_1^2+6x_2^4x_1^2\k^2+2x_1^6x_2^2+x_1^4x_2^4-6x_1^2x_3^4-2x_3^2\k^2+2x_3^2\k^4+\nn\\\vspace{0.15cm}
2x_3^4\k^2+2\k^6x_3^2+2\k^2x_3^6+2\k^4x_3^6-2x_3^2\k^8-2x_3^6\k^6-6x_3^4\k^4+x_3^4\k^8+\k^4x_3^8+2x_3^4\k^6-\nn\\\vspace{0.15cm}
2\k^2x_3^8+2x_1^2x_3^2-6x_1^4x_3^2+4x_3^6x_1^2+4x_1^6x_3^2+6x_1^4x_3^4+x_2^4x_3^4+4x_2^2x_1^2x_3^2\k^6-2x_1^2x_2^4x_3^2\k^2-\nn\\\vspace{0.15cm}
6\k^2x_3^2x_2^2x_1^2-6x_2^2x_3^2x_1^2\k^4-2x_3^6+x_3^8+x_3^4-2x_3^2x_1^6\k^2-6x_1^4x_3^4\k^2+10x_1^4x_3^2\k^2-\nn\\\vspace{0.15cm}
4x_3^2x_1^2\k^2-2x_2^6x_3^2\k^2-6\k^2x_3^2x_2^4+10x_2^4x_3^2\k^4-6x_2^2x_3^2\k^2+8\k^4x_2^2x_3^2-6x_2^4x_3^4\k^2+\nn\\\vspace{0.15cm}
\left.6x_2^2x_3^4x_1^2+2x_2^4x_3^2x_1^2+4x_1^2x_2^2x_3^2+6x_1^4x_2^2x_3^2\right)
\left(3x_2^2x_3^4\k^4-2x_2^2x_3^2\k^6-4\k^2x_3^4x_2^2+x_2^4x_1^2\k^4+\right.\nn\\\vspace{0.15cm}
2x_1^4\k^2-x_1^2\k^2+\k^4x_1^2+x_2^2\k^4+2x_2^4\k^4+x_2^6\k^4-x_2^4\k^6-\k^6x_2^2+x_1^4x_2^2+\k^4x_1^2x_3^4-\k^4x_3^2x_1^2-\nn\\\vspace{0.15cm}
x_1^4+x_1^6+x_2^2x_1^2\k^2+2x_1^4x_2^2\k^2+\k^4x_2^2x_1^2+2x_2^4x_1^2\k^2+3x_1^2x_3^4+x_3^2\k^2-2x_3^2\k^4+x_3^4\k^2+\nn\\\vspace{0.15cm}
\k^6x_3^2-2\k^2x_3^6+\k^4x_3^6+x_3^4\k^4-x_3^4\k^6-2x_1^2x_3^2+3x_1^4x_3^2+x_2^2x_3^4-2\k^2x_3^2x_2^2x_1^2+\nn\\\vspace{0.15cm}
2x_2^2x_3^2x_1^2\k^4+x_3^6-x_3^4-4x_3^4x_1^2\k^2-2x_1^4x_3^2\k^2+3x_3^2x_1^2\k^2-2\k^2x_3^2x_2^4+3x_2^4x_3^2\k^4-\nn\\\vspace{0.15cm}
\left.x_2^2x_3^2\k^2+3\k^4x_2^2x_3^2+2x_1^2x_2^2x_3^2\right)\nn\vspace{0.15cm}
\end{array}\end{equation}

\newpage

\begin{equation}\begin{array}{l}
a_1=-48x_1^4x_2^6\k^4+16x_2^8x_3^2\k^6-144x_2^2x_3^6\k^4-96x_2^4x_1^4\k^4-48x_2^2x_3^4\k^4+48x_3^2x_1^2\k^6+\nn\vspace{0.15cm}\\\vspace{0.15cm}
192x_1^4x_3^4\k^6-96\k^6x_1^4x_3^2-16\k^{12}x_2^2x_3^2+48x_2^4x_3^2\k^6+48\k^{10}x_2^4x_3^2-96x_3^6x_2^4\k^4-\nn\\\vspace{0.15cm}
48x_2^6x_3^2\k^8+48x_1^6x_3^4\k^4+48\k^{10}x_2^2x_3^4+16x_2^2x_3^2\k^6-144x_3^4x_2^2\k^8+16x_2^2x_3^8\k^6-\nn\\\vspace{0.15cm}
48x_1^6x_2^4\k^6-16x_1^8x_2^2-16x_2^2x_3^8-48x_2^6x_3^4\k^4+96x_1^6x_2^4\k^4+32x_1^8x_2^2\k^2-96\k^8x_1^4x_2^4-\nn\\\vspace{0.15cm}
48x_2^2x_1^6\k^2+96\k^8x_1^2x_2^4-48x_1^2x_2^4\k^6+192x_1^4x_2^4\k^6-48x_2^6x_1^2\k^6-16x_2^8x_1^2\k^{10}-\nn\\\vspace{0.15cm}
16\k^{10}x_2^2x_1^2-48\k^{10}x_2^4x_1^2-48x_2^2x_1^4\k^4-48\k^8x_2^2x_1^4+96x_2^6x_1^4\k^6-48x_1^6x_2^4\k^2-\nn\\\vspace{0.15cm}
16x_2^2x_1^2\k^6-48x_1^6x_2^2\k^6-16x_1^8x_2^2\k^4+96x_1^4x_2^2\k^6+32\k^8x_2^2x_1^2+32x_2^8x_1^2\k^8+\nn\\\vspace{0.15cm}
96x_2^6x_1^2\k^8-48x_2^6x_1^4\k^8-48x_1^2x_2^6\k^{10}-16x_2^8x_1^2\k^6+96x_1^6x_2^2\k^4+48\k^{10}x_2^2x_3^2-\nn\\\vspace{0.15cm}
96x_3^4x_1^4\k^4-48\k^{10}x_3^4x_1^2-144x_3^6x_1^2\k^8+144x_3^6x_1^2\k^6-96x_1^4x_3^6\k^6+144\k^8x_1^2x_3^4-\nn\\\vspace{0.15cm}
96\k^8x_1^4x_3^4-48x_3^6x_1^2\k^4+16\k^{10}x_1^2x_3^2+48x_1^4x_3^6\k^4+48x_3^2x_1^4\k^4-16x_3^8x_1^2\k^{10}+\nn\\\vspace{0.15cm}
48x_1^4x_3^6\k^8+48\k^4x_1^2x_3^4+48x_1^6x_3^2\k^6-48x_3^8x_1^2\k^6-144x_3^4x_1^2\k^6+16x_1^2x_3^8\k^4+\nn\\\vspace{0.15cm}
48\k^{10}x_1^2x_3^6+48\k^8x_3^2x_1^4-48x_1^6x_3^4\k^6+16x_1^8x_3^2\k^4-16\k^4x_3^2x_1^2+48x_3^8x_1^2\k^8-\nn\\\vspace{0.15cm}
48\k^8x_3^2x_1^2-48x_1^6x_3^2\k^4+48x_3^8x_2^2\k^2-96\k^4x_3^4x_2^4-48x_3^8x_2^2\k^4-96x_3^2x_2^4\k^8+\nn\\\vspace{0.15cm}
48x_2^6x_3^4\k^6-48x_3^6x_2^2\k^8+48x_2^4x_3^6\k^6-48x_2^2x_3^2\k^8-96x_2^4x_3^4\k^8+144x_2^2x_3^4\k^6+\nn\\\vspace{0.15cm}
48x_2^2x_3^6\k^2+48x_2^4x_3^6\k^2+48x_2^6x_3^2\k^6+192x_2^4x_3^4\k^6+144x_2^2x_3^6\k^6-64x_1^6x_2^2x_3^2-\nn\\\vspace{0.15cm}
96x_2^2x_3^4x_1^4-64x_2^2x_3^6x_1^2+32x_1^6x_2^2x_3^2\k^6-96x_2^4x_1^2x_3^4\k^{10}-96x_1^4x_3^4x_2^2\k^4+\nn\\\vspace{0.15cm}
240x_1^4x_3^4x_2^2\k^2-48x_2^6x_3^2x_1^2\k^6+144x_2^6x_1^2x_3^2\k^8+176x_2^2x_1^2x_3^6\k^8-96x_1^4x_2^2x_3^2\k^6-\nn\\\vspace{0.15cm}
112x_2^2x_3^6x_1^2\k^6-64x_2^2x_1^2x_3^6\k^{10}-96x_2^4x_3^4x_1^2\k^6+240x_2^4x_1^2x_3^4\k^8+96x_1^4x_3^2x_2^4\k^4-\nn\\\vspace{0.15cm}
112x_1^2x_3^6x_2^2\k^4+32x_1^2x_2^6x_3^2\k^4-64x_2^6x_1^2x_3^2\k^{10}-48x_1^4x_2^4x_3^2\k^2+176x_3^6x_2^2x_1^2\k^2-\nn\\\vspace{0.15cm}
48x_1^6x_2^2x_3^2\k^4+144x_1^6x_2^2x_3^2\k^2+192x_2^2x_1^2x_3^4\k^6+48x_2^4x_3^4x_1^2\k^2-96x_1^2x_3^4x_2^4\k^4+\nn\\\vspace{0.15cm}
48x_2^2x_3^4x_1^2\k^2-144x_2^2x_3^4x_1^2\k^4+48\k^{10}x_1^2x_2^2x_3^4+48x_1^4x_2^2x_3^4\k^8-144\k^8x_1^2x_2^2x_3^4-\nn\\\vspace{0.15cm}
96x_2^2x_1^4x_3^4\k^6+96x_2^4x_1^4x_3^2\k^6-48x_1^4x_2^4x_3^2\k^8-96x_2^4x_1^2x_3^2\k^6+48\k^8x_1^4x_2^2x_3^2-\nn\\\vspace{0.15cm}
48\k^{10}x_1^2x_2^4x_3^2+96\k^8x_1^2x_2^4x_3^2+48x_2^4x_3^2x_1^2\k^4-48x_2^2x_3^2x_1^4\k^2+96x_2^2x_3^2x_1^4\k^4\nn\\\nn\\\nn\\
a_0=64x_2^2x_3^2x_1^2\k^4(\k-1)^2(\k+1)^2\,.\nn
\end{array}\end{equation}
\newpage

\setcounter{secnumdepth}{-1}

\chapter{Acknowledgements}
\label{acknowledgements}

I would like to thank my supervisor Pierre van Baal for his scientific guidance, encouragement and ideas that
were invaluable for writing this thesis. I am very grateful for his scrupulous reading of the manuscript contributing substantially to a crisp
and sound presentation of my results. I would like to acknowledge him as well as Falk Bruckmann for the years long
collaboration and numerous enlightening discussions on the topics covered in this thesis. 

In addition, I would like to thank Ernst-Michael Ilgenfritz, Zolt\'an K\'ad\'ar,
Bal\'azs Szendr\H oi, Mike Teper and Ariel Zhitnitsky for stimulating discussions
from which I have benefited greatly. Ell\'ak Somfai has been very patient in explaining the secrets of Linux and C, both his
and Michael's help were especially useful while writing the lattice programs.

\end{document}